\documentclass[rmp,twocolumn,amsmath,amssymb]{revtex4}

\usepackage[dvips]{color}

\usepackage{graphicx}
\usepackage{epsfig}
\usepackage{bm}

\begin{document}

\title{Fermi gases in one dimension: From Bethe Ansatz to experiments}

\author{Xi-Wen Guan}
\email{xwe105@physics.anu.edu.au}

\affiliation{State Key Laboratory of Magnetic Resonance and Atomic and Molecular Physics, Wuhan Institute of Physics and Mathematics, Chinese Academy of Sciences, Wuhan 430071, China}
\affiliation{Department of Theoretical Physics, Research School of Physics and Engineering, Australian National University, Canberra ACT 0200, Australia}

\author{Murray T. Batchelor}
\email{murray.batchelor@anu.edu.au}

\affiliation{Centre for Modern Physics, Chongqing University, Chongqing 400044, China}
\affiliation{Mathematical Sciences Institute and Department of Theoretical Physics, Research School of Physics and Engineering, Australian National University, Canberra ACT 0200, Australia}

\author{Chaohong Lee}
\email{lichaoh2@mail.sysu.edu.cn}

\affiliation{State Key Laboratory of Optoelectronic Materials and Technologies, School of Physics and Engineering, Sun Yat-Sen University, Guangzhou 510275, China}
\affiliation{Nonlinear Physics Centre and ARC Centre of Excellence for Quantum-Atom Optics, Research School of Physics and Engineering, Australian National University, Canberra ACT 0200, Australia}

\begin{abstract}
This article reviews theoretical and experimental developments for one-dimensional Fermi gases.
Specifically, the experimentally realized two-component delta-function interacting Fermi gas -- the Gaudin-Yang model -- and its generalisations to
multi-component Fermi systems with larger spin symmetries.
The exact results obtained for Bethe ansatz integrable models of this kind enable the study of the nature and microscopic origin of a wide range of
quantum many-body phenomena driven by spin population imbalance, dynamical interactions and magnetic fields.
This physics includes Bardeen-Cooper-Schrieffer-like pairing, Tomonaga-Luttinger liquids, spin-charge separation,
Fulde-Ferrel-Larkin-Ovchinnikov-like pair correlations,  quantum criticality and scaling, polarons and the few-body  physics of the trimer state (trions).
The fascinating interplay between exactly solved models and experimental developments in one dimension promises to yield further
insight into the exciting and fundamental physics of interacting Fermi systems.
\end{abstract}

\date{\today}

\maketitle

\tableofcontents

\section{Introduction}
\label{sec:intro}

Fundamental quantum many-body systems involve the interaction of bosonic and/or fermionic particles. 
The spin of a particle makes it behave very differently at ultracold temperatures below the degeneracy temperature.
There are thus fundamental differences between the properties of bosons and fermions.
However, as bosons are not subject to the Pauli exclusion principle,
they can collapse under suitable conditions into the same
quantum groundstate -- the Bose-Einstein condensate (BEC).
Remarkably, even a small attraction between two fermions
with opposite spin states and momentum can lead to the formation of  a  Bardeen-Cooper-Schrieffer (BCS) pair that has a bosonic nature.
Such BCS  pairs can undergo the phenomenon  of BEC as  temperature tends to absolute zero.
Over the past few decades, experimental achievements in trapping and cooling atomic gases have revealed the beautiful and
subtle physics of the quantum world of ultracold atoms, see recent review articles
\cite{Dalfovo:1999,Leggett:2001,Regal:2006,Lewenstein:2007,Giorgini:2008, Bloch:2008,Bloch:2012,Chin:2010,Zhai:2009}.

\begin{figure}[t]
{{\includegraphics [width=0.70\linewidth,angle=-0]{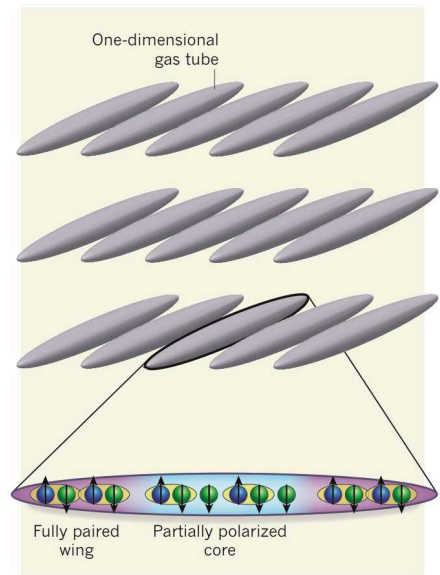}}}
\caption{ Experimental confinement of two-component ultracold ${}^6$Li atoms trapped in an array of 1D tubes \cite{Liao:2010}.
The system has spin population imbalance caused by a difference in the number of spin-up and spin-down atoms.
From  Bloch (2005).}
\label{fig:1D}
\end{figure}

In particular, recent experiments on ultracold bosonic and fermionic atoms confined to one dimension (1D) have provided a better
understanding of the quantum statistical and dynamical effects in quantum many-body systems \cite{Yurovsky:2008,Cazalilla:2011}.
These atomic waveguide particles are tightly confined in two transverse directions and weakly confined in the axial direction.
The transverse excitations are fully suppressed by the tight confinement.
As a result the trapped atoms can be effectively characterised by a quasi-1D system, see Fig.~\ref{fig:1D}.
The effective 1D inter-particle potential can be controlled in the whole interaction regime.
In such a way, the 1D many-body systems ultimately relate to previously considered exactly solved models of interacting bosons and fermions.
This has led to a fascinating interplay between exactly solved models and experimental developments in 1D.
Inspired by these developments, the study of integrable models has undergone a renaissance over the past decade.
Their study has become crucial to exploring and understanding the physics of quantum many-body systems.

\subsection{Exactly solved models}

\subsubsection{The virtuoso triumphs  of the Bethe ansatz }

The study of Bethe ansatz solvable models began when Bethe (1931) introduced a particular form of wavefunction -- the Bethe ansatz (BA) --
to obtain the energy eigenspectrum of the 1D Heisenberg spin chain.
After laying in obscurity for decades, the BA emerged to underpin a diverse range of physical problems,
from superconductors to string theory, see, e.g., Batchelor (2007).
For  such exactly solved models, the energy eigenspectrum of the model Hamiltonian is obtained exactly in terms of the
BA equations, from which physical properties can be derived via mathematical analysis.
From 1931 to the early 1960s there were only a handful of papers on the BA, treating the passage to the thermodynamic limit and
the extension to the anisotropic XXZ Heisenberg spin chain \cite{Hulthen:1938,Orbach:1959,Walker:1959,Cloizeaux:1962, Griffiths:1964}.
Yang and Yang (1966a) coined the term Bethe's  hypothesis and proved that Bethe's solution was indeed the groundstate of the
XXZ spin chain \cite{Yang:1966a,Yang:1966b,Yang:1966c}.

The next development was the exact solution of the 1D Bose gas  with delta-function interaction by
Lieb and Liniger (1963), which continues to have a
tremendous impact in quantum statistical mechanics \cite{Cazalilla:2011}.
They diagonalised the hamiltonian  and derived the groundstate energy of the model.
This study was further extended to the excitations above the groundstate \cite{Lieb:1963b}.
McGuire (1964) considered the model in the context of quantum many-body scattering in which the
condition of non-diffractive scattering appeared.

Developments for the exact solution of the 1D Fermi gas  with delta-function interaction \cite{Gaudin:1967a,Gaudin:1967b,Yang:1967} are discussed in the next subsection.
A key point is Yang's observation \cite{Yang:1967} that a generalised Bethe's hypothesis works for the fermion problem, subject to a set of cubic equations
being satisfied.
This equation has since been referred to as the Yang-Baxter equation (YBE) after the name was coined by Takhtadzhan and Faddeev (1979).
Baxter's contribution was to independently show that such relations also appear as conditions for commuting transfer matrices in
two-dimensional lattice models in statistical mechanics \cite{Baxter:1972a,Baxter:1982}.
Moreover, the YBE was seen as a relation which can be solved to obtain new exactly solved models.
The YBE thus became celebrated as the masterkey to integrability \cite{Perk:1989}.

The study of Yang-Baxter integrable models flourished in the 70's, 80's and 90's in the Canberra, St Petersburg, Stony Brook and Kyoto schools, with
far reaching implications in both physics and mathematics.
During this period the YBE emerged as the underlying structure behind the solvability of a number of quantum mechanical models.
In addition to the XXZ spin chain, examples include the XYZ  spin chain \cite{Baxter:1972b},
the $t-J$ model at supersymmetric coupling \cite{Foerster:1993a,Foerster:1993b,Essler:1992}
and the Hubbard model \cite{Lieb:1968,Shastry:1986a,Shastry:1986b,Ogata:1990,Shiba:1972,Frahm:1990,Frahm:1991,Essler:2005}.
Three collections of key papers have been published \cite{Jimbo:1990,Mattis:1993,Korepin:1994}.

Further examples are strongly correlated electron systems \cite{Tsvelik:1995,Giamarchi:2004,Takahashi:1999,Schollwock:2004},
spin exchange interaction \cite{Sutherland:2004,Montorsi:1992,Essler:2005},
Kondo physics of quantum  impurities coupled to conduction electrons in equilibrium \cite{Tsvelik:1983,Andrei:1983}
and out of equilibrium \cite{Doyon:2007,Mehta:2006,Nishino:2007,Nishino:2009},
the BCS model \cite{Richardson:1963a,Richardson:1963b,Richardson:1964,Richardson:1965,Cambiaggio:1997,vonDelft:2001,Dukelsky:2004,Links:2003,Dunning:2004},
 models with long range interactions \cite{Calogero:1969,Sutherland:1971,Gaudin:1976,Haldane:1988,Shastry:1988},
 two Josephson coupled BECs \cite{Zhou:2002,Zhou:2003a},
 BCS-to-BEC crossover \cite{Ortiz:2005},  atomic-molecular BECs \cite{Zhou:2003b,Foerster:2007}
 and quantum degenerate gases of ultracold atoms  \cite{Korepin:1993,Pethick:2008,Yurovsky:2008,Cazalilla:2011}.

A significant development in the theory of quantum integrable systems is the
algebraic BA \cite{Sklyanin:1979,Kulish:1982,Faddeev:1982}, essential to
the so called Quantum Inverse Scattering method (QISM), a quantized version of the classical inverse scattering method.
The QISM gives a unified description of the exact solution of quantum integrable models.
It provides a framework to systematically construct and solve quantum many-body systems
\cite{Thacker:1981,Korepin:1993,Takahashi:1999,Essler:2005}).

Other related threads are the quantum transfer matrix (QTM) \cite{Suzuki:1985,Klumper:1992,Destri:1992} and
$T$-systems  \cite{Kuniba:1994a,Kuniba:1994b,Kuniba:2011} from which one can derive temperature-dependent
properties in an exact non-perturbative fashion. Applications of this approach include the Heisenberg model \cite{Shiroishi:2002},
higher-spin chains  \cite{Tsuboi:2003,Tsuboi:2004},  and integrable quantum spin ladders \cite{Batchelor:2003, Batchelor:2004a,Batchelor:2004b,Batchelor:2007b}.
$T$-systems and integrability in general also play a fundamental role in the gauge/string theories of high energy physics \cite{Kuniba:2011,Beisert:2012}.

Yang-Baxter integrability has also played a crucial role in initiating and inspiring progress in mathematics.
Particularly to the theory of  knots, links and braids \cite{Jones:1985,Kauffman:1987,Wadati:1989,Wu:1992,Yang:2006}
and the development of quantum groups and representation theory \cite{Chari:1994,Gomez:1996}.

\subsubsection{Fermions in 1D -- historical overview}

In the mid-60's many physicists worked on extending the results obtained by Lieb and Liniger (1963) and McGuire (1964)
for 1D bosons with delta-function interaction to the problem of 1D fermions.
McGuire \cite{McGuire:1965,McGuire:1966}  solved the eigenvalue problem of $N-1$ fermions of  the same spin and
one fermion of opposite spin and studied the low-lying excited states with repulsive and attractive potentials.
The dynamics of this one spin-down Fermi problem has been studied \cite{McGuire:1990}.
The problem of $N-2$ fermions of the same spin with two fermions of opposite spin was solved by Flicker and Lieb (1967).
A further step came when Gaudin (1967a,1967b) and Yang (1967) solved the general problem in terms of a nested BA
for arbitrary spin population imbalance.\footnote{Missing phase factors for the spin sector in Eq.~(4c) of Gaudin (1967a) were corrected in Gaudin (1967b).
 A thorough treatment of the 1D Fermi problem can be found in Gaudin's book on the Bethe wavefunction \cite{Gaudin:1983}.}
Gaudin derived the groundstate energy for the balanced (fully paired) case for attractive interaction,
pointing out that the result is equivalent to that for repulsive bosons \cite{Lieb:1963a}.
The delta-function interacting two-component Fermi gas is commonly referred to as the Gaudin-Yang model.

Yang's concise solution of the problem had a profound impact.
As already remarked above, a key point in the solution is that the matrix operators describing many-body scattering can be
factorised into two-body scattering matrices, provided that a set of cubic equations -- the Yang-Baxter equation -- are satisfied
by the two-body scattering matrices.
This in turn is equivalent to no diffraction in the outgoing waves in three-body scattering processes.
In this sense Yang's solution completes McGuire's formulation of the scattering process in the context of the 1D Bose gas.
Indeed, the $R$-matrix obtained for the 1D Bose gas is known as the simplest nontrivial solution of the
Yang-Baxter equation \cite{Jimbo:1989}.

The solution of the 1D Fermi problem triggered a series of further breakthroughs.
Yang (1968) obtained the $S$-matrix of the delta-function interacting many-body problem for Boltzmann statistics  \cite{Gu:1989}.
The exact solution of the 1D Fermi gas with higher spin symmetry was obtained by  Sutherland \cite{Sutherland:1968,Sutherland:1975}.
The 1D Hubbard model solved  by  Lieb and Wu (1968) is a fundamental model in the theory of  strongly correlated electron systems.
Its solution is a significant example of the factorization condition (the YBE) in which the quasimomenta of particles $k$  is replaced by $\sin k$.
The Lieb-Wu solution thus gives a similar set of integral equations as Yang's  Fredholm equations for the continuum  gas.
This  exactly solved model has been extensively studied  in the literature.
The exact results  for the Hubbard  model not only provide the essential physics of 1D strongly correlated electronic systems
\cite{Essler:2005,Ha:1996,Takahashi:1999}, but also are relevant to phenomena in high $T_c$ superconductivity.
Indeed, the 1D Hubbard model is  an archetypical many-body system featuring Fulde-Ferrel-Larkin-Ovchinnikov (FFLO)  pairing,
universal Tomonaga-Luttinger liquid (TLL) physics, spin-charge separation and quantum entanglement \cite{Essler:2005,Gu:2004,Larsson:2005}.

Although further study \cite{Yang:1970,Takahashi:1970b} of the 1D Fermi gas
was initiated soon after its solution, it was not until much later that this model began to receive more attention
\cite{Fuchs:2004,Tokatly:2004,Astrakharchik:2004,Iida:2005,Batchelor:2006a} as a result of the brilliant experimental progress
in ultracold atom physics.
The fundamental physics of the model  is  determined by the set of   generalised Fredholm integral equations obtained in the thermodynamic limit.
Takahashi (1970a) discussed the analyticity of the Fredholm equations in the vanishing interaction limit.
A thorough study of the  Fredholm equations for the Gaudin-Yang model with attractive and repulsive interactions has been carried out
\cite{Guan:2012a,Guan:2007a,Iida:2007,Wadati:2007,Iida:2008,Zhou:2012}.
The numerical solution of the Fredholm equations  has also been  discussed in the
context of harmonic traps \cite{Orso:2007,Hu:2007,Colome-Tatche:2008,Kakashvili:2009,Ma:2009,Ma:2010a,Ma:2010b}.
In particular, the eigenfunction has been obtained explicitly for the Fermi gas in  the infinitely strong repulsion limit  by
using the hard-core contact boundary condition \cite{Girardeau:1960} and group theoretical methods \cite{Guan-L:2009,Ma:2010c}.

The next major advance with implications for the 1D Fermi problem was the solution of the
finite temperature problem for 1D bosons.
Yang and Yang (1969) showed that the thermodynamics of the Lieb-Liniger Bose gas can be determined from the minimisation conditions of the
Gibbs free energy subject to the BA equations.
Takahashi went on to make significant  contributions to Yang and Yang's grand canonical approach to the
thermodynamics of 1D integrable models  \cite{Takahashi:1971a,Takahashi:1971b,Takahashi:1972,Takahashi:1972b,Takahashi:1973,Takahashi:1974}.

Takahashi gave the general name Thermodynamic Bethe Ansatz (TBA) equations to the Yang-Yang type of equations for the thermodynamics.
He discovered spin strings patterns to the BA equations in addition to those  for the  groundstate of the spin chain \cite{Takahashi:1971a}.
Using a similar spin string hypothesis, Gaudin (1971) studied the thermodynamics of the Heisenberg-Ising chain.
Lai (1971a, 1973) independently derived the TBA equations for spin-1/2 fermions in the repulsive regime.
It turns out that Takahashi's spin string hypothesis  allows one to study the grand canonical ensemble  for many 1D many-body systems with internal degrees of freedom,
e.g.,   the 1D Fermi gas \cite{Takahashi:1971b},  the 1D Hubbard model \cite{Takahashi:1972,Takahashi:1974,Usuki:1990},
the quantum sine-Gordon model \cite{Fowler:1981},  the
Kondo problem \cite{Filyov:1981,Lowenstein:1981} among many other integrable models.

Building on Takahashi's spin string hypothesis,  Schlottmann  derived the TBA equations for $SU(N)$
fermions with repulsive and attractive interactions \cite{Schlottmann:1993,Schlottmann:1994}.
The Yang-Yang method has been revealed to be an elegant way to analytically access not only the thermodynamics, but also correlation functions,
quantum criticality and TLL physics for a wide range of low-dimensional quantum many-body systems \cite{Essler:2005,Ha:1996,Takahashi:1999}.
The Yang-Yang thermodynamics of the 1D Bose gas  has been tested in recent experiments
\cite{van Amerongen:2008,Armijo:2010,Jacqmin:2011,Stimming:2010,Armijo:2011a,Armijo:2011b,Kruger:2010,Sagi:2012}.

Recently, numerical schemes have been developed to solve the TBA equations of the 1D two-component spinor
Bose gas with delta-function interaction \cite{Caux:2009,Klauser:2011}.
The QTM method has also been applied to the thermodynamics of  the 1D Bose and Fermi gases with repulsive delta-function interaction \cite{Klumper:2011,Patu:2012}.
The Canberra group and their collaborators \cite{Zhao:2009,Guan:2010,Guan:2011a,Guan:2011b,He:2010,He:2011} developed  an asymptotic method to
calculate the thermodynamics of strongly interacting bosons and fermions  in an analytic fashion using the polylog function in the framework of the
Yang-Yang and Takahashi methods.
This  approach does away with the need to numerically solve the TBA equations for these systems at quantum criticality,
where the temperature is very low and the inter-particle interaction is strong.

\subsection{Renewed interest in 1D fermions}

The renewed interest over the past decade in 1D fermions has been on a number of related fronts.
Here we give a brief introductory outline of these developments.

\subsubsection{Novel BCS pairing states}

Quantum matter at  low temperatures has already been seen to exhibit some
remarkable physical properties, such as BEC and  superfluidity.
Fermionic quantum matter with mismatched Fermi surfaces has long been
expected to exhibit more exotic behaviour than seen in conventional materials.
The two-component attractive Fermi gas  is particularly interesting  due to its connection
with the  exotic pairing phase -- the FFLO  state -- involving BCS  pairs with nonzero centre-of-mass momenta.
In this phase, where the system is partially polarized, the Fermi energies of spin-up and spin-down electrons become unequal.
Originally,  Fulde and Ferrell (1964) discovered that under a strong external field, superconducting electron pairs
have nonzero pairing momentum and spin polarization.
Larkin and Ovchinnikov (1965) found that the formation of pairs of electrons with different momenta, i.e., $\vec{k}$ and $-\vec{k}+\vec{q}$
with non zero $\vec{q}$ is energetically favoured over pairs of electrons with opposite momenta, i.e.,  $\vec{k}$ and $-\vec{k}$,
when the separation between Fermi surfaces is sufficiently large.
Consequently, the density of spins and the superconducting order parameter become periodic functions of the spatial coordinates.

Theoretical study of the  FFLO state in 1D interacting fermions was initiated by K. Yang (2001),
who used bosonization to study the pairing correlations.
The FFLO-like pair correlations and spin correlations  for the attractive Hubbard model  were later  investigated  numerically by two groups  \cite{Feiguin:2007,Tezuka:2008}.
Both groups  showed the  power-law decay of the form  $n^{\mathrm{pair}}\propto\cos(k_{\mathrm{FFLO}}|x|)/|x|^{\alpha}$
for the pair  correlation, with spatial oscillations depending solely on  the mismatch $k_{\mathrm{FFLO}}=\pi(n_{\uparrow}-n_{\downarrow})$ of the Fermi surfaces.
Thus the momentum pair distribution  has peaks at the mismatch of the Fermi surfaces.
The FFLO state has since been studied by various methods:
density-matrix renormalization group (DMRG) \cite{Rizzi:2008a,Luscher:2008,Tezuka:2010},
quantum Monte Carlo (QMC) \cite{Batrouni:2008,Baur:2010,Wolak:2010},
mean field theory and other methods \cite{Parish:2007,Liu:2007,Liu:2008b,Zhao:2008,Edge:2009,Datta:2009,Pei:2010,Edge:2010,Devreese:2011,Kajala:2011,Chen:2012}.

Very recently,  the asymptotic correlation functions and FFLO signature were analytically studied  using the dressed charge formalism in the context of the
Gaudin-Yang model \cite{Lee:2011a,Schlottmann:2012b}.
However, a convincing theoretical proof for the existence of the 1D FFLO state in  the expansion dynamics of  the 1D polarized Fermi gas 
after its sudden release from the longitudinal confining potential is still rather elusive, see recent further developments  \cite{Lu:2012,Dalmonte:2012,Bolech:2012}. 
So far the spatial oscillation nature of FFLO pairing has not been experimentally confirmed.

\subsubsection{Large-spin ultracold atomic fermions}

It was shown \cite{Ho:1999,Yip:1999} that  large-spin atomic fermions exhibit rich pairing structures and collective modes in low-energy physics.
Further progress towards understanding many-body physics with large-spin Fermi gases was made \cite{Wu:2003, Wu:2005}
on spin-3/2 systems which can be realized with $^{132}$Cs, $^{9}$Be and $^{135}$Ba ultracold atoms \cite{Wu:2006}.
Such systems exhibit a generic $SO(5)$ symmetry (isomorphically, $Sp(4)$ symmetry).
The spin-3/2 system with $SU(4)$ symmetry can exhibit a quartet state (four-body bound state).
More generally,  ultracold atoms offer an exciting opportunity to  investigate  spin-liquid behavior via trapped
fermionic atoms with large-spin symmetry \cite{Honerkamp:2004,Rapp:2007,Zhou:2006,Cherng:2007,ZhaoJ:2006,Zhai:2007,Tu:2007,Corboz:2011,Szirmai:2011,Krauser:2012}.
The trimer state  (``trions") consisting of fermionic $^6$Li atoms in the three energetically lowest substates  has been reported  \cite{Lompe:2010,Williams:2009,Huckans:2009}.

On the other hand, fermionic alkaline-earth atoms display an exact $SU(N)$ spin symmetry with $N=2I+1$ where $I$ is the nuclear spin \cite{Gorshkov:2010,Cazalilla:2009,Xu:2010}.
Such fermionic systems with enlarged $SU(N)$ spin symmetry are expected to display a remarkable diversity of new quantum phases and quantum critical phenomena
due to the existence of multiple charge bound states.
De Salvo {\em et al.} (2010)  have  reported  quantum degeneracy in a gas of ultracold fermionic $^{ 87}$Sr atoms with $I=9/2$  in an optical dipole trap.
An experiment  by  Taie {\em et al.} (2010) dramatically realised the model of fermionic atoms  with  $SU(2)\otimes SU(6)$
symmetry where electron spin decouples from  its  nuclear spin  $I=5/2$ for ${}^{173}$Yb  together with atoms of its spin-1/2 isotope.
This group also successfully realized the $SU(6)$ Mott-insulator state with ultracold fermions of  ${}^{173}$Yb atoms in a 3D optical lattice \cite{Taie:2012}.

In the context of large-spin ultracold atomic fermions, Lecheminant {\em et al.} (2005) considered 1D ultracold atomic systems of fermions with general half-integer spins.
The instability of the  BCS pairing phase and molecular superfluid phase in these systems have been studied by a low-energy approach.
The low-energy physics and competing orders in large-spin fermionic systems in a 1D lattice  were  further investigated
\cite{Capponi:2007,Azaria:2009,Nonne:2010,Nonne:2011}.
In this scenario,  a  new class of integrable models of  ultracold fermions  and bosons with large-spin symmetries  were found  \cite{Cao:2007,Jiang:2009,Jiang:2011}.
They derived the BA solutions for spin-3/2 fermions with $SO(5)$ symmetry and the $Sp(2s+1)$-invariant model of fermions.

From the  integrable model perspective, the study of multi-component attractive fermions was initiated a long time ago
by Yang (1970)  and Takahashi (1970b).
In the light of ultracold higher spin atoms, Controzzi and Tsvelik (2006)
proposed an exact solution of a model describing the low energy physics of spin-3/2 fermionic atoms in a 1D lattice.
The  exact results obtained from 1D many-body systems with higher spin symmetries have provided insight into
understanding the few-body physics of  trions  \cite{Guan:2008a,Liu:2008a,He:2010};
quartet states (four-body charge bound states) \cite{Guan:2009,Schlottmann:2012a,Schlottmann:2012b,Schlottmann:2012c}; and
an arbitrary large spin-neutral bound state of different sizes \cite{Schlottmann:1993,Schlottmann:1994,Guan:2010,Lee:2011a,Yang:2011,Yin:2011a}.
The study of critical phenomena and universal TLL physics in low-dimensional ultracold atomic Fermi gases with large pseudo-spin symmetries is a
rapidly developing  frontier in ultracold atom physics.

\subsubsection{Quantum criticality of ultracold atoms}

Quantum criticality describes a V-shaped phase of quantum critical matter fanning
out to finite temperatures from the quantum critical point (QCP). It is associated
with competition between the two distinct ground states near the QCP.
Near a QCP,  the quantum critical behaviour is characterized by the energy gap $\Delta \sim \xi^{-z}$
and  a diverging length scale $\xi\sim |\mu-\mu_c|^{-\nu}$, where $\mu_c $ is the critical value of the driving parameter $\mu$.
The universality class of quantum criticality is characterized by  the dynamic critical exponent $z$ and the correlation exponent $\nu$ \cite{Fisher:1989,Wilson:1975,Sachdev:1999}.
The many-body system is expected to show universal scaling behaviour in the thermodynamic quantities  at quantum criticality due to the collective nature of many-body effects.
Thus  a universal and scale-invariant description of the system is  expected through the power-law scaling of thermodynamic properties.
However, understanding the various aspects of quantum criticality in quantum systems  represents a major challenge to our knowledge of many-body physics
\cite{Vojta:2003,Sondhi:1997,Sachdev:2011,Coleman:2005,Lohneysen:2007,Gegenwart:2008}.

Ultracold atoms have become the tool of choice to simulate and test universal quantum critical phenomena.
The study of quantum criticality and finite-size scaling in trapped atomic systems is thus attracting considerable
interest \cite{Campostrini:2009,Campostrini:2010a,Campostrini:2010b,Ceccarelli:2012, Kato:2008,Pollet:2010,Fang:2011a,Hazzard:2011}.
The experimental study of  critical behaviour in a trapped Bose gas was initiated by Donner {\em et al.}  (2007).
In particular, significant experimental progress  on quantum criticality and quantum phase transitions in 2D Bose atomic gases  has been made
\cite{Gemelke:2009,Hung:2010,Hung:2011a,Hung:2011b,Zhang:2011,Zhang:2012}.

Some of the remarkable features of criticality in general are the notions of universality class and symmetry.
Using integrable quantum field theory, Zamolodchikov (1989) was able to show that the 2D Ising model
in a magnetic field, or equivalently the quantum Ising chain with a transverse field \cite{Henkel:1989} displays $E_8$ symmetry close to the critical point.
Such exotic quantum symmetry in the excitation spectrum was observed in a recent experiment in the traditional setting of condensed matter physics \cite{Coldea:2010}.

Zhou and Ho (2010) have proposed a precise  theoretical  scheme for mapping out quantum criticality of ultracold atoms.
In this framework,  exactly solvable models of ultracold atoms, exhibiting quantum phase transitions,  provide  a rigorous way to  explore quantum criticality in  many-body systems.
The equation of state has been obtained for a number of key integrable models,
allowing the exploration of TLL physics and quantum criticality.
These include the Gaudin-Yang Fermi gas  \cite{Zhao:2009,Guan:2011a,Yin:2011a},
the Lieb-Liniger Bose gas \cite{Guan:2011b},  the Fermi-Bose mixture  \cite{Yin:2011b} and the spin-1 spinor Bose gas
with antiferromagnetic spin-spin exchange interaction \cite{Kuhn:2012a,Kuhn:2012b}.
The  exact results for the scaling forms of thermodynamic  properties in these systems near the critical point  illustrate the physical origin of quantum criticality in many-body  systems.

\subsubsection{Experiments with ultracold atoms in 1D}

Many remarkable 1D  quantum phenomena have been experimentally observed due to recent  rapid progress in material synthesis and tunable manipulation of ultracold atoms.
These developments have  provided a better understanding of significant quantum statistical effects and strong correlation effects in low-dimensional quantum many-body systems.
The observed results to date are seen to be in excellent agreement with results obtained using the mathematical methods and analysis of exactly solved models.
These include the experimental realization of the Tonks-Girardeau gas \cite{Paredes:2004,Kinoshita:2004} and 
a quantum Newton's cradle, i.e., a demonstration of out-of-equilibrium physics in arrays of trapped 1D Bose gases \cite{Kinoshita:2006} 
and quantum correlations \cite{Tolra:2004,Kinoshita:2005,Betz:2011,Haller:2011,Endres:2011,Guarrera:2012}.
Haller {\em et al.} (2009) made a further experimental breakthrough by realising  a stable highly excited gas-like phase,
called the super Tonks-Girardeau gas, in the strongly attractive regime of bosonic Cesium atoms.

The Yang-Yang thermodynamics  and thermal fluctuations  of an  ultracold Bose gas of $^{87}$Rb atoms were further tested in a series of
publications  \cite{van Amerongen:2008,Armijo:2010,Jacqmin:2011,Stimming:2010,Armijo:2011a,Armijo:2011b,Kruger:2010,Sagi:2012,Jacqmin:2012}.
The universal low-energy physics was demonstrated as hosting a TLL \cite{Haller:2010,Blumenstein:2011}.

The experimental research using ultracold Fermi gases to explore pairing phenomena in a 1D Fermi gas was first reported in 2005 \cite{Moritz:2005}.
In a major breakthrough towards understanding  the exotic  pairing signature and quantum phase diagram of the attractive Fermi gas,
Liao {\em et al.} (2010) measured the finite temperature density profiles of trapped fermionic $^6$Li atoms, see Fig.\ref{fig:1D}.
They confirmed the key features of the  $T = 0$ phase diagram predicted from the
exact solution \cite{Batchelor:2006a,Orso:2007,Hu:2007,Guan:2007a,Feiguin:2007,Parish:2007,Iida:2007,Kakashvili:2009}.

\subsection{Outline of this review}

In light of these recent developments we review the BA  solution of the Gaudin-Yang model  in Sec.~II and discuss the physical
understanding of the solution in terms of BCS pairing, the polaron problem,
molecule states and the super Tonks-Girardeau gas.
In Sec.~III we further discuss many-body phenomena  in the Gaudin-Yang model.
Especially, we discuss 1D fermions in a harmonic trap and review the universal  features of 1D interaction,
including magnetism, FFLO-like pairing,  TLL physics, spin-charge separation, universal thermodynamics and quantum criticality.
In Sec.~IV we review recent progress on mixtures of ultracold atoms and the exact solution of the 1D Fermi-Bose mixture.

Sec.~V reviews  the exotic many-body physics of 1D multi-component interacting fermions, including the three-component Fermi gas,
the $SU(N)$ invariant Fermi gases, and spin-3/2 fermions with $SO(5)$ symmetry.
The discussion in this Section covers  magnetism for systems of large-spin fermions, trions, molecular states of different sizes,
 multi-component TLL phases, universal low-temperature thermodynamics and critical behaviour caused by population imbalance.
 In Sec.~VI, we focus on the asymptotics of various relevant correlation functions for  the Gaudin-Yang model and multi-component Fermi gases.
 The characteristics of the FFLO-like pairing correlations and spin-charge separation correlation functions are discussed in the framework of conformal field theory.

The experimental breakthroughs with quasi-1D ultracold atoms and tests of 1D many-body physics are reviewed in Sec.~VII.
A brief conclusion and an outlook on future developments  are  given in Sec.~VIII.

\section{The Gaudin-Yang model}

The Hamiltonian
\begin{eqnarray}
{\cal H} &=& \sum _{\sigma=\downarrow,\uparrow} \int \phi _{\sigma}^{\dagger}(x) \left
(-\frac{\hbar^{2}}{2m}\frac{d^{2}}{dx^{2}} - \mu_{\sigma}+V(x) \right ) \phi
_{\sigma}^{}(x) dx\nonumber\\
&& +  g_{\rm 1D} \int \phi _{\downarrow}^{\dagger}(x) \phi
_{\uparrow}^{\dagger}(x) \phi _{\uparrow}^{}(x)
\phi _{\downarrow}^{}(x) dx \nonumber\\
&& - \frac12{H} \int \left (\phi _{\uparrow}^{\dagger}(x) \phi
_{\uparrow}^{}(x) - \phi _{\downarrow}^{\dagger}(x) \phi
_{\downarrow}^{}(x) \right ) dx \label{Ham-1}
\end{eqnarray}
describes a 1D $\delta$-function interacting two-component (spin-$\frac12$) Fermi gas of $N$ fermions  with mass $m$
and an external magnetic field $H$ constrained by periodic boundary conditions to a line of length $L$.
The function $V(x)$ is the trapping potential.
The field operators $\phi_{\downarrow}$ and $\phi_{\uparrow}$ describe the fermionic atoms in the states
$| \!\! \downarrow \rangle$ and $| \!\! \uparrow \rangle$, respectively.
The $\delta$-type interaction between fermions with opposite hyperfine states preserves the spin states such
that the Zeeman term in the Hamiltonian (\ref{Ham-1}) is a conserved quantity.

The experimental realization \cite{Liao:2010,Moritz:2005} of this system of interacting fermions is described in Sec.~VI.
The coupling constant  $g_{\rm 1D}=\hbar^{2}c/m$, where $c=-2/a_{\rm 1D}$ can be tuned by Feshbach resonance  \cite{Olshanii:1998,Bergeman:2003}.
For repulsive interaction $c>0$ and for attractive interaction $c<0$.
Where appropriate, we use units of $\hbar =2m=1$.
A  dimensionless coupling constant $\gamma=c/n$ is used to characterize physical regimes, i.e.,
$\gamma\gg 1$ for the strong coupling regime and $\gamma\ll 1$ for the weak coupling regime.
Here $n$ is the linear number density.

\subsection{Bethe ansatz solution}

For a homogeneous gas, i.e., $V(x)=0$, the eigenvalue problem for Hamiltonian (\ref{Ham-1}) reduces to the 1D $N$-body delta-function interaction  problem
\begin{equation}
{\cal H}=-\frac{\hbar^{2}}{2m}\sum_{i=1}^{N}\frac{\partial^{2}}{\partial
x_{i}^{2}}+g_{1D}\sum_{1\leq i<j\leq N}\delta(x_{i}-x_{j})\label{Ham}
\end{equation}
solved by Gaudin and Yang.
Bethe's hypothesis states that the wavefunction  of such a  many-body system is a superposition of plane waves.
In the domain $0<x_{Q1}<x_{Q2}<\ldots<x_{QN}<L$, the wave function is given by
\begin{eqnarray}
\psi&=&\sum_{P}\left[P,Q\right] \exp
\textrm{i}(k_{P1}x_{Q1}+\ldots+k_{PN}x_{QN}) \label{Wave}
\end{eqnarray}
where both $P=P_{1},\ldots,P_{N}$ and $Q=Q_{1},\ldots,Q_{N}$ are permutations of the integers $\{1,2,\ldots,N\}$.
The sum runs over all $N!$ permutations $P$. 
The $N!\times N!$  coefficients $\left[P, Q\right]$ of the exponentials can be  arranged as an $N!\times N!$  matrix.  
The columns are denoted by $N!\times N!$  dimensional  vectors $\xi_{P_1,\ldots,P_N}$ \cite{Yang:1967,Takahashi:1999}.
For example,  for two fermions with  one spin-up and one spin-down, the wave function is written as
\begin{eqnarray}
\psi&=&\theta_{12}\left([12,12]e^{\mathrm{i} (k_1x_1+k_2x_2)}+[21,12]e^{\mathrm{i} (k_2x_1+k_1x_2} \right) \nonumber\\
&&+\theta_{21}\left([12,21]e^{\mathrm{i} (k_2x_1+k_1x_2)}+[21,21]e^{\mathrm{i} (k_2x_2+k_1x_1} \right)\nonumber
\end{eqnarray}
where  $\theta_{ij}$  denotes  the step function $\theta_{ij}({x_j-x_i})$.
A plane wave repeatedly reflected  from the hyperplanes $x_{Q_i}=x_{Q_j}$ gives a total of $N!$ plane waves.
The idea in setting up such an ansatz is an attempt at a hypothetical solution followed by demonstrating 
that it gives the eigenfunction of the many-body problem, rather than solving the problem directly.

The derivative of the wavefunction  is discontinuous when two particles are infinitesimally close to one another.
This property can be derived by considering the eigenvalue problem  $\mathcal{H}\psi=E\psi$ in the center of mass coordinate $X=(x_{Q_i}+x_{Q_j})/2$
and the relative coordinate $Y=(x_{Q_i}-x_{Q_j})$ of the two adjacent particles involved.
This discontinuity in the first derivative of the wave function and the  continuity of the wave function at $x_{Q_i}=x_{Q_j}$
give  a two-body scattering relation between the adjacent vector coefficients  $\xi_{\cdots ij\cdots}=Y_{P_jP_i}^{ij} \xi_{\cdots ji \cdots }$.

The matrix operator  $Y_{ij}^{ab}$ is defined  by
\begin{eqnarray}
Y_{ij}^{ab}=\frac{-\textrm{i}(k_{i}-k_j)P_{ab}+cI}{\textrm{i}(k_i-k_j)-c}
\end{eqnarray}
where $I$ is the identity and $P_{ab}$ is the permutation operator acting on the vector $\xi_{\cdots ij \cdots}$.
Due to the Fermi statistics, $P_{ab}=-1$ for all $a$ and $b$.
Yang denoted the unequal indices $a,\, b, \, c,$  in the three particle scattering process
as the interchanges with coordinates $x_a, x_b$ and $x_c$ under the permutation $Q_{a},Q_b,Q_c$.
The consistency  condition for factorizing the many-body scattering matrix into the product of two-body scattering matrices
$Y_{ij}^{ab}$ leads to  the celebrated YBE
\begin{equation}
Y_{jk}^{ab}Y_{ik}^{bc}Y_{ij}^{ab}=Y_{ij}^{bc}Y_{ik}^{ab}Y_{jk}^{bc} \label{YBE}
\end{equation}
where  $Y_{ij}^{ab}Y_{ji}^{ab}=1$.
Defining the $R$-matrix by $R_{ij}=P_{ij}Y_{ij}^{ij}$ and spectral paramters $u=k_{2}-k_{1}$ and $v=k_{3}-k_{2}$
the YBE  is often written in the  form
\begin{equation}
R_{12}(u)R_{23}(u+v)R_{12}(v)=R_{23}(v)R_{12}(u+v)R_{23}(u).
\end{equation}

Returning to solving the problem of $N$ particles in a periodic box of length $L$, the second step is to apply the periodic boundary condition
$\psi(x_{1},\ldots,x_{i},\ldots,x_{N})=\psi(x_{1},\ldots,x_{i}+L,\ldots,x_{N})$
on the wavefunction with period $L$ for every $1\leq i\leq N$.
The two-column Young tableau 
$[2^{N_{\downarrow}},1^{N_{\uparrow}-N_{\downarrow}}]$ \cite{Yang:1967,Oelkers:2006} encodes the spin symmetry, 
where  $N_{\uparrow}$ and $N_{\downarrow}$  are the numbers of fermions in the hyperfine levels
$| \!\! \uparrow \rangle $ and $| \!\! \downarrow \rangle$  such that $N_{\uparrow}\geq N_{\downarrow}$.
This gives the second  eigenvalue problem
\begin{equation}
\mathfrak{R}_{i}(k_{i})A_{E}(P|Q)=\exp(\textrm{i}k_{i}L)A_{E}(P|Q)\label{BA2}
\end{equation}
where  $A_E(P|Q)$  is an abbreviation of the amplitude of the wave function which provides the eigenvector of the $N$ operators  $\mathfrak{R}_{i}$ with $i=1,\ldots, N$
\begin{eqnarray}
\mathfrak{R}_{i}(k_{i})&=&R_{i+1,i}(k_{i+1}-k_{i})\ldots
R_{N,i}(k_{N}-k_{i})\nonumber \\
&&\times R_{1,i}(k_{1}-k_{i})\ldots
R_{i-1,i}(k_{i-1}-k_{i}). \label{TM}
\end{eqnarray}

Using Bethe's hypothesis again, Yang  solved the eigenvalue problem  (\ref{BA2}) by the ansatz
\begin{eqnarray}
A_{E}(P|Q)=\sum\alpha_{P_1\ldots P_M}F(\lambda_{P_1},y_1)\ldots F(\lambda_{P_M},y_M),\label{Func-a}
\end{eqnarray}
where $y_1<y_2\ldots < y_M$ are the coordinates of the down-spin fermions and $\lambda_1,\ldots,\lambda_M$ are the spin rapidities
within the function   $F(\lambda,y)=\prod_{j=1}^{y-1}\frac{k_j-\lambda+\mathrm{i}c'}{k_{j+1}-\lambda-\mathrm{i}c'}$.
By the symmetry of the Young tableau $\left[2^{N_{\downarrow}},1^{N_{\uparrow}-N_{\downarrow}}\right]$, the vector $A_E$ describes a spin system with
a number of $N_{\downarrow}$ spins on  an $N$-site  lattice.

The generalized ansatz (\ref{Func-a}) plays an important role in solving multi-component many-body problems \cite{Sutherland:1968}.
Alternatively, the eigenvalue problem (\ref{BA2}) can be worked out in a straightforward  way in terms of the QISM,
where the operator $\mathfrak{R}_{i}(k_{i})$ can be written in terms of the quantum transfer matrix \cite{Ma:1993,Oelkers:2006,Li:2003,Korepin:1993,Jiang:2009}.
This approach was introduced in the study of the 1D Hubbard model \cite{Ramos:1997,Martins:1998,Essler:2005}.

The energy eigenspectrum is given in terms of the quasimomenta $\left\{k_i\right\}$ of the fermions via
\begin{equation}
E=\frac{\hbar ^2}{2m}\sum_{j=1}^Nk_j^2
\label{Ek}
\end{equation}
subject to the BA equations
which in terms of the function $e_b(x)=\frac{x+\mathrm{i}{bc}/{2}}{x-\mathrm{i}{bc}/{2}}$ are
\begin{eqnarray}
\exp(ik_{i}L)&=&\prod_{\alpha=1}^{N_{\downarrow}}e_1\left(k_{i}-\lambda_{\alpha}\right),\nonumber\\
\prod_{j=1}^{N}e_1\left(\lambda_{\alpha}-k_{j}\right)&=&-\prod_{\beta=1}^{N_\downarrow}
e_2\left(\lambda_{\alpha}-\lambda_{\beta}\right),\label{BE}
\end{eqnarray}
for $i=1,2,\ldots,N$ and $\alpha=1,2,\ldots,N_{\downarrow}$.
All wave numbers $k_i$ are distinct and uniquely define the wave function (\ref{Wave}) \cite{Gu:1989}.

The fundamental physics of the model is determined by the BA equations (\ref{BE}).
For repulsive interaction, the quasimomenta $\left\{ k_i\right\}$ are real, but the rapidities $\left\{ \lambda_{\alpha }\right\}$ are real only for  the groundstate.
The complex roots $\lambda_{\alpha}$ are the spin strings for excited states.
In the thermodynamic limit, i.e., $L,N \to \infty$, where $N/L$ is finite, the BA equations (\ref{BE}) can be written as generalized Fredholm equations
\begin{eqnarray}
{r_1}(k)&=&\frac{1}{2\pi}+ \int_{-B_2}^{B_2}K_1(k-k'){r_2}(k')dk', \nonumber\\
{r_2}(k)&=&\int_{-B_1}^{B_1}K_1(k-k'){r_1}(k')dk \nonumber\\
&&- \int_{-B_2}^{B_2}K_2(k-k'){r_2}(k') dk'\label{BE-r}
\end{eqnarray}
where the integration boundaries $B_1$ and $B_2$   are determined by
$n= N/L=\int_{-B_1}^{B_1}{r_1}(k)dk,   \, \, n_{\downarrow}= N_{\downarrow}/L=\int_{-B_2}^{B_2}{r_2}(k')dk'$.
In the above equations,  the kernel function
$K_{\ell }(x)=\frac{1}{2\pi}\frac{\ell c}{(\ell c/2)^2+x^2}$.
The functions  $r_m(k)$ denote  the Bethe root  distributions, with $r_1(k)$ the quasimomenta distribution function and
 $r_2(k)$  the spin  rapidity parameter distribution function.
The groundstate energy per unit length is given by
$E=\int_{-B_1}^{B_1}k^2{r_1}(k) d k$.

For the attractive regime, the quasimomenta $\left\{ k_i\right\}$ of fermions with different spins form two-body bound states,
i.e., the wave numbers are complex with $k_i=\lambda_i'\pm   \mathrm{i}  c/2$ in the thermodynamic limit \cite{Yang:1970,Takahashi:1971b}.
Here $i=1,\ldots,N_{\downarrow}$.
The excess fermions have real quasimomenta $\left\{ k_j\right\} $ with $j=1,\ldots, N-2N_{\downarrow}$.
In the thermodynamic limit,  the density of unpaired fermions $\rho_1(k)$ and the density of pairs $\rho_2(k)$
satisfy the Fredholm equations
\begin{eqnarray}
\rho_1(k)&=&\frac{1}{2\pi}+\int_{-A_2}^{A_2}K_1(k-k' )\rho_2(k')dk' \nonumber \\
\rho_2(k)&=&\frac{1}{\pi}+\int_{-A_1}^{A_1}K_1(k-k')\rho_1(k')dk'\nonumber\\
&&+\int_{-A_2}^{A_2}K_2(k-k')\rho_2(k')dk'. \label{Fermi2-a}
\end{eqnarray}
The distribution $\rho_2(k)$ coincides with the distribution function of the real parts of the  bound states.
The linear densities are defined by
$N/L= 2\int_{-A_2}^{A_2}\rho_2(k)dk+\int^{A_1}_{-A_1}\rho_1(k)dk$ and   $N_{\downarrow}/L=\int_{-A_2}^{A_2}\rho_2(k)dk$.
The groundstate energy per unit length is given by $E=\int_{-A_2}^{A_2}\left(2k^2-c^2/2\right)\rho_2(k)dk+\int_{-A_1}^{A_1}k^2\rho_1(k)dk$.

In the context of magnetism, the magnetization per unit length is defined by $M^z=(n-2n_{\downarrow})/2$.
By definition,   the groundstate energy  can be expressed as a function of total particle density $n$ and magnetization $M^z$.
In the grand canonical ensemble,  the magnetic field $H$ and the chemical potential $\mu$ can be obtained via the relations
\begin{equation}
H=2\frac{\partial E(n,M^z) }{\partial M^z},\qquad  \mu =\frac{\partial E(n,M^z)}{\partial n}, \label{mu-h}
\end{equation}
which have been used to work out the phase diagrams of the attractive Fermi gas \cite{Orso:2007,Hu:2007} and the repulsive Fermi gas \cite{Colome-Tatche:2008,Guan:2012a}.
We now turn to extracting such information from both the discrete and continuum versions of the BA  solution.

\subsection{Solutions to the discrete Bethe ansatz equations}

The 1D Fermi gas (\ref{Ham-1}) with spin population imbalance exhibits an unconventional pairing order that presents a major subtlety of many-body correlations in the Gaudin-Yang model.
Starting with  the discrete  BA equations (\ref{BE}), we will review how the exact solution enables us to understand precisely
such subtle many-body physics driven by the interaction of quantum statistics and dynamics.
In particular, we will see that for the weakly and strongly attractive coupling regimes 1D interacting fermions
give significantly different phenomena: weakly bound BCS-like pairs vs tightly bound molecules.

\subsubsection{BCS-like pairing and  tightly bound molecules}
\label{ground-state}

For  weakly attractive interaction,  i.e.,  $L|c| \ll 1$, two fermions with spin-up and
spin-down form a weakly bound pair with a small binding energy $\epsilon_{\rm b}=-{\hbar^2|c|}/{mL}$,
where the two-body binding energy is less than the kinetic energy  \cite{Batchelor:2006a}.
The complex conjugate pair leads to an exponential decay of the wave function with a factor $e^{-|c||x_i-x_j|/2}$.
Thus the balanced case has a BCS-like fully paired state where the size of the Cooper pairs is much larger than the mean average distance between the fermions.
In this weakly attractive regime, the energy gap separating  the first triplet excited state from the groundstate is found
to have an asymptotic behaviour $\Delta \approx 2n^2 \sqrt{\pi |\gamma |}\exp\left(-\pi^2/(2|\gamma|)\right)$
as $|\gamma|\to 0$ \cite{Fuchs:2004,Krivnov:1975}.
In fact, the BA equations (\ref{BE}) for weak attraction give  an explicit  relation
$H\approx \frac{\hbar^2 n^2}{2m}\left(2\pi^2 m^z+4|\gamma | m^z \right)$
between the external field and magnetization in the thermodynamic limit.
The lower critical field gives the energy gap at $m^z=0$.
This relation indicates a vanishing energy gap $\Delta=H_c\to 0$  for  $\gamma \to 0$.
Here the magnetization is defined by $m^z := M^z/n=(N_{\uparrow}-N_{\downarrow})/2N$ \cite{He:2009,Iida:2007}.

\begin{figure}[t]
\includegraphics[width=0.95\linewidth]{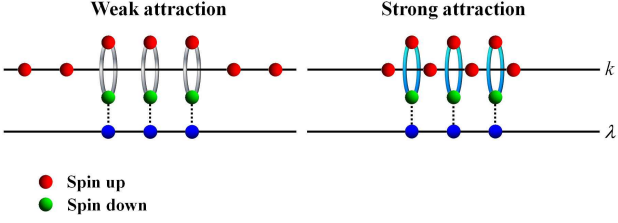}
\caption{Schematic BA root configurations for pairing and depairing in the Gaudin-Yang model.
For weakly attractive interaction, the unpaired roots sit in the outer wings due to Fermi statistics.
For strongly attractive interaction, the unpaired roots can penetrate into the central region,
occupied by the bound pairs \cite{Batchelor:2006c,Iida:2007}.}
\label{fig:distribution}
\end{figure}

For a polarized gas with weak attraction, Fermi statistics lead to segmentation in quasimomentum space, i.e.,
the excess fermions are located at the two outer wings in quasimomentum space, see Fig.~\ref{fig:distribution}.
For a finite size system with arbitrary polarization $P=(N_{\uparrow}-N_{\downarrow})/N$,  the BA equations (\ref{BE})
determine  $N_{\downarrow}$ weakly bound  BCS pairs with
$k_{\alpha}^{\rm p}\approx \lambda _{\alpha}\pm \mathrm{i}\sqrt{|c|/L}$ and $N-2N_{\downarrow}$ unpaired fermions with real $k_i$ \cite{Batchelor:2006c}.
In this case, $\left\{\lambda_{\alpha}\right\}$ and  $\left\{ k_j\right\}$ are symmetrically distributed around zero in the quasimomentum parameter space, see Fig.\ref{fig:distribution}.

Assuming that $N_{\downarrow}$ is odd and $N$ is even,  the first few leading orders of the positive roots $\left\{\lambda_{\alpha}\right\}$ and $\left\{ k_j\right\}$
are  determined by the equations
\begin{eqnarray}
k_j&\approx & \frac{2n_j\pi}{L}+\frac{c}{Lk_j}+\frac{c}{L}\sum_{\alpha=1}^{(M_{\downarrow}-1)/2}\left[\frac{2k_j}{k_j^2-\lambda_{\alpha}^2}\right],\nonumber\\
\lambda_{\alpha}&\approx  & \frac{2n_{\alpha}\pi}{L}+\frac{3c}{2L\lambda_{\alpha}}+\frac{c}{L}\sum_{\begin{array}{l}
\beta =1\\ \alpha \neq \beta\end{array} }^{(M_{\downarrow}-1)/2}\left[ \frac{2\lambda_{\alpha} }{\lambda_{\alpha}^2-\lambda_{\beta}^2}\right]\nonumber\\
&&+\frac{c}{2L}\sum_{j=1}^{(N-2M_{\downarrow})/2}\left[ \frac{2\lambda_{\alpha}}{\lambda_{\alpha}^2-k_j^2}\right],\label{BAd-2}
\end{eqnarray}
where $n_j=\frac{M_{\downarrow}+1}{2},\frac{M+3}{2},\ldots, \frac{N-M_{\downarrow}-1}{2}$,  and $n_{\alpha}=1,2,\ldots, \frac{M_{\downarrow}}{2}$.
These case ${\alpha}=\beta$ is excluded in Eq.~(\ref{BAd-2}).

The root patterns  reveal  the cooperative nature of many-body effects, i.e., an individual quasimomentum depends on that of all the particles.
Here the momenta of unpaired fermions and bound pairs depend on the scattering energies between pair and between paired and unpaired fermions.
This  indicates that the quantum  statistics of the weakly interacting fermions  is mutual according to exclusion statistics \cite{Haldane:1991}.
From Eq. (\ref{BAd-2}),  the groundstate energy  per unit length is given by \cite{Batchelor:2006c}
\begin{eqnarray}
\frac{E}{L}=\frac{1}{3}n_{\uparrow}^3\pi^2+\frac{1}{3}n_{\downarrow}^3\pi^2+2cn_{\uparrow}n_{\downarrow}+O(c^2).\label{e-dis-w}
\end{eqnarray}
Here the groundstate energy  (\ref{e-dis-w}) is also valid for weakly repulsive interaction, i.e., for $Lc\ll 1$.
This leading order correction to the  interaction energy indicates a mean field effect.

On the other hand, for strong attraction, i.e.,  $L|c| \gg 1$  (or $c\gg k_F$),
 the discrete BA equations  (\ref{BE})   give  the root patterns
  $k_{i}^{\rm b} \approx \lambda_{i}\pm   \mathrm{i} \frac{1}{2} c$ for bound pairs and real $k_j^{\rm u}$ for unpaired fermions  \cite{Yang:1970,Takahashi:1971b}.
 Here $i=1,\ldots, N_{\downarrow}$ and $j=1,\ldots, N_1$, the number of  unpaired fermions $N_1=N-2N_{\downarrow}$.
The binding energy $\varepsilon_{\rm b}=-\frac{c^2}{2}$ is the largest energy scale  than the kinetic energies of pairs or excess fermions.
For a strong attraction (up to order $O(1/c^3)$), $N$ fermions have root patterns  \cite{Batchelor:2006c}
  \begin{eqnarray}
  k^u& \approx & \frac{(2n^u+1)\pi }{L}\alpha_{\rm u},\,\,\,
  \lambda \approx  \frac{(2n^b+1)\pi}{2L}\alpha_{\rm b},\nonumber
  \end{eqnarray}
  where the effective statistical parameters are given by
\begin{equation}
\alpha _{\rm b} \approx \frac{1}{2}\left(1-\frac{2N-2N_{\downarrow}}{L|c|}\right)^{-1}, \,\,  \alpha _{\rm u}  \approx \left(1-\frac{4N_{\downarrow}}{L|c|}\right)^{-1},\nonumber
\end{equation}
and $n^u=-N_1/2,-N_1/2+1,\ldots, N_1/2-1$ with $n^b=-N_{\downarrow}/2, -N_{\downarrow}/2+1,\ldots, N_{\downarrow}/2-1$.
 In this strong attraction limit,  the groundstate energy per unit length is given by $E/L=E_0^u+2E_0^b+n_{\downarrow}\varepsilon_{\rm b}$, where  the energies of excess  fermions and pairs
 are given by
\begin{equation}
E_0^{u}=\frac{1}{3}n_1^3\pi^2\alpha _{\rm u}^2,\qquad E_0^{b}=\frac{1}{3}n_2^3\pi^2\alpha _{\rm b}^2.
\end{equation}

The bound states behave like hard-core bosons which can be viewed as ideal particles with fractional exclusion statistics.
However, the bound pairs have tails and they interfere with each other.
It is impossible to separate the intermolecular forces from the interference between  molecules and  single fermions.
From this explicit form of the groundstate energy, we  see that for
$n_{\downarrow}\gg x=n_{\uparrow}-n_{\downarrow}$,  the single atoms are repelled by the molecules, i.e.,
\begin{eqnarray}
E(n_{\downarrow}, x)-E_0 \approx \frac{1}{6}n_{\downarrow }^3\pi^2\left[\frac{4x}{|c|} +\frac{12x(x+n_{\downarrow})}{c^2}\right]>0.
\end{eqnarray}
Here $E_0$ is the groundstate energy per unit length of the balanced gas.
This result indicates that the single atoms are repelled by the molecules on the tightly bound dimer limit.
The atom-dimer scattering problem of three fermions in a quasi-one-dimensional trap has been studied \cite{Mora:2004,Mora:2005b}.

\subsubsection{Highly polarized fermions:  polaron vs  molecule}

In higher dimensions,  a  spin-down fermion immersed in a fully polarized spin-up Fermi sea
gives rise to  the quasiparticle phenomenon called Fermi polaron
\cite{Combescot:2007,Combescot:2008,Bruun:2010,Mathy:2011,Parish:2011,Klawunn:2011,Prokofev:2008a,Prokofev:2008b,Schmidt:2011}.
The Fermi polaron is a spin-down  impurity  fermion dressed  by the surrounding scattered fermions in the spin-up Fermi sea.
In particular, recent  observations of Fermi Polarons in a 3D or 2D  tunable Fermi liquid of ultracold atoms
\cite{Schirotzek:2009,Nascimbene:2009,Koschorreck:2012,Kohstall:2012} provide insightful understanding of quasiparticle physics  in many-body systems.
For an attractive polaron,  with increasing  attraction, the single spin-down  fermion
undergoes a polaron-molecule transition in the fermionic medium   \cite{Schirotzek:2009,Nascimbene:2009}.

For repulsive interaction,  theoretical studies have suggested the existence of such novel quasiparticles --
repulsive polarons \cite{Pilati:2010,Massignan:2011,Schmidt:2011,Schmidt:2012,Ngampruetikorn:2012}.
The properties of repulsive polarons, such as the energy, life time, and quasiparticle residue,
give a fundamental understanding of the coherent nature of the quasiparticle.
The repulsive polaron is metastable and eventually decays  to either a molecule state or an attractive polaron
with particle-hole excitations in the majority Fermi sea.
This quasiparticle  phenomenon  was experimentally observed by a magnetically tuned Feshbach resonance
on the BEC side with positive scattering length \cite{Koschorreck:2012,Kohstall:2012}.

So far, most  studies concerning the first order nature of the  polaron-molecule transition in a
3D fermionic medium \cite{Combescot:2007,Combescot:2008,Bruun:2010,Mathy:2011}
involve a variational ansatz with some approximations that are ultimately not justified  in low dimensions \cite{Parish:2011,Giraud:2009}.
It is generally accepted  that  there actually do not exist quasiparticle excitations in 1D systems  due to the
collective nature of the 1D many-body effect.
The elementary excitations in 1D are still eigenstates, where all particles are involved in a low energy nature.
Therefore, we cannot find a simple operator,  acting on the groundstate, to get a quasiparticle excitation, unlike for higher dimensional  systems.
But this does not rule out a well-defined quasi-particle like behaviour, e.g., a polaron, which is a typical example of the collective nature  of the
1D many-body effect.
The quantum impurity problem in 1D trapped ultracold atoms has shed new insight on the collective nature of particles \cite{Palzer:2009,Catani:2012}.

\begin{center}
\begin{figure}[htb]
\includegraphics[width=0.9\linewidth]{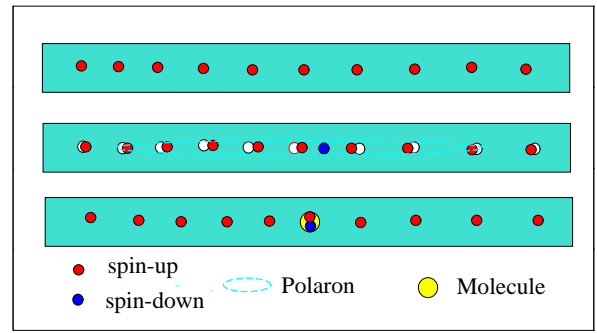}
\caption{Schematic BA root configuration of polaron-molecule crossover in the 1D attractive Fermi gas. 
The upper panel shows the free fermion distribution.
In the weakly attractive limit (middle panel),  the single impurity fermion (blue) dressed by  the surrounding scattered  spin-up fermions (red)
from the medium behave like  a polaron (dashed oval) with a mean field binding energy and an effective mass $m^*\approx m$.
For strong attraction (lower panel), the single impurity fermion binds with one spin-up fermion from the  Fermi sea to form a tightly bound molecule (yellow circle)
of a two-atom with a sole molecule binding energy and an effective mass $m^*\approx 2m$. }
\label{fig:Cartoon}
\end{figure}
\end{center}

The  BA  solvable models are likely to provide a rigorous treatment of polaron-like phenomena
in different mediums \cite{McGuire:1966,Leskinen:2010,Guan:2012c,Li-G:2012}.
In particular, a 1D attractive polaronic phenomenon  does  occur if one (or a few) spin-down fermion (fermions) is (are) immersed
into  a large spin-up Fermi sea \cite{Parish:2011,Giraud:2009,Klawunn:2011,McGuire:1966,Leskinen:2010,Guan:2012c,Massel:2012}.
The excitation energy of a system with one spin-down fermion has a certain momentum-dependent relation, which includes a mean field attractive
binding energy plus a classical kinetic energy of polaron with effective mass $m^*$.
A cartoon picture is shown in Fig.~\ref{fig:Cartoon}.

McGuire (1965,1966) studied the exact eigenvalue problem of $N-1$ fermions of  the same spin and  one fermion of the opposite spin.
He calculated the energy shift caused by  this extra spin-down fermion by solving the  equation
$az_i+1/\tan z_i= \mathrm{constant}$ for the quasi momentum $k_i=2z_i/L$ with $i=1,\ldots, N$ and $ a=4/(gL)$.
Here $g>0$ for an attractive  interaction strength.
McGuire found  a hermitian conjugate pair  $z_{1,2}= \alpha \pm \mathrm{i} \beta$ and $N-2$ real roots $z_i$.
The energy is given by $E=\frac{2}{L^2} \sum_{i=1}^Nz_i^2$.
This single impurity problem was recently studied  \cite{Guan:2012c} by means of the BA equations (\ref{BE}).
For an attractive interaction,  the quasimomenta $k_{\downarrow,\uparrow}=p\pm \mathrm{i} \beta$ of a pair and
$N-2$ real roots $\left\{k_i\right\}$ with $i=1,\ldots,N-2$ are determined by  the equations (\ref{BE}) with $N_{\downarrow}=1$.
It  was  found  \cite{Guan:2012c}  that the imaginary part $\beta$ in the pair  is determined by  the equation
$\beta L = \tanh ^{-1}\frac{\beta |c|}{\beta^2 +c^2/4}$.
For an excited state with total momentum of the system $q$, the spin-down fermion associated with the weakly bound  pair in the fully-polarized Fermi sea
thus has a nonzero  momentum
\begin{equation}
p\approx q/\left(1-2|c|\sum_{i=1}^{\frac{N_{\uparrow}}{2}-1} \frac{1}{L(k_i^2-p^2)}\right)
\end{equation}
which depends  on all  individual  momenta of the spin-up fermions.
This givens a collective signature of the 1D many-body effect.
Thus   the   energy shift is explicitly given by
\begin{eqnarray}
E(q,N,N_{\downarrow}=1) -E_{\uparrow}(N_{\uparrow},0)
&\approx& \epsilon_{p-b}+ \frac{\hbar^2q^2}{2m^{*}}
\label{polaron-e2}
\end{eqnarray}
which  behaves like a polaron quasiparticle.
Here, the   attractive mean field binding energy of the polaron is given by
 $\epsilon_{p-b}\approx  -\frac{6}{\pi^2}e_F|\gamma|$ for weak attraction.
The  Fermi energy is $e_F=\frac{\hbar^2}{2m}\frac{1}{3}n^2\pi^2$.
We see that this  binding energy depends solely on the Fermi energy of the medium and interaction strength in 1D.
 In Eq.~(\ref{polaron-e2}),  the polaron-like state has an effective mass
$m^{*}\approx m(1+O(c^2))$ which is almost the same as the actual mass of the fermions due to the decoupling from the bound pair in the weak coupling  limit.
We point out that the polaron-like state only occurs for few impurity fermions immersed into a fully-polarized Fermi sea.

For a weakly repulsive interaction and in the thermodynamic limit,  the low energy physics of the  1D Fermi gas
is described by a spin-charge separation theory.
The spin rapidity parameters decouple from the quaismomenta of the fermions.
However, using the BA equations (\ref{BE}),   a single spin-down fermion immersed into the 1D fully-polarized Fermi sea with weak repulsion
can form a repulsive Fermi polaron, with energy of the form   (\ref{polaron-e2}) and  an effective mass
$m^{*}\approx m(1+O(c^2))$.
But  here the impurity fermion receives a  positive mean field shift
$\epsilon_{p-b}\approx  \frac{6}{\pi^2}e_F|\gamma|$ from  the fermionic  medium.

In the opposite limit,  a spin-down fermion immersed into a fully polarized spin-up medium with strong attraction,
i.e., with $ L|c|\gg  1$, the bound pair has $k_{\downarrow,\uparrow}=p\pm \mathrm{i} \beta$ and   the
$N-2$ real roots $\left\{k_i\right\}$ with $i=1,\ldots,N-2$ are given by \cite{Guan:2012c}
\begin{equation}
k_i\approx \left(\frac{n_j\pi}{L}-\frac{4p}{L|c|} \right) \left(1-\frac{4}{L|c|}\right)^{-1},
\end{equation}
with $n_j=\pm 1, \pm 3,\ldots, \pm(N_{\uparrow}-1)$.
For an  excited state  with total system momentum $q$,  the relation between the centre-of-mass pair quasimomentum $p$
and the total momentum of the system $q$ is given by  $p\approx q/\left[ 2\left( 1-\frac{2(N_{\uparrow}-2)}{L|c|} \right)\right]$, which
is independent of the individual quasimomenta of the spin-up fermions.
 The energy shift is given by $\Delta E=E_{\rm M}-\mu$ with  the chemical  potential $\mu=n^2\pi^2$, where  the molecule energy is given by
\begin{eqnarray}
E_{\rm M}
&\approx& E_b+ \frac{\hbar^2q^2}{2m^{*}}.
\label{polaron-e2-m}
\end{eqnarray}

 The binding energy of the bound state  is
\begin{equation}
E_b \approx \frac{\hbar^2n^2}{2m}\left( -\frac{\gamma^2}{2} + \frac{8\pi^2}{3|\gamma|}\right)
\end{equation}
which    tends to the binding energy of a sole molecule $\varepsilon_b=-\frac{\hbar^2}{2m}\frac{c^2}{2}$  in the strongly attractive regime $L|c| \to \infty$.
The effective mass of the molecule
\begin{equation}
m^{*}\approx 2m\left(1-\frac{4}{|\gamma|}\right)
\end{equation}
becomes  twice the actual mass of the fermions in this limit.
From the shift energies (\ref{polaron-e2}) and (\ref{polaron-e2-m}), we see clearly that  as the  attractive interaction grows, the spin-down fermion binds
only with one spin-up fermion from the medium to gradually  form a   tightly bound molecule.
The polaron-molecule crossover is regarded as  a change from a mean field attractive binding energy of a polaron with an effective mass $m^{*}=m$
to the binding energy of a  single molecule with an effective mass $m^{*} = 2m$ as the attraction grows from zero to infinity.
The non-equilibrium dynamics of an impurity in a 1D lattice within a harmonic trap have been studied using
numerical methods and the BA solution  \cite{Massel:2012}. 
The numerical simulation of an impurity injected into a 1D quantum liquid has been reported \cite{Knap:2013}.

\subsection{Solutions in the thermodynamic limit}

In the  last section we discussed the solutions to the discrete BA (\ref{BE}) in the limits $|c|\to 0,\,\infty$.
They give rise to different phenomena in the two extreme limits.
Usually, the many-body phenomena of interest refers to the physics of the system in the thermodynamic limit, where $N,L\to \infty$ keeping $N/L$ finite.
This entails considering  the solutions to the two sets of Fredholm equations (\ref{BE-r}) and (\ref{Fermi2-a}) for the repulsive and attractive regimes.

\subsubsection{BCS-BEC crossover and fermionic super Tonks-Girardeau gas }

In order to conceive the physical nature of the super Tonks-Girardeau gas, we first recall  the Lieb-Liniger Bose gas with zero-range delta-function interaction,
where  the Tonks-Girardeau gas was determined by a Fermi-Bose mapping to an ideal Fermi gas \cite{Girardeau:1960}.
For a  strong attractive interaction, McGuire (1964) predicted that the quasimomenta of the bosons are given in terms of a bound state of
$N$-particles, with $k_{\pm j}\approx \pm \frac{1}{2}c\left[ N-2j+1\right]$ with $j=1,\ldots, N/2$.
In this case, the wave function is approximately given by
\begin{equation}
\Psi(x_1,\ldots, x_N)\approx {\cal N} \exp{\left(\frac{c}{2} \sum_{1\le i<j\le N}|x_j-x_i|\right)}
\end{equation}
where ${\cal N}=\left( \sqrt{(n-1)!} /\sqrt{2\pi }\right)|c|^{(n-1)/2}$ is a normalization constant.
The energy of the McGuire cluster state is given by $E_0=-\frac{1}{12}c^2N(N^2-1)$.
However,  if the interaction strength is abruptly changed from strongly repulsive to strongly attractive,
the highly excited  gas-like state  may  be metastable against this cluster-like state  due to the Fermi pressure inherited from the repulsive Tonks-Girardeau gas.
This  gas-like state exhibits  a more exclusive quantum statistics than the free Fermi gas, and is called super Tonks-Girardeau gas.
The  super Tonks-Girardeau gas-like  state was first predicted by Astrakharchik  {\em et al.} (2005) and further proved by Batchelor {\em et al.} (2005b)
using the exact BA solution of the Lieb-Liniger Bose gas.
Remarkably,  such a highly excited  state was realized in a breakthrough experiment  \cite{Haller:2009}.

The equally populated components in an attractive Fermi gas give rise to physics related to the cross-over between a
BCS superfluid and a BEC  \cite{Fuchs:2004,Tokatly:2004,Chen:2010,Iida:2005,Wadati:2007,Feiguin:2012}.
In this context, for a balanced attractive Fermi gas with $N_{\downarrow}=N/2$,
the  discrete BA equations  (\ref{BE}) reduce to
\begin{align}
\exp\left( 2\mathrm{i}\lambda_{\alpha} L\right) & =-\prod_{\beta=1}^{N_{\downarrow}}\left( \frac{
\lambda_{\alpha}-\lambda_{\beta}+ \mathrm{i} c_F}{\lambda_{\alpha}-\lambda_{\beta}-\mathrm{i} c_F}
\right)\label{Lamda}
\end{align}
with $c_F = -2/a^F_{1d}$. This is equivalent to the equations
\begin{equation}
\exp\left( \mathrm{i} k_{j}L\right)  = - \prod_{l=1}^{N_B}\left( \frac{
k_{j}-k_{l}+ \mathrm{i} c_B}{k_{j}-k_{l}- \mathrm{i} c_B} \right), \label{BAEbose}
\end{equation}
with $c_B= -2/a^B_{1d}$ for the Lieb-Liniger gas in the super Tonks-Girardeau phase under the identification
$c_B = 2 c_F$, $N_B=N/2$ and $m_B=2m_F$ \cite{Chen:2010,Wadati:2007}.
Since the bound pair formed by two
fermions with opposite spin has a mass $m_B=2m_F$,  the $M$ bound pairs are equivalently described by the super Tonks-Girardeau
phase of the interacting Bose gas with effective 1D scattering
length $a_{{\rm 1D}}^B=\frac12 {a_{{\rm 1D}}^F}$.  This relation is also obtained by an exact mapping based on
the two-body scattering   problem associated with BEC-BCS crossover \cite{Mora:2005a}.

For the balanced Fermi gas,  the binding energy is subtracted from the energy that gives the energy of the
bosonic pairs of a two-atom, with result  $E_0^{\rm F}=E+N_{\downarrow} \epsilon_b=\frac{\hbar^2}{2m_F} \sum_{\alpha=1}^{N_{\downarrow}} 2\lambda_{\alpha}^2$.
The energy eigenvalues of the bosons are given by $E=\frac{\hbar^2}{2m_B} \sum_{j=1}^{N_B} k_j^2$.
In this regard,  the groundstate of the  balanced attractive Fermi gas can be viewed as the fermionic super Tonks-Girardeau
gas \cite{Chen:2010}.
The identification between the balanced attractive Fermi gas  and the attractive Lieb-Liniger Bose gas suggests  an effective  attraction  between pairs.

\begin{figure}[t]
\includegraphics[width=1.0\linewidth]{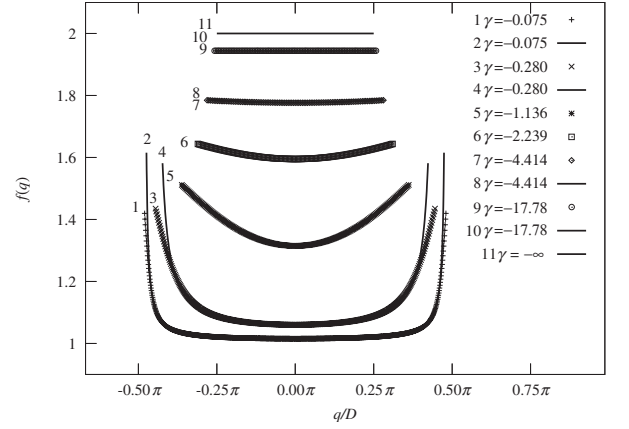}
\caption{The normalized pair quasimomentum distribution function  $f(q)$  for different values of interaction strength $\gamma$ \cite{Iida:2005}.
The analytic result for the distribution function matches the numerical solution. The quasimomentum distribution indicates the fermionization
from mutual statistics to non-mutual generalized exclusion statistics as $\gamma$ increases.
From Iida and Wadati (2005). }
\label{BEC-BCS crossover}
\end{figure}

In the thermodynamic limit, the BA equations (\ref{Lamda}) give a particular Fredholm equation  which can be deduced from (\ref{Fermi2-a}), i.e.,
\begin{equation}
 \rho_2(k)=\frac{1}{\pi}+\int_{-Q_2}^{Q_2}K_2(k-k')\rho_2(k')dk'\label{BEC-BCS}
 \end{equation}
where  the Fermi pair momentum $Q_2$ is determined by   $n=2\int_{-Q_2}^{Q_2}\rho_2(k)$.
It turns out that \cite{Iida:2005,Wadati:2007,Chen:2010} the reduced Fredholm Eq.~(\ref{BEC-BCS}) maps to the Lieb-Liniger integral equation
for 1D spinless bosons on identifying $m_B=2m_F$, $N_B=N_F/2$ and $\gamma_B=4\gamma_F$.
For a weak attraction,   the groundstate is the BCS-like  pairing  state with a pairing correlation length larger than the average interparticle
spacing and the energy is given by $E=\frac{1}{12}n^3\pi^2-\frac{1}{2}n^2|c|+O(c^2)$.
In particular, for strong attraction, the groundstate of bound pairs determined  by (\ref{BEC-BCS}) can be regarded as a particular
super Tonks-Girardeau gas of hardcore bosons \cite{Chen:2010}.
The distribution function of the pair density  plotted in Fig.~\ref{BEC-BCS crossover} provides an understanding of the subtle BEC-BCS  crossover in the
balanced Fermi gas.
In the weak coupling regime, the single quasimomentum  essentially depends on that of the other particles.
This gives a signature of mutual statistics \cite{Aneziris:1991,Wu:1994,Wilczek:1982,Haldane:1991}.
However, in the limit $|\gamma| \to \infty$, the quasimomentum distribution  becomes an equally-spaced separation.
This indicates free Fermi non-mutual statistics.
Further study on the dimer-dimer scattering properties in the confinement induced resonance has been reported \cite{Mora:2005a,Mora:2005b}, see also \cite{Feiguin:2012}.

\begin{figure}[t]
\includegraphics[width=0.90\linewidth]{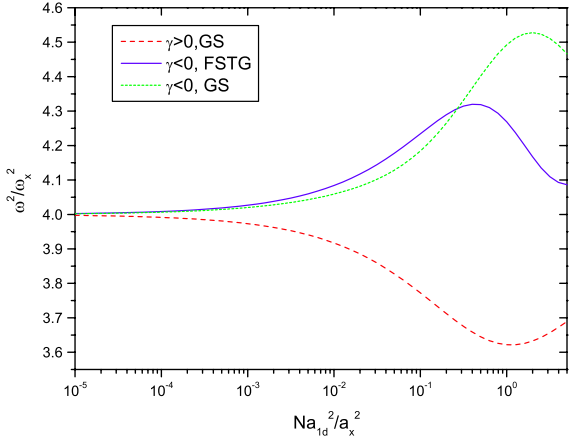}
\caption{The square of the lowest breathing mode frequency vs  the groundstate energy per unit length
vs the rescaled interaction strength $a_{\rm 1d}/\omega_{x}$. The quantum gases are trapped in a 1D harmonic potential
$V_x=\frac{1}{2}m\omega_x^2x^2$.  Here $GS$ and $FSTG$ stand for  the  frequency ratio $\omega^2/\omega_x^2$
for the groundstate and fermionic super Tonks-Girardeau gas, respectively.
From L.-M. Guan and Chen (2010).}
\label{breathering-mode}
\end{figure}

Furthermore,  it was  demonstrated  \cite{Girardeau:2010,Guan-L:2010} that another  metastable highly excited gas-like state
without bound pairs in the strongly  attractive regime can be realized  through a  sudden switch of the interaction from
strongly repulsive to strongly attractive.
In the limit $c\to -\infty$, this gas-like state is still an eigenstate of the system with the energy per particle
$E \approx \frac{1}{3}n^2\pi^2(1+\frac{4\ln2}{|\gamma|} +\frac{12\ln2}{\gamma^2})$, but it is a highly excited state.
From the experimental point of view, these different quantum states can be tested from measuring  the frequencies  of the lowest
breathing mode from the mean square radius of the 1D  trapped gases in a harmonic potential \cite{Menotti:2002,Astrakharchik:2004},
e.g., in the super Tonks-Girardeau Bose gas \cite{Haller:2009}.
The low breathing mode featuring  different states of  the strongly repulsive and attractive Fermi gas  can be analyzed via the
local density approximation.
The lowest breathing mode is given by  the mean square radius of the trapped fermionic Tonks-Girardeau gas
$\omega^2 =-2\langle x^2\rangle/ (d\langle x^2\rangle /d\omega^2_x)$, see Fig.~\ref{breathering-mode}.
Here $\langle x^2\rangle =\int \rho(x) x^2dx /N$.
The frequency ratio $\omega^2/\omega^2_x$ exhibits a peak which is a typical characteristic of the super Tonks-Girardeau phase.
Further evidence for this fermionic super Tonks-Girardeau gas-like state has been seen in the experimental observation of the
fermionization of two distinguishable fermions \cite{ Zurn:2012}.  
It is also particularly interesting that a ferromagnetic transition is likely to occur in 1D strongly interacting fermions across the 
resonance from infinite repulsion to finfinite attraction \cite{Cui:2013a,Cui:2013b}.

 \subsubsection{Solutions to the Fredholm equations and analyticity}

Despite the two sets of Fredholm integral equations (\ref{BE-r}) and (\ref{Fermi2-a}) for the homogeneous gas being derived long
ago \cite{Yang:1967,Yang:1970,Takahashi:1971b},  their analytical study are still restricted to particular regimes, e.g.,
$\gamma \gg 1$, $|\gamma | \ll 1$ and $ |\gamma| \ll  -1$.
The first few terms in the asymptotic expansions of the groundstate energy for the 1D attractive Fermi gas  for  both the strong and
weak coupling cases has been  calculated in terms of power series \cite{Guan:2007a,He:2009,Iida:2005,Iida:2007}
and in terms of Legendre polynomials \cite{Zhou:2012}. The first few terms of the groundstate energy have been
 derived recently  \cite{Guan:2012a} by an asymptotic expansion for (a) strong repulsion, (b) weak repulsion, (c) weak attraction and (d) strong attraction.

\begin{figure}[t]
\includegraphics[width=1.0\linewidth]{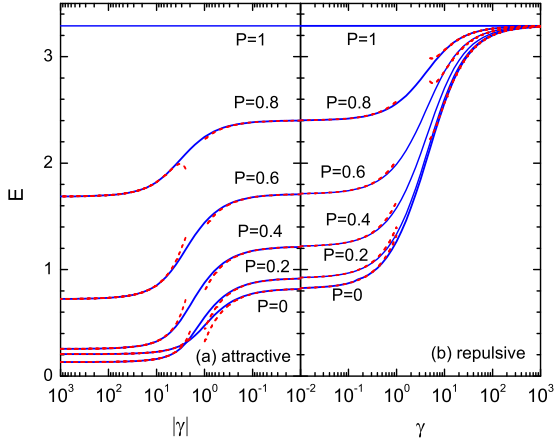}
\caption{The groundstate energy as a function of $\gamma=cL/N$ in units of $\frac{\hbar^2N^3}{2mL^2}$.
The figure shows the  comparison between the asymptotic solutions and numerical solutions of the Fredholm equations for
different polarization. In the attractive regime, the binding energy $\varepsilon_{b} =-c^2/2$ was subtracted.
From Guan and Ma (2012).}
\label{fig:Energy}
\end{figure}

For strong repulsion, the groundstate energy of the Gaudin-Yang model  is given by \cite{Guan:2012a}
\begin{eqnarray}
\frac{E}{L} \approx \left\{ \begin{array}{ll} \frac{1}{3}n^3\pi^2\left[1- \frac{4\ln2}{\gamma} +\frac{12(\ln2)^2}{\gamma^2} \right.  \\
\left.  -\frac{32(\ln2)^3}{\gamma^3}+\frac{\pi^2\zeta(3)}{\gamma^3}  \right], & {\rm for} \,\, P=0,\\
\frac{1}{3}n^3\pi^2\left[ 1-\frac{8n_{\downarrow}}{c} +\frac{48n_{\downarrow}^2}{c^2} \right. \\
\left. -\frac{1}{c^3}\left(256n_{\downarrow}^3-\frac{32}{5}\pi^2n^2n_{\downarrow}\right)  \right], &  {\rm for } \,\, P\ge 0.5.
\end{array}  \right.
\label{energy-SR}
\end{eqnarray}
Here $\zeta(z)$  is the Riemann zeta function.
The leading order ($1/\gamma$)  correction was also found in \cite{Fuchs:2004,Batchelor:2006c}.
Fig.~\ref{fig:Energy} shows that this  groundstate energy  is a  good approximation for the balanced and imbalanced Fermi gas with a  strongly repulsive  interaction.

For strong attraction,  the groundstate energy is given by $E/L=E_0^u+2E_0^b+n_{\downarrow} \varepsilon_b$, where
\begin{eqnarray}
E_0^u& \approx &\frac{(n_{\uparrow}-n_{\downarrow})^3\pi^2}{3}\left[1+\frac{8n_{\downarrow}}{|c|}+\frac{48n_{\downarrow}^2}{c^2}\right.\nonumber\\
&&\left. -\frac{8n_{\downarrow}}{15|c|^3}\left(12\pi^2(n_{\uparrow}-n_{\downarrow})^2-480 n_{\downarrow}^2+5n_{\downarrow}^2\pi^2\right) \right],\label{E0u-sa}\\
E_0^b& \approx& \frac{n_{\downarrow}^3\pi^2}{6}\left[1+\frac{2(2n_{\uparrow}-n_{\downarrow})}{|c|}  +\frac{3(2n_{\uparrow}-n_{\downarrow})^2}{c^2} \right.\nonumber\\
&&\left.  -\frac{4}{15|c|^3} \left(
180n_{\downarrow}n_{\uparrow}^2+20\pi^2n_{\uparrow}^3-90n_{\uparrow}n_{\downarrow}^2-22\pi^2n_{\downarrow}^3\right.\right.\nonumber\\
&&\left.\left.+15n_{\downarrow}^3-120n_{\uparrow}^3+63\pi^2n_{\downarrow}^2n_{\uparrow}-60\pi^2n_{\downarrow}n_{\uparrow}^2\right) \right].\label{E0b-sa}
\end{eqnarray}
This energy is highly accurate for arbitrary polarization as  can be seen in Fig.~\ref{fig:Energy}.
The high precision of expansions for the groundstate energy of the attractive Fermi gas  were also studied \cite{He:2009,Iida:2007,Zhou:2012}.

In contrast to the strong coupling case, it is more difficult to proceed with asymptotic expansion for the two sets of Fredholm equations (\ref{BE-r}) and (\ref{Fermi2-a})
at vanishing interaction strength.
In terms of the polarization $P$, the groundstate energy in weak attraction limit was found to be \cite{Iida:2007}
\begin{equation}
\frac{E}{L}\approx \frac{\pi^2n^3}{12}\left\{(1+3P^2)-\frac{6}{\pi^2}(1-P^2)|\gamma| -B_2\gamma^2\right\}.\label{energy-Iida}
\end{equation}
The coefficient $B_2$ is a  complicated function obtained from  the power series expansions with respect to $\gamma$.
However, it contains divergent sums and the coefficients are singular as $\gamma \to 0$ \cite{Iida:2007}.
So far, only the  leading order correction to the interaction energy is mathematically convincing  and consistent with the result (\ref{e-dis-w})  obtained from the discrete BA equations (\ref{BE}).
Beyond the mean field term, finding the next  leading term in the groundstate energy  is still an open problem.
For zero polarization, Krivnov and Ovchinnikov (1975) found $O(c^2) \approx -\frac{\gamma^2 n^3}{4\pi^2}(\ln |\gamma|)^2$
obtained from  the 1D Hubbard model in dilute limit.
Iida and Wadati (2007) found the term  $O(c^2) \approx -\gamma^2 n^3/12$.
This difference reveals a subtlety of the vanishing interaction limit, i.e., the two limits $c\to 0$ and the thermodynamic limit ($N,\,L \to \infty$ with $N/L$ finite) do not commute.
In fact, the  groundstate energy (\ref{energy-Iida}) only counts the density distributions away from the integration boundaries in the Fredholm equations (\ref{BE-r}) and (\ref{Fermi2-a}),
i.e., $|B_i-k| \sim c$ and $|A_i-k| \sim c$ with $i=1,2$.
At the integration  edges,  these distribution functions are not analytically extractable as $\gamma\to 0$ \cite{Guan:2012a}.

The analyticity of the groundstate energy at $\gamma=0$  is still in question \cite{Guan:2012a,Takahashi:1970c}.
Takahashi showed that (a) the groundstate energy function $f(n_{\uparrow}, n_{\downarrow}; c)$ is analytic on the real $c$ axis
when $n_{\uparrow}\ne n_{\downarrow}$ and (b)  $f(n_{\uparrow}, n_{\downarrow};c)$ is analytic on the real $c$ axis except for
$c=0$ when $n_{\uparrow}= n_{\downarrow}$.
However, the Fredhom equations for weakly repulsive and attractive interactions are identical as long as  $B_1>B_2$ and $A_1>A_2$
where the integration boundaries match each other between the two sides \cite{Guan:2012a}.
In this identical region,  the asymptotic expansions of the energies of the repulsive and attractive fermions  are identical to all orders  as $c\to 0$.
But  the identity of the asymptotic expansions  may not mean that the energy analytically connects due to the divergence of the
Fredholm equations in the region $c\to \mathrm{i}0$.

\section{Many-body physics of the Gaudin-Yang model}

So far, we have only discussed the groundstate properties of the Gaudin-Yang model.
We now survey the wide range of fundamental many-body physics exhibited in the model.

\subsection{1D analog of the FFLO state and magnetism}
\label{Fermi-FFLO}

The particularly interesting  feature of the attractive Fermi gas   is  the   exotic FFLO-like pairing,
where the system is gapless with mismatched Fermi points between the  two Fermi seas.
In the gapped phase,  it is well understood that the
correlation function for the single particle Green's function decays exponentially  \cite{Bogoliubov:1988,Bogoliubov:1989,Bogoliubov:1990}
$\langle\psi_{x,s}^{\dagger}\psi_{1,s}\rangle\rightarrow e^{- x/\xi}$ with $\xi=v_F/\Delta$ and $s=\uparrow,\, \downarrow$,
whereas the singlet pair correlation function decays as a power of
distance, i.e.,
$\langle\psi_{x,\uparrow}^{\dagger}\psi_{x,\downarrow}^{\dagger}\psi_{1,\uparrow}\psi_{1,\downarrow}\rangle\rightarrow x^{-\theta}$.
Here $\Delta$ is the energy gap, and the critical exponents $\xi$ and $\theta$ are both greater than zero.
However,  once the external field exceeds the lower critical field, the system has a gapless phase where  both of these
correlation functions decay as a power of distance and the pairs lose their dominance.
The molecule and excess fermions form  the polarized FFLO-like pairing state, where
the spatial oscillations of the pairing correlation are   caused by an
imbalance in the densities of spin-up and spin-down fermions, i.e., $n_{\uparrow}-n_{\downarrow}$.
In section \ref{CFT-CF},  we will further discuss the pair and spin correlations with the spatial oscillation signature in the context of conformal field theory.

\begin{figure}[t]
{{\includegraphics[width=0.90\linewidth]{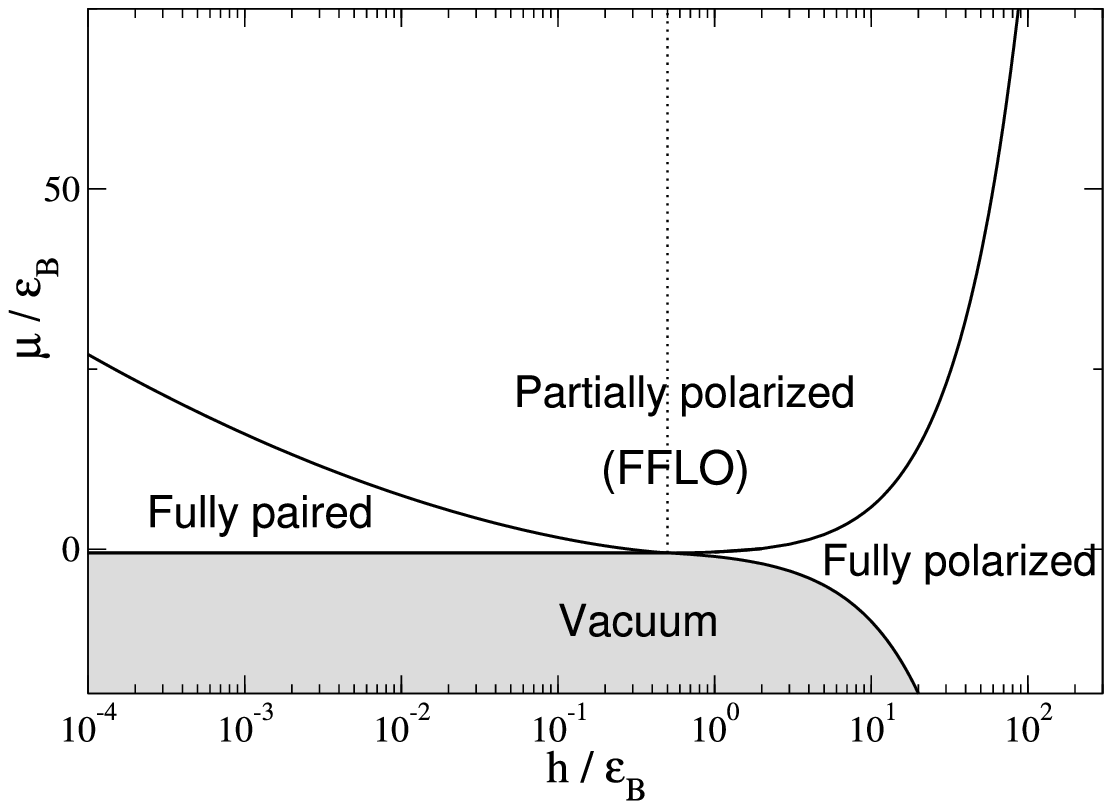}\\
\includegraphics [width=0.780\linewidth,angle=-90]{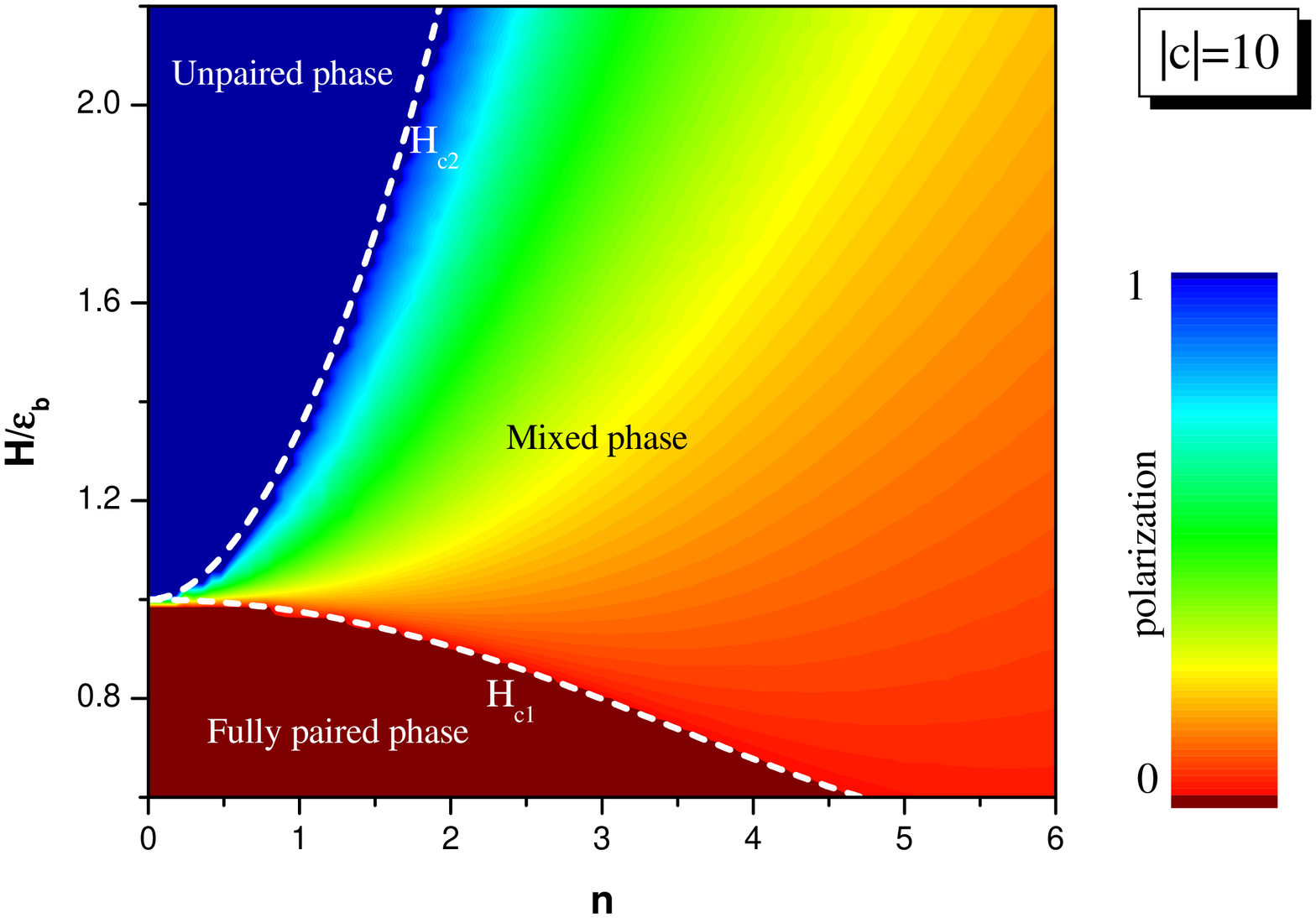}}}
\caption{Upper panel: Phase diagram of the Gaudin-Yang model  in the $\mu-H$ plane.
The phase boundaries are obtained from Eq. (\ref{mu-h}) in terms of the numerical solution of the BA equations (\ref{BE-r}).
From Orso (2007).\\
Lower panel: Phase diagram of the model  in the $H-n$ plane with $|\gamma| =10$ and density $n=1$.
The dashed lines denote the two critical lines  (\ref{Hc1}) and (\ref{Hc2}).
The coloured phases are obtained by numerical solution of the energy-magnetization  (\ref{E-Mz}).
From He {\em et al.} (2009). }
  \label{fig:phase-s}
\end{figure}

In terms of the polarization, the Gaudin-Yang model with attractive interaction exhibits three quantum phases at zero temperature:
the fully paired phase which is a quasicondensate with zero polarization,  the fully polarized normal Fermi gas with  $P= 1$,
and the partially polarized FFLO-like phase with polarization $0 < P < 1$, see the phase diagram in $\mu-H$ plane and in $H-n$ plane, see Fig.~\ref{fig:phase-s}.  The two phase diagrams   describe the quantum phases in terms of the  grand canonical  and canonical ensembles. 
The phase boundaries in the two phase diagrams can be mapped onto each other \cite{Orso:2007,Guan:2011a}.
In this gapless phase,  the magnetic properties  can be exactly described by the  external field-magnetization relation
\begin{equation}
\frac12{H}=\frac12{\epsilon_b}+\mu^u-\mu^b\label{E-Mz}
\end{equation}
where $\mu^b=\mu+\epsilon_b/2$ and   $\mu^u=\mu+H/2$
 are given by (\ref{mu-h}).
This relation reveals an important energy transfer relation among the binding energy, the variation of
Fermi surfaces and the external field.

For fixed density and strong attraction,  the paired phase with magnetization $M^z=0$ is stable when the field $H<H_{c1}$,
where the lower critical field is given by
\begin{equation}
H_{c1}\approx \frac{\hbar^2n^2}{2m} \left[ \frac{\gamma^{2}}{2}
-\frac{\pi^2}{8} \left( 1-\frac{3}{4|\gamma|^2}-\frac{1}{|\gamma|^3}\right) \right].
\label{Hc1}
\end{equation}
When the external field exceeds the upper critical field
\begin{equation}
H_{c2}\approx  \frac{\hbar^2n^2}{2m} \left[ \frac{{\gamma}^{2}}{2}+2{\pi
}^{2}\left( 1-\frac{4}{3|\gamma|}+\frac{16\pi^2}{15|\gamma|^3}\right) \right],
\label{Hc2}
\end{equation}
a phase transition from the FFLO-like  phase into the normal  gas  phase occurs, see Fig.~\ref{fig:phase-s}.
The lower critical field gives the energy gap in the spin sector.

\begin{figure}[t]
{{\includegraphics [width=0.750\linewidth,angle=-90]{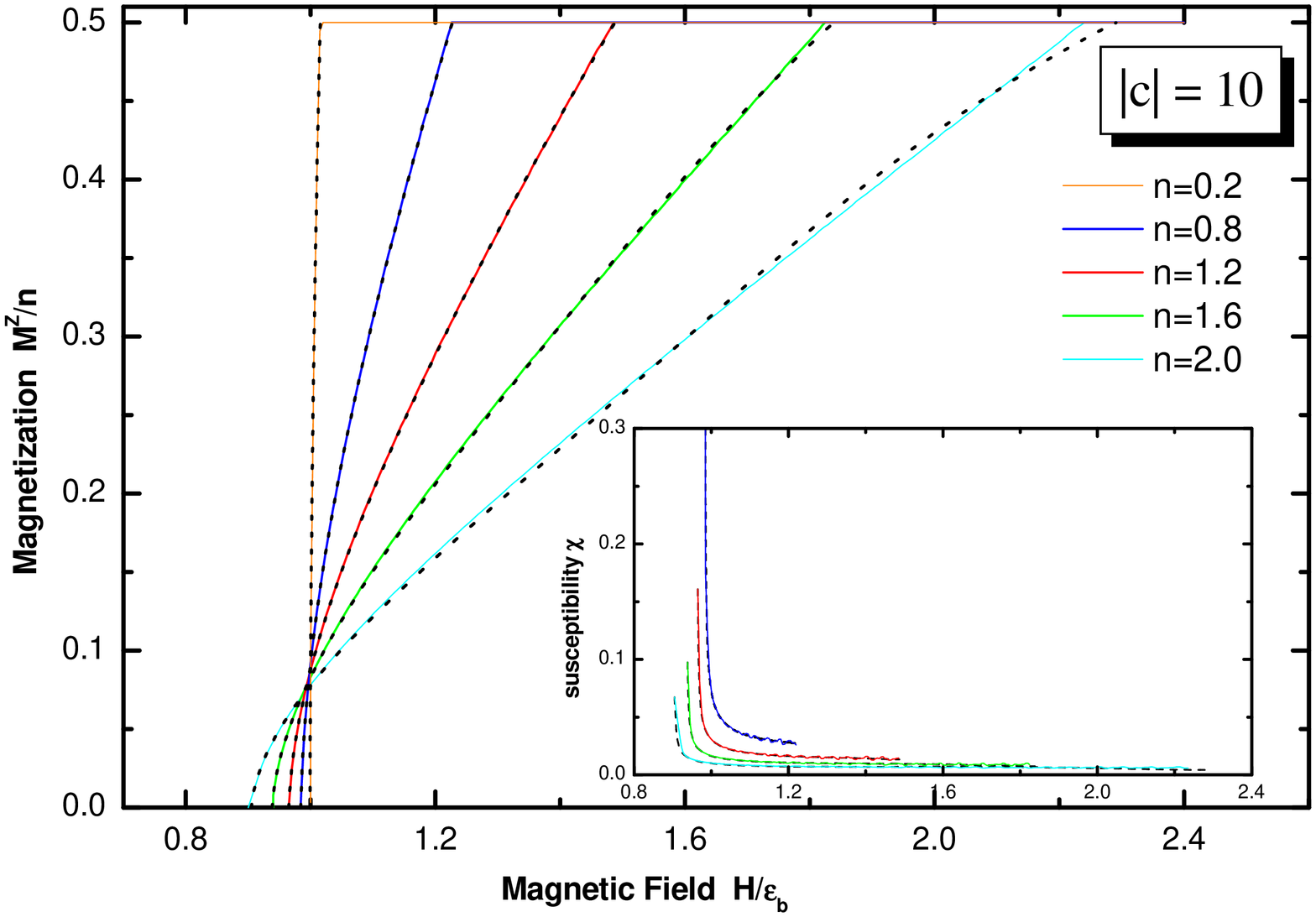}}}
\caption{Magnetization  vs the external field $H/\epsilon_b$ for
$c=-10$ in the units $2m=\hbar=1$ for different densities $n$. The dashed lines are plotted
from the analytic result (\ref{F2-Mz}). The solid curves are obtained from numerical solutions of
the dressed energy equations. The inset shows similar comparison between analytic and numerical results for
the susceptibility vs external field $H/\epsilon_b$.  From He {\em et al.} (2009).}\label{fig:mz-s}
\end{figure}

The magnetization can be obtained from (\ref{E-Mz}), see Fig.~\ref{fig:mz-s}.
It was  found  \cite{Guan:2007a,Iida:2007,He:2009,Woynarovich:1991a}  that in the vicinity of the critical fields $H_{c1}$ and $H_{c2}$,
the system exhibits a linear field-dependent magnetization
\begin{eqnarray}
M^z\approx \left\{ \begin{array}{l} \frac{2(H-H_{c1}) }{n {\pi }^{2}}
\left(1+ \frac{2}{|\gamma|}+  \frac{11}{2{\gamma}^2}+\frac{81-\pi^2}{6|\gamma|^3}\right), \\
 \frac{n}{2}\left[ 1- \frac{H_{c2}-H}{4n^2\pi^2} \left( 1 + \frac
{4}{|\gamma|}+\frac{12}{{\gamma}^{2}}
-\frac{16({\pi
}^{2}-6)}{3{|\gamma|}^{3}} \right)    \right]
\end{array}\right.\label{F2-Mz}
\end{eqnarray}
with a finite susceptibility.
For fixed total number of particles, or say in a canonical ensemble,  the magnetic field driven phase transitions in the 1D  Fermi gases with
an attractive  interaction are linear-field-dependent, which was also found in the $SU(N) $ attractive Fermi gas \cite{Guan:2010,Lee:2011b}.

The magnetism of the attractive Fermi gases has been discussed  by
Schlottmann \cite{Schlottmann:1993,Schlottmann:1994,Schlottmann:1997,Schlottmann:2012a,Schlottmann:2012b}.
However,  the argument which was made on  the initial slope of the magnetization in these papers  \cite{Schlottmann:1993,Schlottmann:1994,Schlottmann:1997}
does not appear to be correct  for a fixed total number of particles.
The reason has been discussed \cite{Woynarovich:1991b}:  ``the bound pairs which have to be broken up to yield the particles with  uncompensated spins
form a Fermi sea, their density of states is finite at the Fermi level, and that keeps the initial susceptibility finite".
It is also shown \cite{Vekua:2009} that the curvature of free dispersion at the Fermi points couples the spin and change modes and leads to a linear
critical behaviour and finite susceptibility for a wide range of models.
They showed that when the magnetic field $H\to H_c$, the magnetization $m^z \sim \sqrt{H-H_c}$ for a fixed chemical potential.
However, for fixed density, the magnetization $m^z \sim (H-H_c)/(\pi v_N^{b})$ as $H= H_c+0^+$.
This leads to a finite onset susceptibility given by $\chi =1/(\pi v_N^b)$ with the  pair density stiffness  $v_N^b=v_F/4$  in the  strong attraction limit $\gamma \to \infty$.
Here we further remark that for finitely strong attraction the onset  susceptibility  $\chi =K^{(b)}/(\pi v_N^{ b})$ where $K^{(b)}\approx (1+\frac{3}{|\gamma|}+\frac{33}{4\gamma^2})$ is the
Tomonaga-Luttinger liquid parameter at the critical point and $v_N^b=\frac{v_F}{4}\left(1-\frac{2}{|\gamma|} -\frac{3}{2\gamma^2}\right)$
is the stiffness of  bound pairs in the limit  $H\to H_c+0^+$.

The magnetization in the Hubbard model with a half-filled band gives rise to the square root dependence  on the field \cite{Takahashi:1969},
where low density  solitons appearing in the spin sector above the critical field behave like free fermions \cite{Japaridze:1978,Japaridze:1981,Pokrovsky:1979}.
More rigorously speaking, the linear field-dependent magnetization  is clearly seen from the energy transfer relation (\ref{E-Mz}),
where the effective chemical potentials $\mu^u \propto (2m^z)^2$ and $\mu^b\propto (n-2m^z)^2$.
Thus the linear term $m^z$ in the relation  (\ref{E-Mz}) gives a finite susceptibility at the onset of magnetization.

\subsection{Fermions in a 1D  harmonic trap}

In experiments, 1D quantum atomic gases are prepared by loading ultracold atoms in an anisotropic harmonic trap with strong transverse confinement and weak longitudinal confinement.
In general, interacting many-body systems trapped in a harmonic potential is a rather complicated problem.
The problem of  the 1D Hamiltonian (\ref{Ham-1})  trapped in a harmonic potential $\frac{1}{2}m\omega _{x}^{2}x^{2}$
has been studied by various methods \cite{Orso:2007,Hu:2007,Yin:2011b,Gao:2008,Colome-Tatche:2008,Yang:2009,Ma:2009,Girardeau:2007,Girardeau:2010,Cui:2012b}.
For $N=2$, the eigenvalue problem of the trapped gas has been studied analytically in \cite{Busch:1998,Idziaszek:2006}.
The energy shift for $N=2$ \cite{Busch:1998} is given by $\sqrt{2}\frac{\Gamma(-E/2+3/4)}{\Gamma(-E/2+1/4)}=1/a_{1D}$, where $a_{1D}$ is  a scattering length
and $\Gamma(x)$ is the Euler gamma function \cite{Busch:1998}.
The system of two fermions with arbitrary interaction in a 1D harmonic potential has been experimentally investigated \cite{Zurn:2012}.
In this experiment, the Tonks-Girardeau state  and the metastable super Tonks-Girardeau state have been observed.

This problem for arbitrary number of particles was studied  analytically \cite{Guan-L:2009,Yang:2009,Ma:2009},
where the limiting cases $c\to \pm \infty$ and $c=0$ have been studied using group theory.
In particular, Yang (2009) gave an analysis  of  the groundstate energy of fermions
in a 1D trap with $\delta$-function interaction.
In the  light of Yang's argument, for any value of interaction strength, the eigenvalue problems:
(a) the trapped Hamiltonian with symmetry $Y=[N-M,M]$ in full $\infty^N$ space and
(b) the Hamiltonian in region $R_Y$ with the boundary condition that the wave function vanishes on its surface,  are equivalent.
Here the region $R_Y$ is bounded by $C_{N-M}^2\times C_{M}^2$ planes at which the wave function $\Psi_Y$ vanishes.
For any value of $g$, the groundstate wave function for problem (b) has no zeros in the interior of $R_Y$ and is not degenerate.
This thus suggests that the groundstate energy of the system with total spin $J=N/2-M$ increases monotonically and
approaches to the energy $E_{J=N/2}$.  The Lieb and Mattis theorem \cite{Lieb:1962}  further suggests $E_{J}>E_{J'}$ if $J>J'$.

For $c\to \infty$, the groundstate energy of the trapped gas with total spin $J$ is given by $E_J=\sum_{n=0}^{N-1}\left(\frac{1}{2}+n \right)=\frac{1}{2}N^2$,
which is independent of the total spin $J$. For $c=0$ and $J=N/2-M$, the energy is given by  $E_J=\frac{1}{2}\left([N/2+J]^2+[N/2-J]^2 \right)$.
Ma and Yang (2009) argued  that $E_J/N^2\to f_J\left(\frac{g}{\sqrt{N}} \right)$ with
\begin{eqnarray}
f_J\left(t\right)=\left\{ \begin{array}{ll} 1/2& {\rm for }\, t\to \infty \\
1/4+\left( J/N\right)^2 & {\rm for}\,  t=0\\
-\left(1/2-J/N \right)t^2/4 & {\rm for }\, t\to -\infty\end{array} \right.\nonumber
\end{eqnarray}
where $t=\frac{g}{\sqrt{N}}$.
In particular, for $c\to \infty$, the exact wave function of the system $\Psi=\psi_A\psi_J$
where the spatial wave function and symmetric spin wave function have been derived explicitly \cite{Guan-L:2009}
\begin{eqnarray}
\psi_A(x_1,\ldots x_N)&=&\frac{1}{(N!)} {\rm det}\left[\phi_j(x_i) \right]_{i=1,\ldots,N}^{j=1,\ldots,N}\nonumber \\
\psi_J&=& \sum _{\alpha =1}^{N!/((N-M)! M!)}\left\{{\cal Y} _\alpha ^{[N-M,M]} Q_\alpha \right\}Z_\alpha \nonumber
\end{eqnarray}
for the symmetry $R=\left[N-M, M \right]$.
Here $Q_\alpha =P_\alpha Q_1$ with $Q_1=\prod_{i=1}^\ell \prod _{j=M+1}^N{\rm sgn}(x_i-x_j)$.
The basis tensor function ${\cal Y} _\alpha^{[M,M]}$ was constructed explicitly from group theory \cite{Guan-L:2009}.

The fermion density distribution for 1D interacting fermions with harmonic trapping has strong oscillations on top of a uniform
density cloud \cite{Gao:2006,Gao:2008,Ma:2009,Guan-L:2009,Rigol:2003}.
These oscillations can be described by an analytical form of the density distribution \cite{Gleisberg:2000,Butts:1997,Soffing:2011}
\begin{equation}
n(x)\approx n_0(x)-\frac{(-1)^{\frac{N}{2}}}{\pi L_F}\frac{\cos (2k_F(x) x)}{1-x^2/L_F^2}
\end{equation}
for $x\le L_F$, where the density cloud is given by the Thomas-Fermi profile, i.e., $n_0(x)=\frac{2\omega L_F}{\pi} \sqrt{1-x^2/L_F^2}$
with a Thomas-Fermi radius $L_F =\sqrt{ N/\omega}$.
If the longitudinal confinement is weak enough, the atomic density varies smoothly along the longitudinal direction
and so the atomic gases can be treated as locally homogeneous systems \cite{Kheruntsyan:2005,Orso:2007,Hu:2007}.
This type of approximate treatment is known as the local density approximation (LDA).  
In this way density functional theory has been used to study 1D interacting fermions 
\cite{Gao:2006,Gao:2007,Magyar:2004,HuH:2010}.

To ensure the validity of the LDA, the correlation length $\xi (z)$ should be much smaller than the characteristic inhomogeneity length
$\xi_{inh}=\frac{n(z)}{\left| dn(z)/dz\right|}$, i.e., $\xi (z) \ll \xi_{inh}$.
The two length scales $\xi (z)$ and $\xi_{inh}$ are determined by the local chemical potential $\mu(z)$ and the local density $n(z)$.
Obviously, from the definition of $\xi_{inh}$, the LDA becomes invalid near the edge of an atomic cloud where the density drops rapidly.
However, in real measurements, almost all signal strengths are proportional to the density.
Therefore, due to the very small density at the edge, the central region of large density dominates the measurement signals.
For a large number of particles, $N\to \infty$, the density profiles of the trapped gas can be precisely analyzed within the LDA.

In a harmonic trap, the equation of state (\ref{mu-h})  can be reformulated within the LDA by the replacement $\mu \left( x\right)
=\mu \left( 0\right) -\frac{1}{2}m\omega _{x}^{2}x^{2}$ in which $x$ is the
position and $\omega _{x}$ is the frequency within the trap.
Using the LDA for the 1D Bose gas in a harmonic trap, its global chemical potential reads \cite{Kheruntsyan:2005}
\begin{equation}
\mu_{g} = \mu_{0}[n(z)]-V(z) = \mu_{0}[n(z)] -\frac{1}{2}m\omega_{z}^{2} z^2,
\end{equation}
where the local chemical potential $\mu_{0}[n(z)]$ at position $z$ is given by the chemical potential for a
homogeneous system of an uniform density $n=n(z)$.
The total number of atoms $N$ is given as $N = \int n(z) dz$.

Similarly, for a 1D two-component Fermi gas in a harmonic trap, the global chemical potential is  \cite{Orso:2007,Hu:2007,Ma:2010a,Heidrich-Meisner:2010a}
\begin{equation}
\mu_{\rm hom}[n(x),P(x)]=\mu_0-\frac{1}{2}m\omega_x^2x^2\nonumber
\end{equation}
where the chemical potential  $\mu_{\rm hom}[n,P]$ can be obtained from  the homogenous gas (\ref{mu-h}).
$n(x)$ is the total linear number density and $P(x)$ is the local spin polarization.
They can be determined from restriction  on the total particle
number $N=\int_{-\infty }^{\infty }n\left( x \right) dx$ and polarization $P=\int_{-\infty }^{\infty }n^{u}\left( x\right) dx/N$
which are rewritten as   \cite{Orso:2007,Hu:2007,Yin:2011b,Gao:2008}
\begin{eqnarray}
{Na_{1D}^{2}}/{a_{x}^{2}} &=&4\int_{-\infty }^{\infty }\tilde{n}\left(
x\right) d\tilde{x}  \nonumber \\
\left( {Na_{1D}^{2}}\right)\,P &=&4\int_{-\infty }^{\infty }\tilde{n}^{u}\left( x\right) d\tilde{x}\times
{a_{x}^{2}}. \label{N_P}
\end{eqnarray}
Here $\tilde{n}\left( x\right) =1/\left\vert \gamma \left( x\right) \right\vert $,  $a_x=\sqrt{\hbar/(m\omega_x)}$  and $\tilde{n}^{u}\left( x\right) =n^{u}\left( x\right) /|c|$.

\begin{figure}[t]
\includegraphics[width=0.95\linewidth]{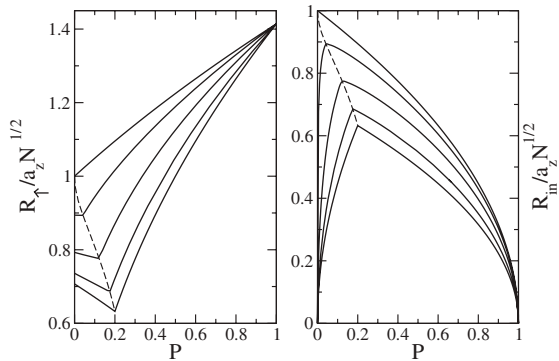}
\caption{The cloud radii of outer shell $R_{\uparrow} $ and inner shell $R_{\rm in}$ are theoretically predicted by means of the 
BA equations (\ref{BE-r}) within the LDA. This figure shows the radii vs polarization $P$ for the values of the parameter 
$Na_{1D}^{2}/a_{x}^{2}=\infty,\,10,\, 1,\, 0.1,\,0$. From Orso (2007).}
\label{fig:phase-attraction}
\end{figure}

If the trapping potentials are the same for the two spin components, calculations for the integrable homogenous attractive gas confined to a 1D trapping potential  thus lead to a two-shell structure composed  of a partially polarized 1D FFLO-like state in the trapping centre surrounded by wings composed of either a fully paired state
or a fully polarized Fermi gas \cite{Orso:2007,Hu:2007,Gao:2008},  see Fig.~\ref{fig:phase-attraction}.
This prediction was verified by Liao {\em et al.} (2010) by the observation of three distinct phases in experimental measurements
of ultracold $^6$Li atoms in an array of 1D tubes.
The analytical study of the phase diagram of the 1D attractive Fermi gas has been  presented in \cite{Guan:2007a,Guan:2011a,Iida:2007}.

\subsection{Tomonaga-Luttinger liquids}

The TLL  \cite{Tomonaga:1950,Luttinger:1963}, describing the collective motion of bosons,
has played an important role in the novel description of universal low energy physics for
low-dimensional many-body physics \cite{Giamarchi:2004,Gogolin:1998}.
In 1D systems of interacting bosons, fermions or spin systems, the effect of quantum fluctuations is
strong enough to yield striking anomalous quantum phenomena.
In this approach, for example, the low energy physics of a 1D  interacting fermion system can be  described by
a bilinear form of bosonic creation and annihilation operators.
The TLL  is phenomenologically  treated by bosonization techniques \cite{Gogolin:1998,Tsvelik:1983,Giamarchi:2004,Cazalilla:2011}
based on a linearization of the dispersion relation of the particles in the collective motion, i.e., $\omega(q)=v_s\,|q|$, here $v_s$
is the sound velocity of the collective  motion.
In contrast to the Fermi liquid, this thus leads to a power law density of states for the TLL at the Fermi energy $E_F$, i.e., $|E-E_F|^\alpha$,
where the exponent $\alpha =\left(K+1/K-2\right)/4$ depending on the so-called TLL  parameter $K$.

In general, the correlation functions of such 1D systems at zero temperature show a power-law decay determined by the TLL parameter $K$ and the velocity $v_s$.
These critical systems not only have global scale invariance but exhibit local conformal invariance.
With the help of exact BA solutions, a wide class of 1D interacting systems can be mapped onto TLLs in the low-energy limit,
including the electronic systems  with spin degrees of freedom like spin-charge separation
\cite{Voit:1994,Kawakami:1990,Schulz:1990,Schulz:1991,Solyom:1979,Giamarchi:2004,Essler:2005}.
Moreover, progress in treating such collective motion of particles beyond the low-energy limit was made by
Imambekov and Glazman (2009a,2009b) and Imambekov {\em et al.} ( 2011).
This method can be applied to a wide variety of 1D systems with collective motion of particles.
This generalized TLL theory could be possibly justified through exact BA results for 1D integrable models in
ultracold atoms and correlated electronic systems.

In contrast to the conventional quasiparticles carrying both spin and charge degrees,
the elementary excitations form spin and charge waves that propagate with different velocities in 1D \cite{Gogolin:1998}.
The relativistic dispersion relation for each one of these excitations is written as $\omega_{\mu}(p)=\sqrt{\Delta^2_{\mu}+v^2_{\mu}p^2}$,
where $\Delta_{\mu}$ is the energy gap and $v_{\mu }$ is the velocity.
For a gapless excitation with vanishing energy gap $v_{\mu}=\partial _p \omega_{\mu} (p)$.
For 1D interacting systems, this gives a phonon dispersion that leads to conformal invariance in the excitation spectrum.
However, for a large  energy gap,  the dispersion can be rewritten as
$\omega_{\mu}(p) =\Delta_{\mu} +v^2_{\mu}p^2/(2\Delta_{\mu}):= \Delta_{\mu}+p^2/(2m^*_{\mu})$
which is the  classical dispersion of a free particle with an effective mass $m_{\mu}^*$.

\begin{figure}[t]
{{\includegraphics [width=0.90\linewidth]{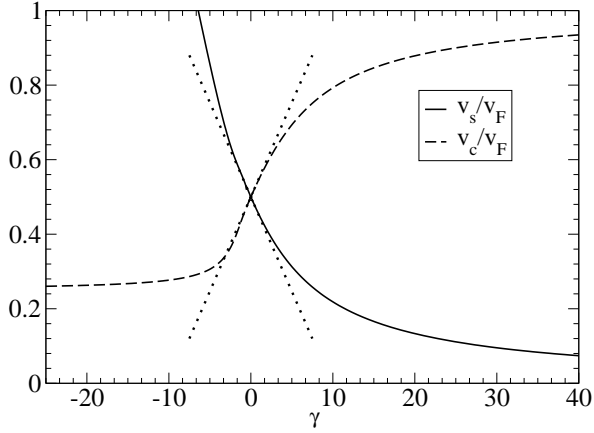}}}
\caption{Charge and spin velocities vs dimensionless interaction strength $\gamma$ for the 1D balanced Fermi gas.
The solid lines are the  velocities obtained by numerically solving the BA equations (\ref{BE}).
The dotted lines denote the analytical result for the velocity in the weak coupling regimes.
From Batchelor {\em et al.} (2006a). }
   \label{fig:velocities}
\end{figure}

From the BA solution (\ref{BE}) with $N_{\uparrow}=N_{\downarrow}$,
the charge and spin velocities are $v_{c,s}=\frac{1}{2}v_F\left(1\pm \gamma/\pi^2 \right)$
for the weak coupling regime \cite{Fuchs:2004,Batchelor:2006a,Batchelor:2006c}. Here the Fermi velocity $v_F=\hbar \pi n/m$.
For strong attraction, the charge and spin  velocities are given by
$v_c=\frac{1}{4}v_F(1-1/\gamma)$ and $v_s=\sqrt{\Delta}(1-2/\gamma)$  \cite{Fuchs:2004,Batchelor:2006a,Batchelor:2006c}
with an energy gap $\Delta \approx \frac{\hbar^2 }{2m}\frac{c^2}{2}$.
This gap increases with increasing interaction strength $\gamma$  so that the spin velocity is divergent in the strongly attractive limit.
However, for strong repulsion the charge velocity   $v_c=v_F(1-4\ln 2/\gamma)$  tends to the Fermi velocity and the spin velocity goes to
zero $v_s=\frac{v_F\pi^2}{3\gamma}(1-6\ln 2 /\gamma)$ \cite{Lee:2012a} due to suppression of spin transportation due to the strong repulsion,
see Fig.~\ref{fig:velocities}.
We shall discuss TLLs and spin-charge separation phenomena in the attractive Fermi gas in the following two sections.

\subsection{Universal thermodynamics and Tomonaga-Luttinger liquids in attractive fermions}

The Yang-Yang formalism with  its generalisation for the study of thermodynamics of BA integrable systems \cite{Takahashi:1999} is a
convenient tool for the study of universal thermodynamics and quantum criticality in the presence of external fields.
At finite temperatures and in the thermodynamic limit, the densities in the Fredholm equations (\ref{Fermi2-a})
evolve into occupied and unoccupied roots in the whole parameter spaces, namely $A_1,A_2 \to \infty$.
In particular, the roots in the spin sector form complicated string patterns that characterize the
spin excitations, i.e., spin wave bound states.
The density distribution functions of pairs, unpaired fermions and spin strings involve the densities of
`particles' $\rho_i(k)$ and `holes' $\rho_i^h(k)$ ($i=1,2$).
Following the Yang-Yang grand canonical ensemble method,  the grand partition function is written as
$Z={\mathrm {tr}} (\mathrm{e}^{-\cal{H}/T})=\mathrm{e}^{-G/T}$, in terms of the Gibbs
free energy $G = E - HM^z - \mu n - TS$ and the magnetic field $H$,
the chemical potential $\mu$ and the entropy $S$ \cite{Takahashi:1999}.
In terms of the dressed energies $\epsilon^{\rm b}(k) := T\ln( \rho_2^h(k)/\rho_2(k) )$ and
$\epsilon^{\rm u}(k) := T\ln( \rho_1^h(k)/\rho_1 (k) )$ for paired and unpaired fermions,
the equilibrium states are determined by the minimization condition of
the Gibbs free energy, which gives rise to a set of coupled nonlinear
integral equations -- the TBA equations \cite{Takahashi:1999}.
For the attractive Gaudin-Yang model, these equations are
\begin{eqnarray}
\epsilon^{\rm
  b}(k)&=&2(k^2-\mu-\frac14{c^2})+TK_2*\ln(1+\mathrm{e}^{-\epsilon^{\rm b}(k)/T} )
  \nonumber\\
& &+ \, TK_1*\ln(1+\mathrm{e}^{-\epsilon^{\rm u}(k)/{T}})\nonumber\\
\epsilon^{\rm
  u}(k)&=&k^2-\mu-\frac12{H}+TK_1*\ln(1+\mathrm{e}^{-\epsilon^{\rm b}(k)/{T}})\nonumber\\
& &-T\sum_{\ell=1}^{\infty}K_\ell*\ln(1+\eta_\ell^{-1}(k)),\label{TBA-Full-1}
\end{eqnarray}
\begin{eqnarray}
\ln \eta_\ell (\lambda)&=&\frac{\ell H}{T}+K_\ell *\ln(1+\mathrm{e}^{-\epsilon^{\rm u}(\lambda)/{T}})\nonumber
\\&&+\sum_{m=1}^{\infty}T_{\ell m}*\ln(1+\eta^{-1}_m(\lambda)).\label{TBA-Full}
\end{eqnarray}

The function $\eta_\ell (\lambda) := \xi^h_\ell(\lambda)/\xi_\ell  (\lambda ) $ is the ratio of the
string densities.
Here $*$ denotes the convolution integral
$(f*g)(\lambda) = \int_{-\infty}^\infty f(\lambda-\lambda') g(\lambda') d\lambda'$.
The function $T_{\ell m}(k)$ is given, e.g., in \cite{Takahashi:1999,Guan:2007a}.
The Gibbs free energy per unit length is given by $G=p^b+p^u$ where the effective pressures of the bound pairs and unpaired fermions are given by
\begin{eqnarray}
p^b&=&-\frac{T}{\pi}\int_{-\infty}^{\infty}dk\ln(1+\mathrm{e}^{-\epsilon^{\rm
      b}(k)/{T}}),\nonumber\\
      p^u&=&-\,\frac{T}{2\pi}\int_{-\infty}^{\infty}dk \ln(1+\mathrm{e}^{-\epsilon^{\rm u}(k)/{T}}).\label{pressure}
\end{eqnarray}

In the grand canonical ensemble, the total number of particles associated with the chemical potential $\mu$ can be changed.
The Fermi sea of unpaired fermions can be lifted by the external field.
The entropy $S$ is a measure of the thermal disorder.
The spin fluctuations (spin strings) are ferromagnetically coupled to the Fermi sea of unpaired fermions.
The direct numerical computation of the TBA equations was presented by Kakashvili and Bolech (2009).
The TBA  equations for this model  involve an infinite number of coupled nonlinear integral equations that
impose a number of challenges to accessing the physics of the model.

For zero external field, the lowest excitations split into collective excitations
carrying  charge and collective excitations carrying spin.
This leads to the phenomenon of spin-charge separation.
The charge excitations are described by sound modes with a linear dispersion.
However, for the external field in excesses of the lower critical field the spin gap vanishes.
In contrast to the spin-charge separation formalism, the spin-charge coupling drastically
changes the critical behaviour  in  the attractive regime of  the Fermi gas.
The TBA equations (\ref{TBA-Full}) indicate  that the spin fluctuations
(the spin wave bound states) are ferromagnetically coupled to the Fermi sea
of unpaired fermions \cite{Zhao:2009}.
In contrast to the antiferromagnetic coupling  $J_{\rm AF}=-\frac{2}{|c|}p^{\rm u}(T,H)$ for
repulsive regime \cite{Guan:2008b}, the spin-spin exchange interaction in the
spin sector is described by an effective spin-1/2 ferromagnetic chain
with a coupling constant $J_{F}\approx \frac{2}{|c|}p^{\rm u}(T,H)>0$ in the strong coupling regime $|c|\gg1$.
The ferromagnetic spin wave fluctuations are
produced due to the thermal fluctuation in the Fermi sea of unpaired fermions.
However, $J_F$ tends to zero for $\gamma \to \infty$.
Therefore the spin transportation becomes weaker and weaker until it vanishes as $|\gamma|\to \infty$.
At zero temperature all unpaired fermions are polarized and spin strings are fully suppressed.
In this gapless phase, excitations involve particle-hole excitations and spin-string excitations.
The TBA equations (\ref{TBA-Full}) can be greatly simplified in the strong coupling  regime
due to the suppression of spin fluctuations,  where $\eta^{-1}_\ell  \sim  e^{-\ell H/T} \to 0$ as  $T\to 0$.
Thus one can extract the universal TLL physics using
Sommerfeld  expansion for temperatures less than chemical potential and magnetic field.

In fact,  in this  spinless phase, the spin fluctuation is suppressed in the limit  $T\to 0$ and $|\gamma| \gg 1$.
Thus  the bound pairs and unpaired fermions form a two-component TLL.
Conformal invariance predicts that the energy per unit length has a
universal finite-size scaling form that is characterized by the
dimensionless number $C$, which is the central charge of the
underlying Virasoro algebra \cite{Blote:1986,Affleck:1986}.
The finite-size corrections to the groundstate energy have been analytically derived  \cite{Lee:2011a}
\begin{equation}
\varepsilon_{0}=\varepsilon_{0}^{\infty}-\frac{C\pi}{6L^{2}}\sum_{\alpha=u,b}v_{\alpha},
\label{eq:energy_Lfinalground}
\end{equation}
where $C=1$ with $v_{u}$ and $v_{b}$ the velocities of unpaired fermions and bound pairs, respectively.
For strong interaction, they are given explicitly by
\begin{eqnarray}
v_{\rm b}&\approx &\frac{\hbar}{2m}\pi n_2\left(1+\frac{2A_2}{|c|}+\frac{3A^2_2}{c^2} \right)\nonumber \\
v_{\rm u}&\approx &\frac{\hbar}{2m} 2\pi n_1\left(1+\frac{2A_1}{|c|}+\frac{3A^2_1}{c^2} \right),\label{vb-vu}
\end{eqnarray}
where $A_1=4n_2$, $A_2=2n_1+n_2$ and $n_2=n_{\downarrow}$.
We will describe universal behaviour of the macroscopic properties of this Fermi gas in the following subsections.

\begin{figure}[tbp]
\includegraphics[width=1.1\linewidth]{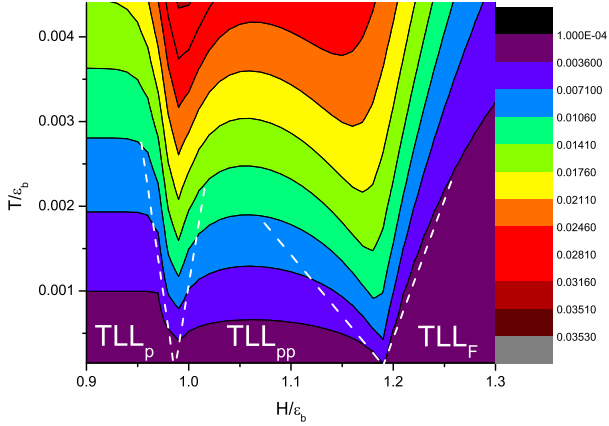}
\caption{Quantum phase diagram of the Gaudin-Yang model  in the $T-H$ plane showing a contour plot of the entropy in the
strong interaction regime. The dashed lines are determined from the deviation from linear temperature dependent entropy
obtained from the result (\ref{FF-C}). The universal crossover temperatures separate the TLLs from quantum critical regimes.
From  Yi {\em et al.}  (2012).
}
\label{fig:entropy1}
\end{figure}

Although a phase transition in 1D many-body systems at finite temperatures does not exist,  the system does exhibit
universal crossover from relativistic dispersions to quadratic dispersions.
Thus  at low temperatures, the bound pairs, normal Fermi gas, and the FFLO phase become relativistic TLLs
of bound pairs ($TLL_P$), unpaired	fermions ($TLL_F$),	and a two-component TLL  ($TLL_{PP}$), respectively, see  Fig.~\ref{fig:entropy1}.
A detailed discussion has been given \cite{Zhao:2009,Yi:2012}.
For the temperature $k_BT \ll E_F$, where $E_F$ is the Fermi energy, the leading low temperature correction to the free energy
of the polarized gas can be calculated explicitly using Sommerfeld expansion with the pressures (\ref{pressure}),
namely \cite{Guan:2007a,He:2009,Zhao:2009,Batchelor:2010}
\begin{equation}
F(T,H)\approx \left\{ \begin{array}{ll} E_0(H)-\frac{\pi  C k_B^2T^2}{6\hbar}\left(\frac{1}{v_{\rm b}}+\frac{1}{v_{\rm u}}\right), & {\rm for}  \, TLL_{PP}\\
E_0(H)-\frac{\pi  C k_B^2T^2}{6\hbar}\frac{1}{v_{\rm b}}, & {\rm for}  \, TLL_P\\
E_0(H)-\frac{\pi  C k_B^2T^2}{6\hbar}\frac{1}{v_{\rm F}}, & {\rm for}  \, TLL_F
\end{array}\right.
\label{FF-C}
\end{equation}
which belongs to the universality class of the Gaussian model with central charge $C=1$.
For strong attraction,  the velocities are given in (\ref{vb-vu}).
In the above equation, the groundstate energy $E_0(H)$ is as given in Sec.~\ref{ground-state}.
In fact, from the TBA equations (\ref{TBA-Full}), the universal thermodynamics (\ref{FF-C}) can be shown to be valid for arbitrary interaction strength.

The two branches of gapless excitations in the 1D FFLO-like phase form collective motions of particles.
The low energy  (long wavelength) physics of the strongly attractive Fermi gas is  described by an effective Hamiltonian
\begin{eqnarray}
H_{\rm eff} &=& \frac{v_{u}}{2}\left[(\partial _x \phi_{u})^2+(\partial _x\theta_{u})^2 \right]+\frac{v_b}{2}\left[(\partial _x \phi_{b})^2+(\partial _x\theta_{b})^2 \right]\nonumber\\
&&-\frac{h}{2}\frac{\partial _x \phi_{u}}{\sqrt{\pi} }-\mu \frac{(\partial _x \phi_u+2\partial_x \phi_b)}{\sqrt{\pi}}
\end{eqnarray}
as long as the spin fluctuation is frozen out \cite{Vekua:2009,Zhao:2009}.
Here the fields $\partial_x \phi_i,\partial_x \theta_i$ with $i=b,u$ are the density and current fluctuations for the pairs and unpaired fermions.
However, in the spin gapped phase, i.e., for $H<H_{c1}$, the energy gap in the spin sector leads to an exponential decay of spin correlations,
whereas the singlet pair correlation and charge density wave correlations have a power-law decay \cite{Gao:2007,Cazalilla:2005}.
Thus the system in the spin gapped phase forms a so-called Luther-Emery liquid \cite{Luther:1974}.

\subsection{Quantum criticality and universal scaling}

As we have seen, the 1D attractive Fermi gas  exhibits  various phases of strongly correlated quantum liquids and is
thus particularly valuable to investigate quantum criticality.
Near a quantum critical point, the many-body system is expected to show universal scaling
behaviour in the thermodynamic quantities due to the collective nature of the many-body effects.
In the framework of Yang-Yang TBA thermodynamics, exactly solvable models of ultracold atoms, exhibiting quantum phase transitions,
provide a rigorous way to treat quantum criticality in archetypical quantum many-body systems, such as the Gaudin-Yang Fermi gas \cite{Guan:2011a},
the Lieb-Liniger Bose gas \cite{Guan:2011b}, a mixture of bosons and fermions \cite{Yin:2012}, and
the spin-1 Bose gas with both delta-function interaction and antiferromagnetic interaction \cite{Kuhn:2012a,Kuhn:2012b}.

At zero temperature, the quantum phase diagram in the grand canonical ensemble can be analytically determined from the
so-called dressed energy equations \cite{Guan:2011a,Guan:2011b,Takahashi:1999}
\begin{eqnarray}
\epsilon^{\rm b}(\Lambda)&=&2\left(\Lambda^2-\mu
-\frac{c^2}{4}\right)-\int_{-A_2}^{A_2}K_2(\Lambda-\Lambda'){\epsilon^{\rm
    b}}(\Lambda')d\Lambda' \nonumber\\ &
&-\int_{-A_1}^{A_1}K_1(\Lambda-k){\epsilon^{\rm u}}(k)d k,\nonumber\\
\epsilon^{\rm u}(k)&=& k^2-\mu -\frac{H}{2} -\int_{-A_2}^{A_2}K_1(k-\Lambda){\epsilon^{\rm
    b}}(\Lambda)d\Lambda,
\label{TBA-F}
\end{eqnarray}
which are obtained from the TBA equations  in the limit $T\to 0$.
The integration boundaries $A_2$ and $A_1$ characterize the Fermi surfaces for bound pairs and unpaired fermions, respectively.
It is convenient to use dimensionless quantities where energy and length are measured  in units of binding energy $\varepsilon_{b}$ and $c^{-1}$ respectively.
In terms of the dimensionless quantities
$\tilde{\mu}:= \mu/\varepsilon_{b}$, $h:= H/\varepsilon_{b}$, $t:= T/\varepsilon_{b}$,
$\tilde{n}:= n/|c|=\gamma^{-1}$, $\tilde{p} := P/|c\varepsilon_{b}|$,
the phase boundaries have been  determined analytically from (\ref{TBA-F}), see \cite{Guan:2011a}.
There are four phases denoted by vacuum $(V)$, fully paired phase $(P)$, ferromagnetic phase $(F)$, and
partially paired $(PP)$ or ($FFLO$-like) phase presenting the same phase diagram as in Fig.~\ref{fig:phase-attraction}.

The low density and strong coupling limits are particularly important to study quantum criticality.
In fact, the TBA equations (\ref{TBA-Full}) can be converted into a dimensionless form with the above rescaling.
Following the notation used  in \cite{Guan:2011a}, the phase boundaries between $V-F$, $V-P$, $F-PP$ and $P-PP$
are denoted by $\mu_{c1}$ to $\mu_{c4}$ respectively.
The closed forms of the critical fields
\begin{eqnarray}
\mu_{c1}&=&-\frac{h}{2}; \qquad
\mu_{c2}=-\frac{1}{2}, \nonumber \\
\mu_{c3}&=&
 -\frac{1}{2}\left(1-\frac{2}{3\pi}(h-1)^{\frac{3}{2}}-\frac{2}{3\pi^2}(h-1)^2\right),\nonumber \\
 \mu_{c4}&=&
-\frac{h}{2}+\frac{4}{3\pi}(1-h)^{\frac{3}{2}}+\frac{3}{2\pi^2}(1-h)^2. \label{CR-S-4}
\end{eqnarray}
are needed to determine scaling functions of thermodynamic properties.
Here  $\mu_{c1}, \mu_{c2}$  applies to all regimes and $\mu_{c3}, \mu_{c4}$ are expressions in the strongly interacting regime.
The above critical fields $\mu_{c3}, \mu_{c4}$  correspond to the upper and lower critical fields in  the $h-n$  plane,
which were found in \cite{Guan:2007a,He:2009,Iida:2007}.

The TBA equations  (\ref{TBA-Full-1})  and (\ref{TBA-Full})  encode the microscopic roles of each single particle that lead to a global coherent state -- quantum criticality.
The quantum criticality is manifested  by universal scaling of thermodynamic properties near the critical points.
The key input to obtain critical scaling behaviour is to derive the form of the equation of state which takes full
thermal and quantum fluctuations at low temperatures into account.
The dimensionless form of the pressure \cite{Guan:2011a}
 \begin{equation}
 \tilde{p}(t, \tilde{\mu}, h) := p/(|c|\varepsilon_b)=\tilde{p}^b+\tilde{p}^{u}, \label{EOS}
 \end{equation}
serves as equation of state, where, to $O(c^4)$,  the pressures of the bound pairs and unpaired fermions are given by
 \begin{eqnarray}
\tilde{p}^b&=&-\frac{t^{\frac{3}{2}}}{2\sqrt{\pi}}F_{3/2}^{b}\left[1+\frac{\tilde{p}^b}{8}+ 2\tilde{p}^u\right] + O(c^4),\nonumber\\
\tilde{p}^u&=&-\frac{t^{\frac{3}{2}}}{2\sqrt{2\pi}}F_{3/2}^{u}\left[1+ 2\tilde{p}^b\right] + O(c^4) 
\end{eqnarray}
with in addition 
\begin{eqnarray}
 \frac{X_b}{t}&=&\frac{\nu_b}{t}-\frac{\tilde{p}^b}{t}-\frac{4\tilde{p}^u}{t}-\frac{t^{\frac{3}{2}}}{\sqrt{\pi}}\left( \frac{1}{16}f_{5/2}^{b}+\sqrt{2}f_{5/2}^{u}\right)\label{Xb},  \nonumber\\
\frac{X_u}{t}&=&\frac{\nu_u}{t}-\frac{2\tilde{p}^b}{t}-\frac{t^{\frac{3}{2}}}{2\sqrt{\pi}}f_{5/2}^{b}+\mathrm{e}^{-{h}/{t}} \mathrm{e}^{-K}I_0(K). \nonumber
\end{eqnarray}
In these equations the functions $F_n^b$, $F_n^u$, $f_n^b$, and $f_n^u$ are defined by
$ F_n^{b,u} := {\rm Li}_n\left(-{\rm e}^{{X_{b,u}}/{t}}\right)$ and $f_n^{b,u} := {\rm Li}_n \left(  -{\rm e}^{{\nu_{b,u}}/{t}}\right)$,
with the notation $\nu_{b} = 2{\tilde{\mu}}+1$, $\nu_{u} =  \tilde{\mu} + h/2$.
The function ${\rm Li}_{s}(z) = \sum_{k=1}^{\infty}z^{k}/k^{s}$ is the polylog function and $I_0(x)=\sum_{k=0}^{\infty}\frac{1}{(k!)^2}(\frac{x}{2})^{2k}$.
Despite the equation of state (\ref{EOS}) having only  a few leading terms in expansions with respect to the
interaction strength, it contains thermal fluctuations in contrast to the TLL thermodynamics (\ref{FF-C}).

The TLL  thermodynamics (\ref{FF-C}) has been derived from low temperature expansion along $T\ll |\mu-\mu_c|$.
This universal thermodynamics is a consequence of the linearly dispersing phonon modes \cite{Maeda:2007}, i.e.,
the long wavelength density fluctuations of  two weakly coupled gases or a gas of bound pairs or single fermions.
The quantum critical regime lies  beyond  $T\gg |\mu-\mu_c|$.
In this limit, the equation of state (\ref{EOS}) provides closed forms for the scaling functions of thermodynamic quantities,
such as density, magnetization and the compressibility.
Near the critical point, the thermodynamic functions can be cast into a universal scaling form \cite{Fisher:1989,Sachdev:1999}.
The explicit universal scaling form of  the density for $T\gg |\mu-\mu_c|$ is
 \begin{eqnarray}
 \begin{array}{ll}
(V{\rm -}F) & \tilde{n} \approx - \frac{\sqrt{t}}{2\sqrt{2\pi}} {\rm Li}_{\frac{1}{2}} \left(-\mathrm{e}^{{(\tilde{\mu}-\mu_{c1})}/{t}}\right),\\
  (F{\rm -}PP) &   \tilde{n}  \approx  n_{o3} -\lambda_{1}\sqrt{t} \, {\rm Li}_{\frac{1}{2}} \left(-\mathrm{e}^{{2(\tilde{\mu}-\mu_{c3})}/{t}}\right), \\
 (V{\rm -}P)&  \tilde{n} \approx  -\frac{\sqrt{t}}{\sqrt{\pi}} {\rm Li}_{\frac{1}{2} }\left(-\mathrm{e}^{{2(\tilde{\mu}-\mu_{c2})}/{t}}\right), \\
(P{\rm -}PP) & \tilde{n} \approx  n_{o4} -\lambda_{2}\sqrt{t} \, {\rm Li}_{\frac{1}{2}} \left(-\mathrm{e}^{{(\tilde{\mu}-\mu_{c4})}/{t}}\right).
\end{array} \label{QC-n}
 \end{eqnarray}
Here the constants  $n_{o3}$ and $n_{o4}$  are the background densities near the critical points $\mu_3$ and $\mu_4$.
These constants, together with $a$ and $b$,  are known  explicitly  in terms of $h$  \cite{Guan:2011a}.

The  universal scaling form for the compressibility is
 \begin{eqnarray}
 \begin{array}{ll}
(V{\rm -}F) & \tilde{\kappa} \approx- \frac{1}{2\sqrt{2\pi t}} {\rm Li}_{-\frac{1}{2}} \left(-\mathrm{e}^{{(\tilde{\mu}-\mu_{c1})}/{t}}\right), \\
 (F{\rm -}PP) &  \tilde{\kappa} \approx \kappa_{o3} -\frac{\lambda_{4}}{\sqrt{t}} {\rm Li}_{-\frac{1}{2}} \left(-\mathrm{e}^{{2(\tilde{\mu}-\mu_{c3})}/{t}}\right),\\
(V{\rm -}P)  & \tilde{\kappa}= -\frac{2}{\sqrt{\pi t}}
{\rm  Li}_{-\frac{1}{2} }\left(-\mathrm{e}^{{2(\tilde{\mu}-\mu_{c2})}/{t}}\right),\\
(P{\rm -}PP) & \tilde{\kappa} =  \kappa_{o4} -\frac{\lambda_{5}}{\sqrt{t}} {\rm Li}_{-\frac{1}{2}} \left(-\mathrm{e}^{{(\tilde{\mu}-\mu_{c4})}/{t}}\right),
\end{array}
\end{eqnarray}
where $\kappa_{o3}, \kappa_{o4}, \lambda_{4}, \lambda_{5}$ are also known \cite{Guan:2011a}.

In the Gaudin-Yang model, the above density and compressibility can be cast into the universal scaling forms
\begin{eqnarray}
n(\mu,T,x)&=&n_0+T^{\frac{d}{z}+1-\frac{1}{\nu z}}{\cal G}\left(\frac{\mu(x)-\mu_c}{T^{\frac{1}{\nu z}}}\right),\label{Scaling-n}\\
\kappa(\mu,T,x)&=&\kappa_0+T^{\frac{d}{z}+1-\frac{2}{\nu z}}{\cal F}\left(\frac{\mu(x)-\mu_c}{T^{\frac{1}{\nu z}}}\right),
\label{Scaling-kappa}
\end{eqnarray}
with dimensionality $d=1$.
Here the scaling functions  are ${\cal G}(x)=\lambda_{\alpha} {\mathrm{Li}}_{1/2}(-{\rm e}^{x})$ and ${\cal F}(x)=\lambda_{\beta} {\mathrm{Li}}_{-1/2}(-{\rm e}^{x})$
from which one can read off  the dynamical critical exponent $z=2$ and correlation length exponent $\nu =1/2$ for different phases of the spin states.
Such results  illustrate the microscopic origin of the quantum criticality of different spin states, i.e.,  the singular parts in (\ref{Scaling-n}) and  (\ref{Scaling-kappa})
characterize sudden changes of the density of state of either excess fermions or bound pairs.
The TLL  is maintained below the cross-over temperature  $T^*$ which  indicates a universal crossover from a relativistic  dispersion into a
nonrelativistic  dispersion \cite{Maeda:2007}.
The quantum criticality driven by the external field $H$ gives rise to the same universality class.

\begin{figure}[t]
\includegraphics[width=1.1\linewidth]{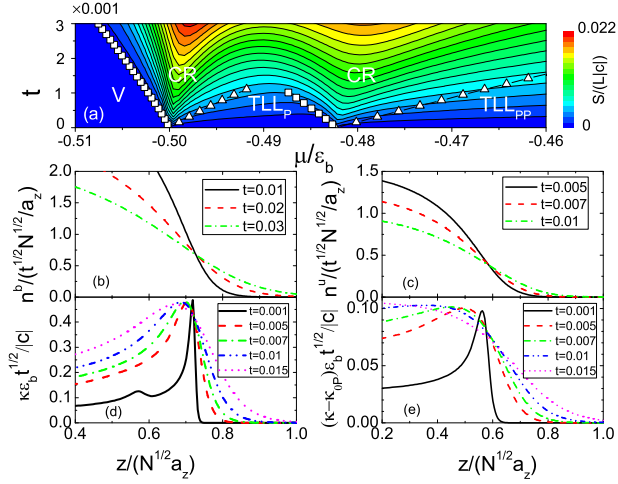}
\caption{Quantum criticality of the 1D Fermi gas in a harmonic trap for low polarization: (a)
contour plot of the entropy in the $t-\mu$ plane.
Symbols   indicate the crossover temperature $T^*$
separating the quantum critical regimes from vacuum, single component $TLL_P$
of paired fermions and two-component $TLL_{PP}$ of FFLO-like states.
The intersection of the density curves at different temperatures
can map out  the critical points (b) $\mu_{c2}$ and (c) $\mu_{c4}$.
The corresponding  compressibility curves (d) and (e)  intersect at the same critical points after a
subtraction of the background compressibility.  From Yin {\em et al.} (2011).
}
\label{fig:QC-1}
\end{figure}

Using  the LDA presented in (\ref{N_P}), quantum criticality of the bulk system can be mapped out through finite temperature density profiles in the trapped gas.
For small polarization, the chemical potential passes the
lower critical point $\mu_{c2}=-\frac{1}{2}$ from the vacuum into the fully paired phase then passes the upper critical point
$\mu_{c4}$ from the fully paired phase into the FFLO-like phase.
At finite temperatures, quantum criticality of the Fermi gas can be seen clearly from contour plots of entropy in $T-\mu$ plane
 see Fig.~\ref{fig:QC-1}(a).
The typical $V$-shape crossover temperature $T^*$ separates the quantum critical regimes where  $T^*\propto |\mu-\mu_c|$.
The crossover temperatures are determined by minimums or maximums of the magnetization or by the breakdown of the
linear-temperature-dependent entropy \cite{Zhao:2009}.
Fig.~\ref{fig:QC-1} (b) and (c) show that the unpaired density curves for different
temperatures intersect at the critical points $\mu_{c2}$ and  $\mu_{c4}$, respectively.

The phase boundary separating the fully paired phase from the FFLO-like phase can be
mapped out from the density profiles of unpaired fermions in the trapped gas
at finite  temperatures.
As the temperature decreases, the compressibility evolves a round peak sitting in the phase of the higher density of state.
It diverges at zero temperature.
Similarly, for the high polarization case, the density profiles of unpaired and paired atoms can be used to map out the phase
boundaries $\mu_{c1}$ ($V\to F$ ) and $\mu_{c3}$ ($F\to PP$), respectively \cite{Yin:2011b}.
This signature can be used to confirm the quantum critical law, as per the recent experimental measurements \cite{Zhang:2012}.

The universal scaling behaviour of the homogenous system can be mapped out through the density profiles of the trapped gas
at finite temperatures.
However, inhomogeneity caused by the  finite-size scaling effect is evident in the scaling analysis
\cite{Campostrini:2009,Campostrini:2010a,Campostrini:2010b,Zhou:2010,Ceccarelli:2012}.
It has been proved that quantum criticality of the bulk system can be revealed from the singular part of a thermodynamic quantity
near the trapping centre $x=0$.
The scaling behaviour exists  in the limit of large trapping size.
E.g., under a scale change $b$, the singular part of the density below the critical dimension $d_c$ can be written \cite{Zhou:2010}
\begin{equation}
n(\mu,T,\omega^2,x)=b^{-(d+z)+1/\nu}{\cal G}\left(\bar{\mu}b^{1/\nu},Tb^z,\omega^2b^y,x/b\right)\nonumber
\end{equation}
with $\bar{\mu}=\mu-\mu_c$ and $y=2+1/\nu$.
Choosing $Tb^z=1$, the scaling function with finite-size trapping is $\bar {{\cal G}}(\mu, T| D,x )={\cal G}\left(\bar{\mu}/T^{1/(\nu z)},D,xT^{1/z}\right)$
with $D=\omega^2/T^{y/z}$.
The scaling behaviour is revealed through plotting $\bar{\cal G}$ against $\mu$ at $x=0$ for different temperatures  with a fixed $D$.
This means that the scaling behaviour of the homogeneous system could be extracted from the trapping centre in a small window \cite{Zhou:2010}.
Nevertheless, the finite-size error lies within the current experimental accuracy \cite{Zhang:2012}.
One can either lower the temperature or increase the interaction strength such that all data curves for the physical properties at
different temperatures  collapse  into a single curve with a proper scaling in the trapped gas.

\subsection{Spin-charge separation  in repulsive fermions }
\label{S-C-Separation}

Landau's Fermi liquid theory provides a universal description of  low energy physics of interacting electron systems in higher dimensions,
where interactions only lead to finite renormalizations of physical properties \cite{Landau:1957a,Landau:1957b,Landau:1958}.
The quasiparticle excitations have a divergent lifetime when the excitation energy goes to zero.
Renormalization of individual quasiparticles leads to a similar Fermi liquid behaviour, e.g., a finite density of states and
a step-like singularity in momentum distribution at zero temperature.
The deviations  from  the values  of  physical properties of  noninteracting systems present the interaction effect.
However, the low energy physics of   1D  interacting many-body systems  does not have  such quasiparticle-type excitations.
In 1D many-body systems, all particles participate in the low energy excitations  and form collective motions of the charge and spin densities with
 different velocities \cite{Gogolin:1998,Tsvelik:1983,Giamarchi:2004,Cazalilla:2011}.
 Thus the low energy physics only depends on the TLL parameter and the velocities of collective charge and spin oscillations.
 Introducing  charge and spin  Boson fields
 $\phi_{c,\sigma}=\left( \phi_{\uparrow}\pm \phi_{\downarrow}\right)/\sqrt{2},\, \Pi_{c,\sigma}=\left( \Pi_{\uparrow}\pm \Pi_{\downarrow}\right)/\sqrt{2}$,
 the low energy physics of the 1D spin-$1/2$ repulsive Fermi gas can be described by an effective Hamiltonian \cite{Schulz:1991,Giamarchi:2004}
\begin{equation}
H=H_c+H_{\sigma}+\frac{2g_1}{(2\pi \alpha)^2}\int dx \cos (\sqrt{8\phi_{\sigma}}). \label{effective}
\end{equation}

The fields $ \phi_{\nu}$ and $\Pi_{\nu}$ obey the standard Bose communication relations
$\left[ \phi_{\nu},\Pi_{\mu} \right]=\mathrm{i} \delta_{\nu\mu}\delta(x-y)$ with $\nu,\mu=c,\sigma$.
The parameter $\alpha$ is a short-distance cutoff.
The last term in the  effective Hamiltonian (\ref{effective})  characterizes the backscattering  process,
i.e, it corresponds to  a $2k_F$ scattering. The 1D interacting Fermi gas separates into charge and spin parts
\begin{equation}
H_{\nu}=\int dx \left(\frac{\pi v_{\nu} K_{\nu}}{2} \Pi^2_\nu+\frac{v_{\nu}}{2\pi K_{\nu} }\left(\partial _x \phi_{\nu} \right)^2\right).
\end{equation}
The coefficients  for different processes are given phenomenologically \cite{Giamarchi:2004}.
The coefficient $v_c/K_c$ is the energy cost for changing the particle density while $v_{\sigma}/K_{\sigma}$
determines the energy for creating a nonzero spin polarization.
The compressibility  and susceptibility are given by $\kappa=2K_c/(\pi v_c)$ and  $\chi=2K_{\sigma} /(\pi v_{\sigma})$, respectively.

 At fixed point $g_1=0$ the specific heat is given by a linear temperature dependent relation $c =\gamma_cT$
 where $\gamma_c/\gamma_0=\left(v_F/v_c+v_F/v_{\sigma} \right)/2$.
 Here $\gamma_0$ is the specific heat coefficient of noninteracting fermions.
 The susceptibility is given by $\chi/\chi_0=v_F/v_{\sigma}$ where the noninteracting susceptibility reads $\chi_0=1/(\pi v_F)$.
 Thus the Wilson ratio at the fixed point is given by
 \begin{equation}
 R_W=\frac{2v_c}{v_c+v_{\sigma}}.
 \end{equation}
This ratio gives a universal feature of collective motions of particles.
The TLL parameter $K_{\nu }$ and charge and spin velocities $v_{\nu}$
determine universal behaviour of correlation functions \cite{Schulz:1991,Giamarchi:2004,Cheianov:2004}.

\begin{figure}[t]
{{\includegraphics [width=1.00\linewidth]{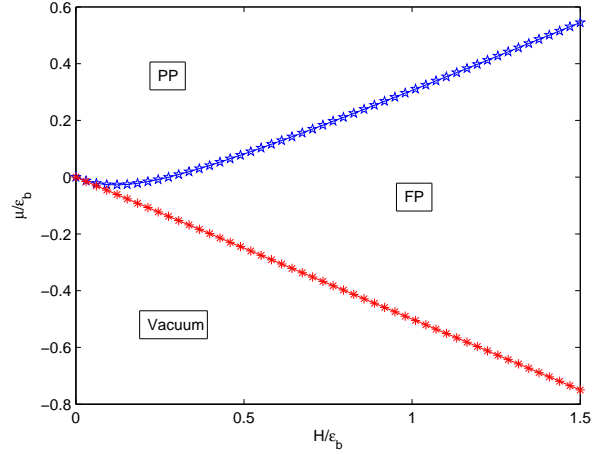}}}
\caption{The zero temperature  phase diagram of the Gaudin-Yang model in the repulsive regime.
Here the chemical potential and magnetic field are rescaled by $\epsilon_b=\frac{\hbar^2}{2m}c^2$.
$PP$ denotes partially polarized phase with finite magnetization. $FP$ denotes the fully polarized phase. }
   \label{fig:phase-r}
\end{figure}

On the other hand, in terms of the TBA formalism, the
physical quantities are described by the set of nonlinear
integral equations \cite{Lai:1971a,Lai:1973,Takahashi:1971b}
\begin{eqnarray}
\varepsilon(k)&=&k^{2}-\mu-\frac{H}{2}-T\sum_{\ell=1}^{\infty}K_{\ell}\ast\ln\left(1+\mathrm{e}^{-\phi_{\ell}(k)/T}\right)\nonumber\\
\phi_{j}(\lambda)&=&jH-TK_{j}\ast\ln\left(1+\mathrm{e}^{-\varepsilon(\lambda)/T}\right)\nonumber\\
&&+T\sum_{m=1}^{\infty}T_{jm}\ast\ln\left(1+\mathrm{e}^{-\phi_{m}(\lambda)/T}\right)
\label{TBA-lambda}
\end{eqnarray}
with $j=1,\ldots, \infty$. Here  $T_{jm}(k)$ is given in Takahashi (1999).
The  free energy per unit length $F$ and the pressure $P$ follow as
\begin{equation}
F\approx \mu n- P, \qquad P= \frac{T}{2\pi}\int_{-\infty}^{\infty}
\ln(1+\mathrm{e}^{-\varepsilon(k)/T}) dk, \label{Pressure-repulsive}
\end{equation}
where $n$ denotes the particle density.
The spin wave bound states give an antiferromagnetic ordering  at low temperatures.
The population imbalance associated with   the Fredholm equations (\ref{BE-r}) lead to three
distinguishable phases: the spin singlet groundstate with magnetization $m^z=0$ (zero magnetic field);
a magnetic phase with finite magnetization ($0<H<H_s$) and a fully polarized phase with $m^z=1/2$ ($H>H_s$).
The critical field, at which the density of down-spin atoms is zero, is given by \cite{Lee:2012a}
\begin{equation}
H_s=\left(\frac{c^2}{2 \pi }+2 \pi  n^2\right) \tan^{-1}\left(\frac{2 \pi n}{c}\right)-c n.
\end{equation}
The phase diagram in the chemical potential-magnetic field plane is shown in Fig.~\ref{fig:phase-r}.

Spin charge separation is a hallmark of the TLL physics for 1D interacting fermions.
For arbitrary repulsive coupling $c>0$ in arbitrary magnetic field $H\le H_c$,
the TBA equations  (\ref{TBA-lambda}) yield \cite{Lee:2012a}
\begin{equation}
F=E_0-\frac{\pi T^2}{6}\left(\frac{1}{v_s}+\frac{1}{v_c}\right)
\label{free-dress}
\end{equation}
where $v_c$ and $v_s$ are, respectively, the holon and spinon
excitation velocities
\begin{equation}
v_c=\frac{\varepsilon_c'(k_0)}{2\pi \rho_c(k_0)}, \qquad
v_s=\frac{\varepsilon_s'(\lambda_0)}{2\pi \rho_s(\lambda_0)}.
\label{fermi-v}
\end{equation}
However, the velocities $v_c$ and $v_s$  can be analytically calculated only for strong and weak interactions.
The numerical solutions of spin and charge velocities have been discussed in the literature
\cite{Recati:2003,Fuchs:2004,Batchelor:2006a,Batchelor:2006c}, see Fig.~\ref{fig:velocities}.
In a harmonic trap,  there is an imbalanced mixture of two-component fermions  in the trapping center and
fully polarized fermions at the edge \cite{Abedinpour:2007,Colome-Tatche:2008,Ma:2009,Ma:2010a}.
The exact analytical solution of quasi-one-dimensional spin-$1/2$ fermions with infinite repulsion for an arbitrary confining potential was present in \cite{Guan-L:2009}.

\begin{figure}[t]
{{\includegraphics [width=1.00\linewidth]{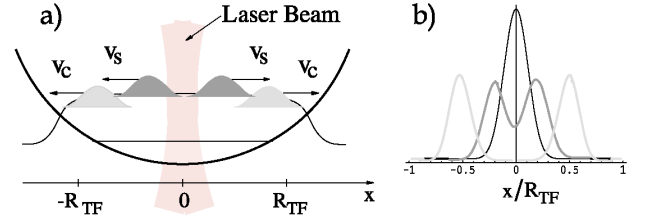}}}
\caption{A short laser pulse near the trap centre  could excite the charge density and spin density wave packets  in   a harmonic trapped Fermi gas.
Charge and spin velocities can be manifested in a spatial separation of the spin (solid curve) and density (dashed curve) wave packets.
From Recati {\em et al.} (2003). }
   \label{fig:S-C-separation}
\end{figure}

In the context of exact solutions, considerable work has been done to derive low temperature analytic results for BA solvable models.
These include the work on  the free energy of spin chains at low temperatures under a small magnetic field \cite{Mezincescu:1993a,Mezincescu:1993b}
and the calculation of the leading temperature dependent terms in the free energy of the massive Heisenberg model by Johnson and McCoy (1972).
Later, Filyov {\em et al.} (1981) derived an exact solution to the s-d exchange model expressed as a series in terms of the temperature.
Following the method proposed \cite{Mezincescu:1993a,Mezincescu:1993b} for small external field $H\ll 1$, Lee {\em et al.}, \cite{Lee:2012a}
solved the TBA equations (\ref{TBA-lambda})  by the  Weiner-Hopf  method, where the dressed energy potential is simplified as $\varepsilon(k)\approx k^2-A$
with the cutoff  chemical potential  $A:=\mu+2p\ln 2/c+cH^{2}/4\pi^{2}p+cT^{2}/6p$. Here $p$ is the pressure.
The terms in the function $A$ provide insights on spin and charge density fluctuations.
With the help of this potential and using  Sommerfeld's lemma \cite{Pathria:1996},
the free energy of the system defined by Eq.~(\ref{Pressure-repulsive})  is given by
\begin{eqnarray}
F&\approx&\frac{1}{3}\pi^{2}n^{3}\left(1-\frac{4\ln 2}{\gamma}\right)-\frac{3\gamma
H^{2}}{8\pi^{4}n}\left(1+\frac{6\ln 2}{\gamma}\right)\nonumber\\
&& -\frac{\gamma
T^{2}}{4\pi^{2}n}\left(1+\frac{6\ln 2}{\gamma}\right)-\frac{T^{2}}{12n}\label{Free-E}
\end{eqnarray}
which gives the universal low temperature form of spin-charge separation theory (\ref{free-dress})
with  the excitation velocities
\begin{equation}
v_c\approx 2\pi n\left(1-\frac{4\ln2}\gamma \right), \qquad
v_s\approx\frac{2\pi^3 n}{3\gamma}\left(1-\frac{6\ln
2}{\gamma}\right). \nonumber
\end{equation}

For the external field approaching the saturation field $H_s$,  the charge and spin velocities can be derived  from the
relations (\ref{fermi-v}).
The leading terms in the velocities are then found to be
\begin{eqnarray}
v_{c}&=&2\pi
n\left(1-\frac{12}{\pi\gamma}\sqrt{1-\frac{H}{H_{c}}}\right),\,\,
v_{s}=\frac{H_{c}}{n}\sqrt{1-\frac{H}{H_{c}}}.\nonumber
\end{eqnarray}
The susceptibility values for different values of the chemical potential are consistent with the field theory  prediction
$\chi  v_s=\theta/\pi$ with $\theta=1/2$ \cite{Giamarchi:2004} for strong repulsion.
On the other hand, the TBA equations (\ref{TBA-lambda})  show that the spin-spin interaction for the system of the polarized fermions with strong
coupling  $\gamma\gg 1$ can be effectively described by the isotropic spin-1/2 Heisenberg chain with a weak antiferromagnetic coupling
$J=-\frac{2}{|c|}p^{\rm u}(T,H)$.
For the isotropic spin-1/2 Heisenberg chain, the susceptibility at $H=0$ is given by
$\chi =1/(J\pi^2)$ which coincides with the  field theory  prediction $\chi v_s=\theta/\pi$ \cite{Lee:2012a}.

Furthermore,  for strong repulsion,  the pressure  of the gas with a weak magnetic field is given by \cite{Lee:2012a}
\begin{equation}
p=-\sqrt{\frac{m}{2\pi\hbar^2}}T^{\frac{3}{2}}\mathrm{Li}_{\frac{3}{2}}\left(-\mathrm{e}^{A/T}\right). \label{Pres-Rep-L}
\end{equation}
This result provides the low temperature thermodynamics which extends beyond the range covered by spin-charge separation theory.
 With the low temperature expansion, the pressure (\ref{Pres-Rep-L}) significantly evolves into the thermodynamics of two free Gaussian fields at criticality.
The result for the free energy at low temperature gives  a universal signature of TLLs where the leading low temperature contributions are
solely dependent on the spin and charge velocities.
This opens a way to experimentally explore  how the low temperature thermodynamics of a 1D many-body system naturally
separates into two free Gaussian field theories.
This may possibly be used to test the spin and charge velocities in experiment with an ultracold atomic 1D Fermi gas in a harmonic trap
\cite{Recati:2003,Cheianov:2004}, where the spin and charge of the ultracold atoms refer to two internal atomic hyperfine states and the
atomic mass density, respectively, as per the  theoretical scheme to explore spin-charge separation waves in Fig.\ref{fig:S-C-separation}.
This phenomenon has already been experimentally observed in electron liquids \cite{Deshpande:2010,Jompol:2009}.

\section{Fermi-Bose mixtures in 1D}
\label{Section:mixture}

The experimental advances  in trapping and cooling ultracold quantum  gases have also led  to realizations of
degenerate Fermi-Bose mixtures with various combinations of fermionic and bosonic atoms such as
${}^6$Li-${}^7$Li \cite{Truscott:2001},
${}^6$Li-${}^{23}$Na \cite{Hadzibabic:2002,Stan:2004},
${}^{40}$K-${}^{87}$Rb \cite{Roati:2002,Inouye:2004,Ospelkaus:2006},
${}^6$Li-${}^{87}$Rb \cite{Silber:2005},
${}^{173}$Yb-${}^{174}$Yb \cite{Fukuhara:2009},  and
${}^{6}$Li-${}^{174}$Yb \cite{Hansen:2011} and ${}^{6}$Li-${}^{133}$Cs \cite{Tung:2013,Repp:2013}.
This experimental success opens up a further gateway for exploring  striking  quantum many-body phenomena
though tuning interactions between inter- and intra-species of atoms.

In light of the experimental realizations of Fermi-Bose mixtures,
various theoretical methods have been used to study phases of superfluids and Mott insulators,
instabilities of collapse and demixing  and quantum  correlations of the  1D Fermi-Bose mixtures, such as
the mean-field approach \cite{Das:2003}, TLL
theory \cite{Cazalilla:2003,Orignac:2010,Rizzi:2008b,Lewenstein:2004,Mathey:2004,Mathey:2007}
and  numerical methods \cite{Varney:2008,Zujev:2008,Pollet:2008,Takeuchi:2007}.
The TLL field theory \cite{Cazalilla:2003,Rizzi:2008a,Mathey:2004} predicts that the binary mixtures of bosons and
spin-polarized fermions with population imbalance present competing ordering --
i) strong attraction between the two species leads to  collapse;
ii) a strong repulsion leads to  demixing;
iii) subtle tuning of the intra-and inter-particle scattering leads to pairing and two-component TLLs.

On the other hand, the 1D Fermi-Bose mixture with equal masses of bosons and spin-polarized fermions
and with the same strength of delta-function interaction between boson-boson,  boson-fermion and fermions with different spins
was solved a long time ago  \cite{Lai:1971b,Lai:1974b}.
The exactly solved model provides a benchmark towards understanding  various quantum many-body effects in 1D Fermi and Bose mixtures.
Recently,   particular theoretical interest has been paid  to the
groundstate properties \cite{Imambekov:2006a,Imambekov:2006b,Hu:2006,Chen:2010,Girardeau:2007},
magnetism \cite{Imambekov:2006b,Batchelor:2005a,Guan:2008b},
correlation functions \cite{Imambekov:2006b,Frahm:2005,Fang:2011b},
thermodynamics and quantum criticality \cite{Yin:2009,Yin:2012}.

\subsection{Groundstate}

The  1D $\delta$-function interacting mixture of $M_b$ spinless bosons and $M_1$ fermions with spin-up and $M_2$  fermions  with  spin-down
is described by the Hamiltonian (\ref{Ham}) with Zeeman term $\frac12 H(M_1-M_2)$.
$H$ is an external magnetic field.
The Bethe wave function for this model requires symmetry under exchange of spatial and internal spin coordinates between
two bosons or fermions with different spin states and antisymmetry for two fermions with the same spin state.
The BA equations for this Fermi-Bose  mixture with
an irreducible representation $[2+M_{b}, 2^{M_{2}-1}, 1^{M_{1}-M_{2}}] $  are  \cite{Lai:1971b}
\begin{eqnarray}
&&\exp(\textrm{i}k_{j}L)=\prod_{\alpha=1}^{M}e_1(k_j-\lambda_\alpha), \nonumber\\
&&\prod_{j=1}^{N} e_1(\lambda_{\alpha}-k_{j}) \prod_{b=1}^{M_{b}}e_{1}(\lambda_{\alpha}-A_{b})= -\prod_{\beta=1}^{M}e_2(\lambda_{\alpha}-\lambda_{\beta})
\nonumber\\
&&\prod_{k=1}^{M}e_{1}(A_{b}-\lambda_{k})=1, \label{BA-M}
\end{eqnarray}
where  $M=M_2+M_b$.
In these equations $k_j$, with $j=1,\ldots, N$, are the quasimomenta of the
particles and $\lambda_{\alpha}$, with $\alpha=1,\ldots, M$, are parameters for
fermions with spin-down and the bosons. $A_b$ with $b=1,\ldots, M_b$ are
the parameters for the bosons.

Lai and Yang (1971) showed by numerically solving the set of coupled integral equations that
the energy of the system for the mixture of bosons and
polarized fermions is a monotonic decreasing function with respect to the ratio of
bosons and the total number of particles in the system.
The  groundstate properties of the system have been further studied \cite{Batchelor:2005a,Frahm:2005}.

The magnetic properties of the mixture of bosons and polarized fermions provide further insight into the competing ordering.
Application of the external magnetic field to the polarized fermions causes Zeeman splitting of the
spin-up and spin-down fermions into different energy levels.
The groundstate can only accommodate fermions that are in the lower energy level.
Therefore it is expected that when the direction of the magnetic field is along the spin-up ($H>0$) direction,
spin-down fermions can no longer populate the groundstate.
The free energy of the strongly coupled mixture is given in the form $F=-(n-m_b)H/2+E_0$
where the groundstate energy per unit length  is given by \cite{Guan:2008b,Imambekov:2006a,Imambekov:2006b}
\begin{eqnarray}
E_0=\frac{1}{3}\pi^{2}n^{3}\left[\begin{array}{l} 1-\frac{4}{\gamma}\left(\frac{m_b}{n}+\frac{\sin(\frac{m_b\pi}{n})}{\pi}\right)\\
+\frac{12}{\gamma^2}\left(\frac{m_b}{n}+\frac{\sin(\frac{m_b\pi}{n})}{\pi}\right)^2\end{array}
\right].\label{E-FB-M}
\end{eqnarray}
Here $n_b$  is the boson number density.
If the external field exceeds the critical field $H_c=8p_0/c$, where the pressure per unit length is given by
$P_0 \approx \frac{2}{3}n^3\pi^2\left[1-\frac{6}{\gamma}\left(\frac{m_b}{n}+\frac{\sin(\frac{m_b\pi}{n})}{\pi}\right)\right]$,
the system enters a phase of fully-polarized fermions.
The susceptibility in the vicinity of the critical field $H_c$ diverges as
\begin{eqnarray}
\chi &\approx& \frac{n}{2\pi}\frac{1}{H^{1/2}(H_{c}-H)^{1/2}}.
\end{eqnarray}
This van Hove type of singularity is subtly  different from the linear field-dependent magnetization
in the two-component  attractive Fermi gas with polarization,
where the effective interaction between the  bosonic pairs is weakly attractive in the Tonks-Girardeau limit \cite{Chen:2010}.

In contrast, for weak repulsion the groundstate energy presents a mean field effect, i.e., the energy per unit length is
\begin{equation}
E_0 \approx \left[\frac{M_1^3}{L^3}+\frac{M_b^2c}{L^2}+\frac{2M_bM_1c}{L^2}\right].
\end{equation}
The magnetization $m^z\approx \frac{1}{4\pi}(\sqrt{2H}+\frac{2c}{\pi})$ and  the susceptibility
$\chi \approx \frac{\sqrt{2}}{8\pi \sqrt{H}}$ follow from this equation.
These results show that in the weak coupling regime the square-root field-dependent behavior of
magnetization emerges for finite external field.

\begin{figure}[t]
{{\includegraphics [width=1.0\linewidth]{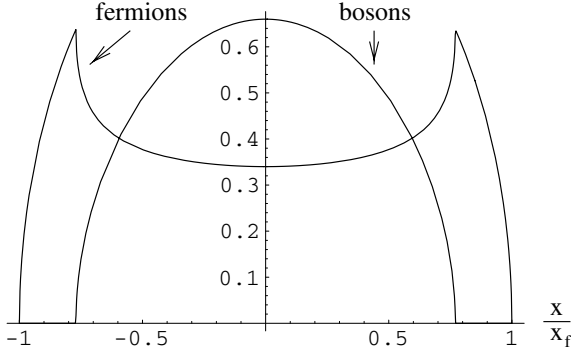}}}
\caption{Density profile of Fermi-Bose mixture in strong coupling regime $\gamma\gg1$.
At zero temperature, fermions sit at the wings while the trapping centre comprises of a  mixture of fermions and bosons. From  Imambekov and Demler (2006b).  }
\label{fig:F-B-density}
\end{figure}

Furthermore, using CFT, Frahm and Palacios \cite{Frahm:2005} have  computed the
asymptotics of boson and fermion Green's functions for a mixtures of bosons and polarized and unpolarized fermions.
The Fourier transform of equal time boson Green's function has  a form $n_b(k) \sim |k-k_0|^{\nu_b}$.
The response function of fermions follows a form $n_{\sigma}(k) \sim {\rm sin}(k-k_0)|k-k_0|^{\nu _f}$.
The exponents $n_b$ and $\nu_f$ are determined by finite-size energies and momenta.
The presence of spin population imbalance of fermions gives rise to different critical exponents.
In fact, the exact result indicates that no demixing occurs in the Fermi-Bose mixture with a  repulsive interaction.
A thorough study of the ground state properties, including density distributions and   the single particle correlation functions was reported in   \cite{Imambekov:2006b}.
Using  exact solution with local density approximation in a harmonic trap, the density profiles of the Fermi-Bose mixture were predicted.  In the weakly interacting regime, bosons can condense to the trapping centre whereas fermions spread out due to  the Fermi pressure.  For strong repulsion,  the boson density distribution is extended  in a wider region and  the fermion  density shows strong non-monotonous behavior, see Fig.\ref{fig:F-B-density}.

\subsection{Universal thermodynamics}

The fully-polarized Fermi-Bose mixture is described by the Hamiltonian
\begin{align}
\cal{H}& =\int_{0}^{L}dx\left( \frac{\hbar ^{2}}{2m_{b}}\partial _{x}\Psi
_{b}^{\dag }\partial _{x}\Psi _{b}+\frac{\hbar ^{2}}{2m_{f}}\partial
_{x}\Psi _{f}^{\dag }\partial _{x}\Psi _{f} + \right. \nonumber  \\
& \left.  \frac{g_{bb}}{2} \Psi _{b}^{\dag }\Psi _{b}^{\dag }\Psi
_{b}\Psi _{b}+g_{bf}\Psi _{b}^{\dag }\Psi _{f}^{\dag }\Psi _{f}\Psi
_{b} - \mu_f \Psi _{f}^{\dag }\Psi _{f} - \mu_b \Psi _{b}^{\dag
}\Psi _{b}\right). \label{H_second_form}
\end{align}
where $\Psi _{b}$, $\Psi _{f}$ are boson and fermion field  operators,
$m_{b}$, $ m_{f}$ are the masses, $\mu_{b}$, $ \mu_{f}$ are chemical
potentials of bosons and fermions, and $g_{bb}$, $g_{bf}$ are
boson-boson and boson-fermion interaction strengths, respectively.

For bosons and fermions of equal mass and equal interaction strength $g_{bb}=g_{bf}$,
a different  set of BA equations from (\ref{BA-M})  were derived
\cite{Imambekov:2006a,Imambekov:2006b}  with
\begin{eqnarray}
\exp(\textrm{i}k_{j}L)=\prod_{\alpha=1}^{M_b}e_1(k_j-\lambda_\alpha), \,\,\,  \prod_{i=1}^{N}e_{1}(k_i-\lambda_{\alpha})=1,\label{BA-M-F}
\end{eqnarray}
where $j=1,\ldots,N$ and $\alpha =1,\ldots, M$. Here $M_b$ is the number of bosons.
The BA equations (\ref{BA-M}) with a  particular choice of fermion spin polarization are physically equivalent to (\ref{BA-M-F}).
This can be seen from the fact that the groundstate energy of the model (\ref{H_second_form})
is given by the same expression as (\ref{E-FB-M}), where the spin-down fermions are gapfull.

At finite temperatures, the equilibrium states become degenerate.
The thermodynamics of the model are determined from the integral equations \cite{Lai:1974a,Yin:2009}
\begin{align}
&& \epsilon ( k) =k^{2}-\mu _{f}  -T\int_{-\infty }^{\infty }K_1\left( \Lambda -k\right) \ln
\left( 1+\mathrm{e}^{ -{\varphi( \Lambda)}/{T}} \right) d\Lambda ,  \notag\\
&&  \varphi ( \Lambda) =\mu _{f}-\mu _{b} -T\int_{-\infty }^{\infty }K_1\left( k-\Lambda \right) \ln \left( 1+\mathrm{e}^{-{ \epsilon (k)}/{T}} \right) dk,
\label{TBA_nonlinear}
\end{align}
 For fixed temperature $T$ and chemical potential
$\mu _{f}$,
$\mu _{b}$, the pressure is given by $
P=\frac{T}{2\pi }\int_{-\infty }^{\infty }\ln \left( 1+\mathrm{e}^{ -{\epsilon(k)}/{T}} \right) dk$.

\begin{figure}[t]
{{\includegraphics [width=1.0\linewidth]{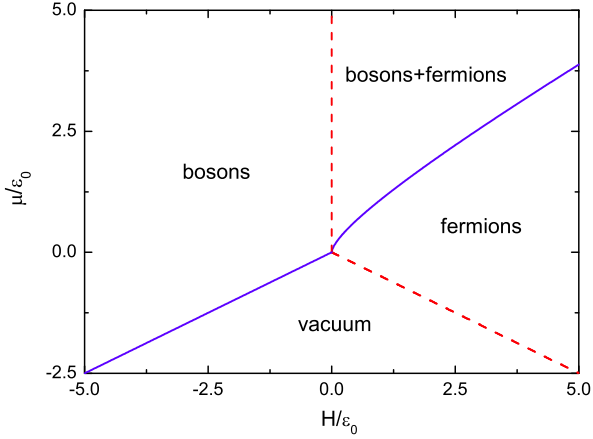}}}
\caption{Phase diagram in the $\protect\mu-H$ plane. Three distinguished  phases result from varying the chemical potential
and the chemical potential difference $H=\mu_f-\mu_b$: the pure boson phase for $H<0$ and $\mu>H/2$; the pure fermion phase
below the phase boundary $\tilde{H_c}$ in the region $\mu >-H/2$ and the mixture of bosons and fermions above the phase boundary
$\tilde{H_c}$  in the region $H>0$.
From Yin {\em et al.} (2012).  }
\label{fig:Phase-mixture}
\end{figure}

In this grand canonical ensemble, it is convenient to use the chemical potential $\mu$ and the chemical bias $H=\mu_f-\mu_b$
to discuss the zero temperature  phase diagram of the model  \cite{Yin:2012}, see Fig.~\ref{fig:Phase-mixture}.
The phase boundaries were determined by analysing the dressed energy equations obtained from the TBA equations (\ref{TBA_nonlinear}) in the limit $T\to 0$.
The critical line
\begin{equation}
\tilde{H}_c = \frac{1}{2\pi }\left[ \left( 4\tilde{\mu}_{f}\allowbreak
+1\right) \arctan \sqrt{4\tilde{\mu}_{f}}-\allowbreak \sqrt{4\tilde{\mu}_{f}}
\right]
\end{equation}
separates the mixed phase and the pure fermion phase.
Here dimensionless units have been used, i.e.,  $\tilde{H}=H/\epsilon_0$ and $\tilde{\mu_f }=\mu_f/\epsilon_0$ with $\epsilon_0=c^2$.
In a harmonic trap, this phase diagram can be presented  within the LDA, i.e., $\mu_b^0(x)+m\omega^2_bx^2/2 =\mu_b^0(0)$ and $\mu_f^0(x)+m\omega^2_fx^2/2 =\mu_f^0(0)$.
It  was found  \cite{Imambekov:2006b,Yin:2009}  that for both strong and weak interactions bosons and fermions coexist in the central part and fermions sit at the wings.
For weak interaction, bosons can condense into the centre while  the fermions spread out in the whole trapping space due to Fermi pressure.
However, for the strong interaction regime, the Fermi density shows strong non-monotonous behaviour \cite{Imambekov:2006b,Yin:2009}.
The significant feature of correlation functions of the mixture in the Tonks-Girardeau limit were discussed in detail \cite{Imambekov:2006b}.
In particular, the Fourier transform of the Bose-Bose correlation function is governed by the TLL parameter
$K_b$ via $n^b(k) \sim |k| ^{-1+1/(2K_b)}$ for $k\to 0$  and the Fourier transform of the Fermi-Fermi correlation function has
singularities at $k=k_f, k_f+2k_b$.  The discontinuity  at $k_f+2k_b$ indicates an interaction effect.

\begin{figure}[htb]
{{\includegraphics [width=1.0\linewidth]{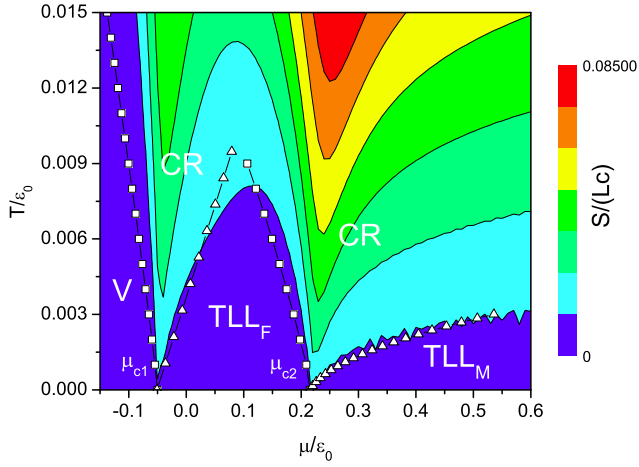}}}
\caption{Contour plot of entropy $S$ vs chemical potential from the exact solution, showing
quantum criticality driven by chemical potential for $H=0.1\varepsilon_0$.
The crossover temperatures (white squares and triangles) separate  the vacuum, $TLL_F$ and $TLL_M$ from the quantum critical regimes.
From Yin {\em et al.} (2012).  }
\label{fig:entropy-mixture}
\end{figure}

The TBA equations  (\ref{TBA_nonlinear}) have been used  to explore scaling behaviour of the thermodynamics in the Fermi-Bose mixture.
The  universal leading order temperature corrections to the free energy  \cite{Yin:2012}
\begin{equation}
F\approx E_0-\frac{\pi CT^2}{6}\left( \frac{1}{v_b} +\frac{1}{v_f} \right) \label{LL}
\end{equation}
indicate a  collective TLL signature at low temperatures.
In the strongly repulsive regime for $H\ll 1$
\begin{equation}
v_s =\frac{4\pi^2n}{3\gamma}\sin \frac{\pi n_b}{n},\,\,
v_f = 2\pi n\left[1-\frac{4}{\gamma} \left(\frac{\pi n_b}{n} +\sin \frac{\pi n_b}{n}  \right) \right]\nonumber .
\end{equation}

As already remarked, the TLL description is incapable of describing quantum criticality since
it does not contain the right  fluctuations in the critical regime.
For the physical  regime,  i.e.,  $c \gg 1$, or $T/\varepsilon_0 \ll 1$,  the pressure is
\begin{equation}
p=-\sqrt{\frac{\beta }{4\pi}}T^{\frac{3}{2}}{\rm Li}_{\frac{3}{2}}\left(-\mathrm{e}^{{A}/{T}} \right).\label{Pressure-FM}
\end{equation}
Here the functions $\beta$ and  $A$ are  determined by
\begin{eqnarray}
\beta &=& 1-\frac{2Tc}{\pi} \int_{-\infty}^{\infty}\frac{4c^2-48\Lambda^2}{\left(c^2+4\Lambda^2 \right)^3}\ln
\left( 1+\mathrm{e}^{-{\varphi(\Lambda)}/{T}}\right)d\Lambda, \nonumber\\
A &\approx& \mu _{f}+T\int_{-\infty }^{\infty }K_1( \Lambda) \ln
\left( 1+\mathrm{e}^{ -{\varphi( \Lambda)}/{T}} \right) d \Lambda. \nonumber
\end{eqnarray}
The function $ \varphi( \Lambda)$ can be obtained from (\ref{TBA_nonlinear}) by iteration  \cite{Yin:2012}.

The equation of state (\ref{Pressure-FM}) can be used to explore the critical behaviour of the model.
E.g.,  the entropy in Fig.~\ref{fig:entropy-mixture}  shows that the TLL is maintained below a crossover temperature
at which the linear temperature-dependent entropy breaks down.
In Fig.~\ref{fig:entropy-mixture}, $TLL_F$ denotes a TLL of fully-polarized  fermions and  $TLL_M$ denotes a
a two-component TLL of the mixture of bosons and fermions described by (\ref{LL}).
As the temperature is tuned over the crossover temperatures the scaling function of thermodynamic properties
give rise to the free fermion  universality class of criticality  that  entirely depends on the symmetry excitation
spectrum and dimensionality of the system.
The equation of state pressure,  determined by (\ref{Pressure-FM}),  contains universal scaling functions
which control the thermodynamic properties in the quantum critical regimes.
Near the critical point, the thermal dynamical properties can be cast into universal scaling forms, e.g.,
(\ref{Scaling-n}) and (\ref{Scaling-kappa}) for  a free Fermi theory of criticality, i.e.,  with $d = 1$, $z = 2$ and
$\nu = 1/2$. The  detailed analysis of quantum criticality of the Fermi-Bose mixture has been presented in
Yin {\em et al.} 2012.
As for the Fermi system, with the help of the exact solutions  the critical properties of the bulk system can be mapped
out from the density profiles of the trapped Fermi-Bose mixture gas at finite temperatures.

\section{Multi-component Fermi gases of ultracold atoms}
\label{Section:Multi}

\subsection {Pairs and trions in three-component systems }

A pseudo spin-1/2 system of interacting atomic fermions  has been experimentally realized by loading atoms
within two lowest hyperfine levels, i.e., states $|1\rangle$ and $|2\rangle$.
The interaction strength can be controlled by tuning the scattering length through Feshbach resonance.
Weakly bound molecules exist in the phase where the scattering length is small and positive.
These molecules can form a BEC.
The tunability  of the scattering length across the Feshbach resonance leads to  divergent scattering length.
As a result the interactions can be effectively enhanced.
In the strong interacting limit, these molecules may continuously transform into BCS pairs
such that the system reaches the BEC-BCS crossover \cite{Regal:2004,Bartenstein:2004,Zwierlein:2005}.
Universal many-body behaviour is expected in the crossover (unitarity) regime  \cite{Heiselberg:2001,Ho:2004}.

It is of a great interest that the third pseudo spin state $|3\rangle$ is added to the two-component Fermi gas \cite{Modawi:1997,Bartenstein:2005}.
In contrast to the two-component case, three-component fermions
possess new features \cite{Bedaque:2009,Ho:1999,Cherng:2007,Zhai:2007,Honerkamp:2004,Martikainen:2009,Miyatake:2010,Inaba:2009,Rapp:2007,Ozawa:2010,HeL:2006}.
As a consequence, BCS pairing can be favored by anisotropies in
three different ways, namely atoms in three low sublevels denoted by
$|1\rangle$, $|2\rangle$ and $|3\rangle$ can form three possible
pairs $|1\rangle + |2\rangle$, $|2\rangle + |3\rangle$ and
$|1\rangle + |3\rangle$ \cite{Bartenstein:2005}.
Degenerate three-component fermions with tunable interparticle scattering lengths
$a_{12}$, $a_{23}$ and $a_{13}$ open up the possibility of  novel many-body phenomena.

\begin{figure}[t]
{{\includegraphics [width=1.0\linewidth]{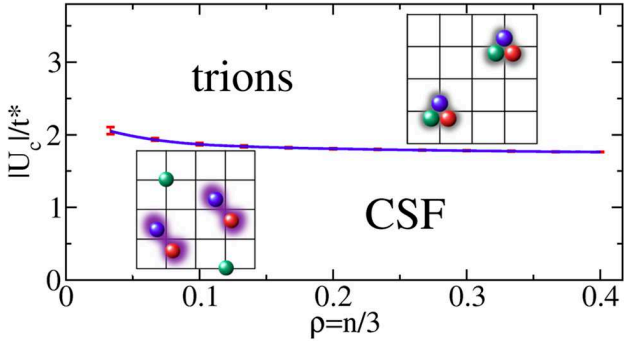}}}
\caption{Three-component fermionic atoms in an optical lattice. The colour pairing phase occurs for on-site
attraction $U<U_c$,  whereas the trionic state occurs  for $U>U_c$. From Rapp {\em et al.} (2007). }
   \label{fig:pair-trion}
\end{figure}

Specifically, strongly attractive three-component atomic fermions can form spin-neutral three-body bound states called  `trions'.
A degenerate Fermi gas  of atoms in three different hyperfine states of $^{6}$Li \cite{Ottenstein:2008}
has been created at a temperature $T=0.37 T_F$, where $T_F$ is the Fermi temperature.
The spin state mixture of ultracold fermionic  atoms has been  further used to study subtle physics of three-body recombination,
atom-dimer scattering, the atomic Efimov  trimer, association and disassociation of the  atom-dimer collisions {\em etc}
\cite{Huckans:2009,Braaten:2009,Ottenstein:2008,Pollack:2009,Wenz:2009,Spiegelhalder:2009,Zaccanti:2009,Ferlaino:2009}.
In particular, the three-component interacting fermions  have received considerable interest
 in the study of the quantum phase transition from a colour superfluid to singlet trions \cite{Rapp:2007,Cherng:2007}, see Fig.~\ref{fig:pair-trion}.
 This study also  sheds light on colour superconductivity of quark matter in nuclear physics.

\subsubsection{Colour pairing  and trions}

Loading three-component fermions with contact interaction on  a 1D optical lattice,  the system, i.e., the 1D multi-component Hubbard model,
is no longer BA solvable  \cite{Choy:1982}.
In terms of bosonization approach,  it has been found  \cite{Capponi:2008,Azaria:2009}  that the low-energy physics of the
1D three-component Hubbard model  shows  the formation of three-atom bound states, i.e.,  a quantum phase transition
from a colour superfluid to the singlet trion state.
Motivated by  such exotic  phases, the  1D integrable three-component Fermi  gas with delta-function interactions
\cite{Sutherland:1968,Sutherland:1975,Yang:1970,Takahashi:1970b} has been studied to
give a  precise understanding of colour pairing and the trionic state \cite{Guan:2008a,Liu:2008a,He:2010}.

The first quantized many-body Hamiltonian of the three-component $\delta$-function  interacting Fermi gas is as defined in Eq.~(\ref{Ham}),
with an additional Zeeman energy term $E_z=\sum_{i=1}^{3}N^{i}\epsilon^{i}_Z(\mu_B^{i},B)$.
In this system,  the fermions can occupy three possible hyperfine levels
($\left\vert 1\right\rangle $, $\left\vert 2\right\rangle $ and $\left\vert 3\right\rangle $)
with particle number $N^{1}$, $N^{2}$ and $N^{3}$, respectively \cite{Sutherland:1968,Yang:1970,Takahashi:1970b}.
The Zeeman energy levels $\epsilon^{i}_Z $ are
determined by the magnetic moments $\mu _{B}^{i}$ and the magnetic
field $B$. By convention, particle numbers in each  hyperfine states
satisfy the relation $N^{1}\geq N^{2}\geq N^{3}$. Thus the particle
numbers of unpaired fermions, pairs, and trions are respectively given by
$N_{1}=N^{1}-N^{2}$, $N_{2}=N^{2}-N^{3}$ and $N_{3}=N^{3}$ for the
attractive regime.
The unequally spaced  Zeeman splitting in the three hyperfine levels can be
characterized by two independent parameters  $H_1 = \bar{\epsilon} -
\epsilon^{1}_Z(\mu_B^{i},B)$ and $H_2 =
\epsilon^{3}_Z(\mu_B^{i},B)- \bar{\epsilon}$. Here
$\bar{\epsilon}=\sum_{\sigma=1}^3\epsilon^{\sigma}_Z/3$ is an
average Zeeman energy.

For the 1D three-component Fermi gas,  the energy eigenspectrum is given in terms of the quasimomenta $\left\{k_i\right\}$  of the fermions via
Eq.~(\ref{Ek}), which satisfy the BA equations  \cite{Sutherland:1968}
\begin{eqnarray}
\exp(\mathrm{i}k_jL)&=&\prod^{M_1}_{\ell = 1} e_1\left(
  k_j-\Lambda_\ell \right) \nonumber \\
\prod^N_{\ell = 1}e_1\left(\Lambda_{\alpha}-k_{\ell}
  \right)&=&-\prod^{M_1}_{ \beta =
  1}e_2\left(\Lambda_{\alpha}-\Lambda_{\beta}
  \right)\prod^{M_2}_{\ell=1}e_{-1}\left(\Lambda_{\alpha}- \lambda_{\ell}\right)\nonumber \\
 \prod^{M_1}_{\ell = 1}e_1\left(\lambda_{m}-\Lambda_{\ell}\right)
&=&-\prod^{M_2}_{\ell = 1}e_2\left(\lambda_{m}-\lambda_{\ell}\right)
\label{BE3}
\end{eqnarray}
Here $j=1,\ldots, N$, $\alpha = 1,\ldots, M_1$ and $m=1,\ldots,M_2$.
The parameters $\left\{\Lambda_{\alpha},\lambda_m\right\}$ are the
rapidities for the internal hyperfine spin degrees of freedom.
It is assumed that there are $M_2$ fermions in state $|3\rangle$,
$M_1-M_2$ fermions in state $|2\rangle$ and $N-M_1$ fermions in state $|1\rangle$.
For the irreducible representation $\left[3^{N_3}2^{N_2}1^{N_1}
\right]$, a three-column Young tableau encodes the numbers of unpaired
fermions, bound pairs and trions given by $N_1=N-2M_1+M_2$,
$N_2=M_1-2M_2$ and $N_3=M_2$.

\begin{figure}
{{\includegraphics [width=1.0\linewidth]{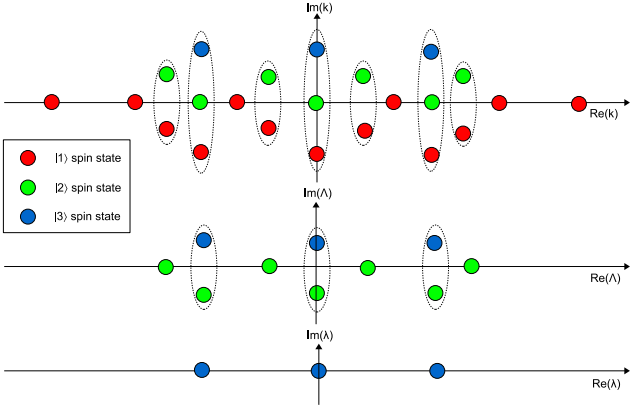}}}
\caption{Schematic root pattern for $3$  trions, $4$ colour pairs and $6$ excess  single fermions in the groundstate.
For strongly attractive interaction, the unpaired and paired quasimomenta can penetrate into the central region occupied by tightly bound trions.
From Lee {\em et al.} (2011).} \label{fig:pattern3}
\end{figure}

In the attractive regime, the BA equations (\ref{BE3}) admit complex string solutions for $k_{j}$,
i.e., three-body bound states (trions) and two-body bound states (colour pairing)
\cite{Yang:1970,Takahashi:1970b,Lee:2011a,Schlottmann:1993,Schlottmann:1994,Guan:2010}.
For  arbitrary spin polarization, (i) there are $N_3$ spin-neutral trions in the quasimomentum
$k$ space accompanied by $N_3$ spin bound states in the
$\Lambda$-parameter space and $N_3$ real roots in the
$\lambda$-parameter space, (ii) $N_2$ BCS bound pairs in $k$ space
accompanied by $N_2$ real roots in $\Lambda$ space and (iii) $N_1$
unpaired fermions in $k$ space.
These root patterns are depicted in Fig.~\ref{fig:pattern3}.

For weak attraction,  we redefine the polarizations $p_i=N^{\rm i }/N$ with $i=1,2,3$ and  the energy  $EL^2/N^3=e_0(\gamma)$ with
dimensionless parameter $\gamma=cL/N$.
The groundstate energy follows from the BA equations (\ref{BE3})  as \cite{Lee:2011b}
 \begin{eqnarray}
 e_0(\gamma) &=&\frac{1}{3}p_1^3\pi^2+\frac{1}{3}p_2^3\pi^2+\frac{1}{3}p_3^3\pi^2\nonumber\\
 &&+ 2\gamma \left[p_1p_2+p_1p_3+p_2p_3 \right]+O(\gamma^2).\label{Energy-w-g}
 \end{eqnarray}
 This result was also obtained from the Fredholm equations  \cite{Guan:2012b}.
 The leading order  of  the interaction energy gives a two-body mean field interaction energy
 between the  particles with different internal spin states.
 The kinetic energy part indicates three free-Fermi gases of different species.

For strong attraction ($|\gamma| \gg 1$) these charge bound states
are stable and the system is strongly correlated.
The corresponding binding energies of the charge bound states are
 given by $\epsilon_{r} = {\hbar^2c^2r(r^2-1)}/ {(24m)}$.
 Explicitly, the BCS pair binding energy is $\epsilon_{\rm b} = {\hbar^2c^2}/ {(4m)}$ and
the binding energy for a trion is $\epsilon_{\rm t} = {\hbar^2c^2}/ {m}$.
Without loss of  generality, the numbers of trions $N_3$, pairs $N_2$ and unpaired fermions $N_1$  are assumed to be even.
The explicit root patterns to the BA equations (\ref{BE3}) for trions, pairs and single fermions are then
\begin{eqnarray}
k^3_i=\left\{ \begin{array}{l}
\lambda_i+\mathrm{i} c\\
\lambda_i\\
\lambda_i  -\mathrm{i}c \end{array} \right.,
\,\,\,k^2_j=\left\{ \begin{array}{l}
\Lambda_j+\mathrm{i} c/2\\
\Lambda_j -\mathrm{i}c/2 \end{array} \right.,
\,\,\, k^1_\ell =k_\ell,
\end{eqnarray}
where $i=1,\ldots, N_3$, $j=1,\ldots, N_2$ and $\ell=1,\ldots, N_1$, see Fig.~\ref{fig:pattern3}.
The real parts  of the bound states have a hard-core signature, explicitly,
\begin{eqnarray}
\lambda_i&\approx& \frac{(2n^{(3)}+1)\pi}{3L}\alpha_3,\qquad
\Lambda_j\approx \frac{(2n^{(2)}+1)\pi}{2L}\alpha_2,\nonumber \\
  k_\ell & \approx & \frac{(2n^{(1)}+1)\pi}{L}\alpha_1,\nonumber
\end{eqnarray}
where $n^{(r)}=-N_r/2,-N_r/2+1,\ldots, N_{r}/2-1$ with $r=1,2,3$.  Here $\alpha_r=(1+A_r+A_r^2)$ with the function
 \begin{eqnarray}
A_{r}=\sum_{j=1}^{r-1}\sum_{i=j}^{\kappa }\frac{4n_{i}\theta(r-2)}{r(i+r-2j)}+
\sum_{i=r+1}^{\kappa}\frac{4n_{i}\theta(\kappa-r-1)}{r(i-r)}, \label{A3}
\end{eqnarray}
where $\kappa =3$ and $\theta(x)$ is the step function.

This result points to  the  interference effects among the molecule states and excess single fermions.
From these roots, the groundstate energy in the thermodynamic limit is given explicitly by  \cite{Lee:2011a,Guan:2008a,Kuhn:2012c}
\begin{equation}
 \frac{E}{L} \approx
\sum_{r=1}^{\kappa }\frac{\pi^2n_r^3}{3r}\left(1+\frac{2}{|c|}A_r+\frac{3}{c^2}A_r^2\right)-\sum_{r=2}^{\kappa}n_{r}
   \epsilon_{r}\label{E}
\end{equation}
with $\kappa=3$.
Here  $n_r=N_r/L$ and $N_r$ is the number of $r$-atom molecule states.
In general the 1D interacting Fermi gases with higher spin symmetries have two distinguishing features: (i) mean field theory for the weak coupling regime,
and (ii)  strongly correlated molecule states of different sizes in the strongly attractive regime.
The fundamental physics of the model is determined by the set of equations (\ref{BE3}) which can be transformed to
generalised Fredholm types of equations in the thermodynamic limit.
The asymptotic expansion solution of the Fredholm equations for arbitrary component Fermi gas has been thoroughly studied \cite{Guan:2012b}.

\begin{figure}[t]
{{\includegraphics [width=0.8\linewidth]{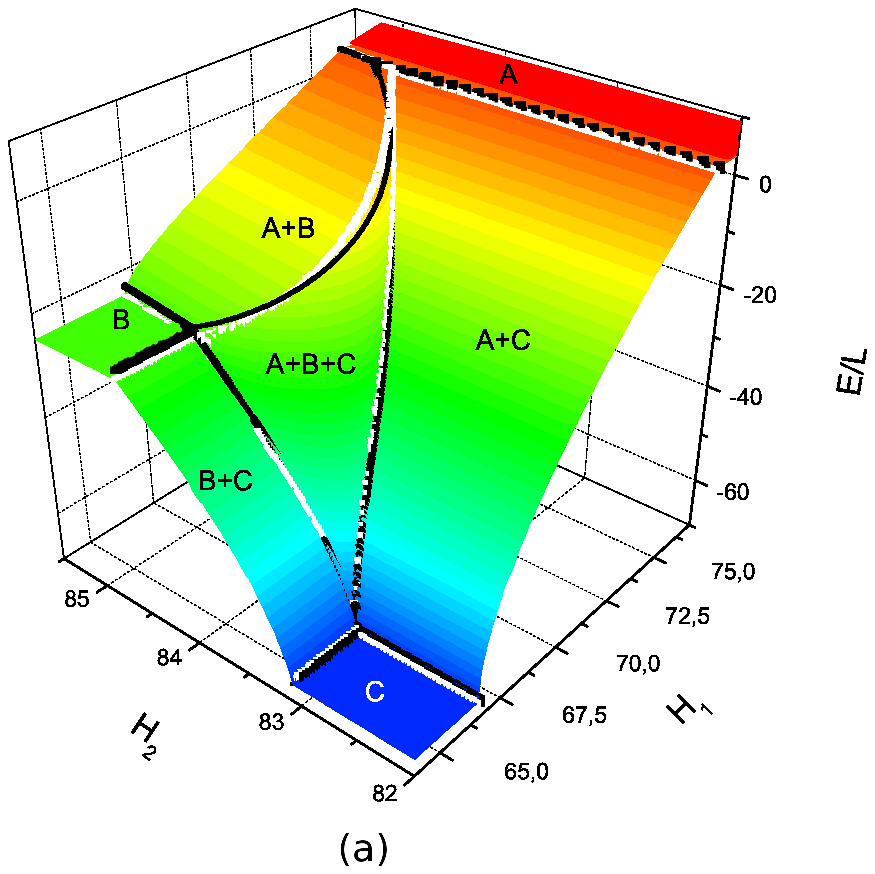}}}
{{\includegraphics [width=0.8\linewidth]{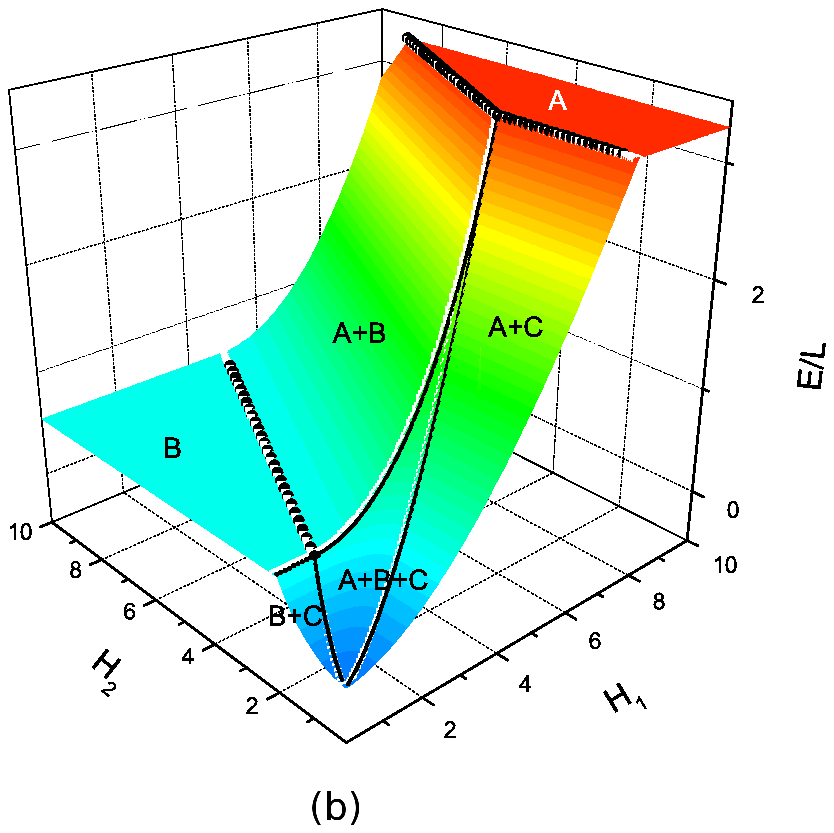}}}
\caption{Ground state energy {\em vs} Zeeman splitting for (a) strong interaction $|c| = 10$ and (b) weak
interaction $|c|=0.5$ with  $n=1$. The figure reveals a trion phase $C$, a
  pairing phase $B$, an unpaired phase $A$ and four different mixtures of these states.
Good agreement is found between the analytical critical fields  (black lines) and the numerical  solutions (white lines) of the
TBA equations (\ref{TBA-F3}).
The pure trion phase $C$ is present in the strong coupling regime,  whereas the colour pairing phase is favoured  in the weak coupling limit.
From Kuhn and Foerster (2012). }
\label{fig:E3}
\end{figure}

\subsubsection{Quantum phase transitions and phase diagrams}

The groundstate energies (\ref{Energy-w-g}) and (\ref{E})  provide full phase diagrams in the $H_1-H_2$ plane for the weak and strong coupling regimes,
see Fig.~\ref{fig:E3}.
The fields $H_1$ and $H_2$ are determined through the relations \cite{Guan:2008a,Kuhn:2012c}
\begin{eqnarray}
H_1 = \frac{\partial E/L}{\partial n_1}\,,\, H_2=\frac{\partial
E/L}{\partial n_2}\label{H1-H2}
\end{eqnarray}
with the constraint condition $n = n_1 + 2 n_2 + 3 n_3$.
For the strong coupling regime  in the absence of Zeeman splitting, i.e., $H_1=H_2=0$,  trions form a singlet groundstate.
However, the Zeeman splitting can lift the $SU(3)$
degeneracy and drive the system into different phases.
For small $H_1$, a transition from a trionic state into a mixture of trions and pairs occurs as $H_2$ exceeds
the lower critical value $H_2^{c1}$.
When $H_2$ is greater than the upper critical value $H_2^{c2}$, a pure  pairing phase takes place.
Trions and BCS pairs coexist when $H_{2}^{c1}<H_2<H_{2}^{c2}$.
These critical fields  are  derived from the relations  (\ref{H1-H2}) \cite{Guan:2008a,Kuhn:2012c}
\begin{eqnarray}
H_{2}^{c1} &\approx&  \frac{\hbar^2n^2}{2m} \left(\frac{5\gamma^2}{6} - \frac{2\pi^2}{81}(1+\frac{8}{27|\gamma|}-\frac{1}{27\gamma^2})\right),\nonumber \\
H_{2}^{c2} &\approx& \frac{\hbar^2n^2}{2m} \left(\frac{5\gamma^2}{6} + \frac{\pi^2}{8}(1+\frac{20}{27|\gamma|}-\frac{1}{36\gamma^2}) \right).\nonumber
\end{eqnarray}
The phase transitions from  the colour paired phase $B$ to the mixed phase $A+B$ of pairs and unpaired fermions
 induced by increasing $H_1$ are reminiscent  of those for the two-component system discussed in Sec.~\ref{Fermi-FFLO}.
This  mixed phase consisting of the  BCS-pairs and unpaired fermions is referred to the FFLO phase, see Fig.~\ref{fig:E3}.

For small $H_2$, a phase transition from a trionic into a mixture of trions and unpaired fermions occur as $H_1$ increases.
Using the relations  (\ref{H1-H2}),  the trionic phase  with zero polarization  forms a singlet  groundstate of trions when the field $H<H_{1}^{c1}$,
whereas when $H_1$ is greater than the upper critical value $H_{1}^{c2} $
all trions are broken and the state becomes a normal Fermi liquid, see Fig.~\ref{fig:E3}.
Here
\begin{eqnarray}
H_{1}^{c1}&\approx &\frac{\hbar^2n^2}{2m} \left( \frac{2\gamma^2}{3} - \frac{\pi^2}{81}\left(1+\frac{4}{9|\gamma|}+
\frac{1}{9\gamma^2}\right)\right),\nonumber\\
 H_{1}^{c2}  &\approx & \frac{\hbar^2n^2}{2m} \left(\frac{2\gamma^2}{3} + \pi^2 \left(1-\frac{4}{9|\gamma|} \right) \right).\nonumber
 \end{eqnarray}
 In addition, for a certain regime of $H_1$ and $H_2$, there is a phase transition from the trionic state into the mixture of trions,
 pairs and unpaired fermions, see Fig.~\ref{fig:E3}.

In the weak coupling regime, the critical fields can be obtained from the relations
\begin{eqnarray}
H_1 &=& \frac{\pi^2}{3}(2n_1^2 + n_2^2 + 4n_1n_2 +4n_1n_3 + 2n_2n_3),\nonumber\\
&+&\frac{2|c|}{3}(2n_1+n_2). \nonumber\\
H_2 &=& \frac{\pi^2}{3}(n_1^2 + 2n_2^2 + 2n_1n_2 +2n_1n_3 + 4n_2n_3)\nonumber\\
&+&\frac{2|c|}{3}(2n_2+n_1).\nonumber
\label{hweak}
\end{eqnarray}
We see  that  either a mixture  of BCS pairs and unpaired fermions or a mixture of trions and unpaired fermions
or a mixture of trions, pairs and unpaired fermions populates the groundstate for  certain
values of $H_1$ and $H_2$.
These asymptotic results indeed agree well with  the full phase diagram  determined from
numerical solutions \cite{Kuhn:2012c}, see Fig.~\ref{fig:E3}.
It is interesting to note that  the pure paired phase can be sustained under certain Zeeman splittings.
In this phase,  the two lowest levels are almost degenerate for certain tuning of $H_1$ and $H_2$.
The persistence of this colour pairing phase is relevant for the study  of phase transition between
the colour BCS-pairing phase  and  the state of trions.

In the thermodynamic limit, the grand partition function
$Z=\mathrm{tr}(\mathrm{e}^{-H/T})=\mathrm{e}^{-G/T}$ is given in terms of the Gibbs free
energy
$G=E-\mu N-H_{1}N_{1}-H_{2}N_{2}-TS$ where the chemical potential $\mu$, the Zeeman energy $E_{\rm Z}$ and the entropy $S$
are given in terms of the densities of unpaired fermions, charge
bound states, trions and spin-strings,  which are all subject to the BA equations (\ref{BE3}).
The equilibrium states are determined by minimizing the Gibbs free energy, which gives rise to
a set of coupled nonlinear integral equations -- the TBA equations
for the dressed energies  $\varepsilon _{a} (a=1,2,3)$  \cite{He:2011,Lee:2011a,Schlottmann:1993,Schlottmann:1994}.
In the thermodynamic limit, the pressure $p$ is defined in terms of the
Gibbs energy  by $p\equiv -(\partial G/\partial L)$,
 including  three parts, $p^{(1)}$, $p^{(2)}$ and $p^{(3)}$,
for the pressure of unpaired fermions, pairs and trions, respectively, with
\begin{equation}
p^{(a)}=\frac{aT}{2\pi }\int dk\ln \left(1+e^{-\varepsilon_{a}\left(
k\right)/{T}}\right).\label{pressure3}
\end{equation}
Here we have set the Boltzmann constant $k_{B}=1$.

\begin{figure}[t]
{{\includegraphics [width=1.0\linewidth]{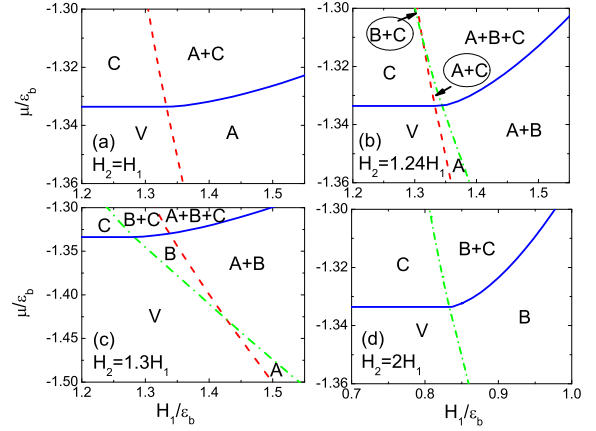}}}
\caption{The $\mu -H$ phase diagrams  for (a) equally spaced Zeeman splitting and
  unequally-spaced Zeeman splitting (b), (c) and (d).
  From He {\em et al.}  (2010). }
\label{fig:mu-h-3}
\end{figure}

The quantum phase transitions in the model with Zeeman splitting
may be analyzed via the dressed energy equations, which are
obtained from the TBA equations  in the limit $T\to 0$ as
\begin{eqnarray}
\epsilon^{(3)}(\lambda)&=&3\lambda^2-2c^2-3\mu-K_2*{\epsilon^{(1)}}(\lambda) \nonumber\\
& &-\left[K_1+K_3\right]*{\epsilon^{ (2)}}(\lambda)-\left[K_2+K_4\right]*{\epsilon^{ (3)}}(\lambda),\nonumber\\
\epsilon^{(2)}(\Lambda)&=&2\Lambda^2-2\mu
-\frac{c^2}{2}-H_2-K_1*{\epsilon^{ (1)}}(\Lambda)\nonumber \\
& &-K_2*{\epsilon^{2}}(\Lambda) -\left[K_1+K_3\right]*{\epsilon^{(3)}}(\Lambda),\nonumber\\
\epsilon^{ (1)}(k)&=&k^2-\mu
-H_1-K_1*{\epsilon^{ (2)}}(k) -K_2*{\epsilon^{(3)}}(k).
\label{TBA-F3}
\end{eqnarray}
Here we denote $K_n*\epsilon^{(a)}(x)=\int_{-Q_a}^{Q_a}K_n(x-y)\epsilon^{(a)}(y)dy$.
The integration boundaries $Q_{a}$ characterize the ``Fermi surfaces'' at $\epsilon^{(a)}(Q_{a})=0$.
The chemical potential and magnetization are determined by
$H_1$, $H_2$, $g_{1D}$ and $n$ through the relations
$
-\frac{\partial G}{\partial \mu} =n\,,\,-\frac{\partial
G}{\partial H_1}=n_1,\,\, -\frac{\partial
G}{\partial H_2}=n_2$. The dressed energy equations (\ref{TBA-F3}) indicate effective  interactions among trions,  pairs and single  fermions.
 If we denote effective chemical potentials $\mu^{\rm t}=\mu+\epsilon_{\rm t}/3$ for trions,
$\mu^{\rm b}=\mu+\epsilon_{\rm b}/2 +H_2/2$ for bound pairs and
$\mu^{\rm u}=\mu+H_1$ for unpaired fermions,
 the energy transfer relations
among the binding energy, the Zeeman energy and the variation of
chemical potentials between different Fermi seas are given by \cite{Guan:2008a}
\begin{eqnarray}
& &H_1=2c^2/3 +(\mu^{\rm u}-\mu^{\rm t}),\,\, H_2=5c^2/6+2(\mu^{\rm
  b}-\mu^{\rm t}),\nonumber\\
& &H_1-H_2/2= c^2/4+(\mu^{\rm u}-\mu^{\rm b}).
\label{E-H-relation}
\end{eqnarray}
These equations determine the full phase diagram and the critical fields triggered by the Zeeman splitting $H_{1}$ and $H_{2}$.
Indeed,  the relations  (\ref{E-H-relation}) give the results (\ref{Energy-w-g}) and (\ref{E}) in the weak and strong coupling regimes.
In the phase diagrams Fig.~\ref{fig:mu-h-3}, the phase boundaries can be also analytically determined by analyzing the
band fillings through the dressed energy equations (\ref{TBA-F3}).

\subsubsection{Universal thermodynamics of the three-component fermions}

At low  temperature regimes, the TBA equations provide  universal thermodynamics of the attractive three-component Fermi gas \cite{He:2010}.
In the grand canonical ensemble, the unequal spaced Zeeman splitting lead to various phases in the $\mu-H$ plane, see Fig.~\ref{fig:mu-h-3}.
This $\mu-H$ phase diagram can be determined by the dressed energy equations (\ref{TBA-F3}).
The quantum phases at zero temperature persist  due to the nature of collective motion (forming TLL phases) for a certain temperature range.
Although there is no quantum phase transition in 1D many-body systems at finite temperatures, quantum criticality
 leads to a crossover from a  relativistic dispersion to a nonrelativistic dispersion between the TLL and the quantum critical regime.
The nature of the TLL physics is revealed  from the universal leading  temperature corrections to the free energy in the
different phases demonstrated in Fig.~\ref{fig:E3}, i.e.,
\begin{equation}
F \approx E_{0}-\frac{\pi T^{2}}{6} \sum_\alpha \frac{1}{v_\alpha},
\end{equation}
where the sum involves a term for each cluster component in the phase.
E.g., $\frac{1}{v_2}+\frac{1}{v_3}$ for phase $B+C$.
For strong attraction, $v_r\approx \frac{\hbar \pi n_r}{mr}(1+\frac{2}{|c|}A_r+\frac{3}{c^2}A_r^2)$ with $r=1,2,
3$ are the velocities for unpaired fermions, pairs and trions, respectively.  Here the function $A_r$ are given by (\ref{A3}).

On the other hand, the pressure $p^{(a )}$ of trions, pairs and excess fermions can be obtained in  an analytical manner
using  the polylog function in the strong attractive regime, with
\begin{equation}
p^{(a )}=-\sqrt{\frac{a }{4\pi }} \, T^{{3}/{2}} \, \mathrm{{Li}}_{{3}/{2}}\left(-\mathrm{e}^{A^{(a )}/{T}}\right),  \label{ppp}
\end{equation}
for $a=1,\,2,\,3$.
Up to a few leading order terms, the functions $A^{(a)}$ are
\begin{eqnarray}
 A^{(1)} &=&\mu +H_{1}-\frac{2}{|c|}p^{(2)}-\frac{2}{3|c|}p^{(3)} \nonumber\\
&& + \, T \,  \mathrm{e}^{-(2H_{1}-H_{2})/{T}} \mathrm{e}^{-{J_{1}}/{T}} I_{0}({J_{1}}/{T}), \nonumber\\
A^{(2)} &=&2\mu+\frac{c^{2}}{2}+H_{2}-\frac{4}{|c|}p^{(1)}-\frac{1}{|c|}p^{(2)}-
\frac{16}{9|c|}p^{(3)} \nonumber\\
&& + \, T \, \mathrm{e}^{-(2H_{2}-H_{1})/{T}} \mathrm{e}^{-{J_{2}}/{T}} I_{0}({J_{2}}/{T}), \\
A^{(3)} &=&3\mu +2c^{2}-\frac{2}{|c|}p^{(1)}-\frac{8}{3|c|}p^{(2)}-\frac{1}{|c|} p^{(3)}, \nonumber
 \label{BBB}
\end{eqnarray}
where $J_a=2p^{(a)}/(a|c|) $.

The total pressure $p=\sum_{a=1}^3 p^{(a )}$
provides a high precision  equation of state through  iterations with (\ref{BBB}).
The thermodynamics and critical behaviour can  be worked out in a
straightforward manner in terms of the  polylog function. The equation of state (\ref{ppp})
covers not only the zero temperature result of the  three-component strongly attractive Fermi gas  but also the TLL thermodynamics.
This result  opens up  further study of quantum criticality  with respect to the phase transitions
when the parameters drive the system across the phase boundaries of Fig.~\ref{fig:mu-h-3}.

For the repulsive  regime, the thermodynamics of the three-component Fermi gas is
determined by another set of TBA equations \cite{Schlottmann:1993,Schlottmann:1994,He:2011}.
In this regime, spin-charge separation is a hallmark of  the 1D three-component Fermi gas.
 In the low-lying excitations,  interacting particles ``split'' into spins
and charges as the temperature tends to absolute zero.
The collective motion of fermions  with only spin or charge, called spinons
and chargons/holons (the antiparticle of a chargon), which have
different velocities. The three-component Fermi gas has $U(1)\times SU(3)$
symmetry that  leads to two sets of spin waves.  It is shown \cite{He:2011} that
 the low temperature thermodynamics  of such
a gas naturally separates into free Gaussian field theories for the
$U(1)$ charge degree of freedom and two spin rapidities.
The free energy gives a universal low temperature TLL behaviour, namely
\begin{eqnarray}
F&\approx& E_0 -\frac{\pi
T^2}{6}\left(\frac{C_s}{v_s}+\frac{C_c}{v_c}\right). \label{F3r}
\end{eqnarray}
The spin and charge velocities can be derived \cite{He:2011} from the relations
$v_c = \varepsilon^\prime\left(k_0\right)/2\pi
\rho_c\left(k_0\right)$ and $v_s =
\phi_1^{(r)\prime}\left(\lambda_0\right)/2\pi
\rho_s\left(\lambda_0\right)$.
For three-component fermions, there are two spin velocities $v_{s1}$ and $v_{s2}$, where
$v_{s1}=v_{s2}$ for pure Zeeman splitting.
The central charge for the spin part is $C_s=2$ and for the charge part $C_c=1$.
The reason for the value $C_s=2$ is because the $SU(3)$-invariant fermion model has two spin ``Fermi
seas'' whose dependence on $H$ are equal, i.e., we considered the case where $H_{1}=H_{2}=H$.

Following  the Wiener-Hopf method developed for the study of the thermodynamics of
Heisenberg spin chains with $SU(2)$ and $SU(3)$ symmetries
\cite{Mezincescu:1993a, Mezincescu:1993b}, the groundstate energy of the
gas in a weak magnetic field is given by
$E_0 = \frac{1}{3} n^3 \pi^2\left(1-\frac{2\pi n}{3\sqrt{3}c}-
\frac{2n\ln3}{c}\right) -\frac{9 c H^2}{4 n^2
\pi^4}\left(1+\frac{\pi n}{\sqrt{3}c}+\frac{3n\ln3}{c}\right)$.
Explicitly, the  spin and charge velocities
\begin{eqnarray}
v_s&=&\frac{4}{9c}n^2\pi^3\left(1-\frac{\pi n}{\sqrt{3}c}-\frac{3n\ln3}{c} \right),\nonumber \\
v_c&=&  2n\pi \left(1-\frac{2n\pi}{3\sqrt{3}c}-\frac{2n\ln3}{c}\right),\nonumber
\end{eqnarray}
are derived in the strong repulsive regime.
The spin velocity  tends to zero while the charge velocity tends to the Fermi velocity as the interaction strength $c\to \infty$.
The low temperature properties  of the equation of state follow from
\begin{equation}
P=-\sqrt{\frac{1}{4\pi}}T^{ \frac{3}{2}} \mathrm{{Li}}
_{\frac{3}{2}}(-\mathrm{e}^{{A}/{T}})
\end{equation}
where  the potential function $A=\mu+2\pi P\left( \frac{1}{6\sqrt{3} c}+\frac{\ln 3}{2\pi c}\right)+\frac{3cH^2}{2 \pi^2 P}+\frac{cT^2}{2 P}$.
The thermodynamics obtained from this pressure  covers the universal thermodynamics of the  TLL given by (\ref{F3r}).

\subsection {Ultracold fermions with higher spin symmetries }

\subsubsection{Bosonization for spin-3/2 fermions with  $SO(5)$ symmetry}

Spinor Bose gases with spin-independent short-range interactions have a ferromagnetic groundstate,
i.e., the groundstate is always fully polarized.
In contrast to the spinless Bose gas,  the spinor Bose gases with  spin-exchange interactions can
display a different groundstate, i.e., either a ferromagnetic or an antiferromagnetic groundstate solely
depending on the spin-exchange interaction \cite{Ho:1998,Ho:2000,Ohmi:1988}.
In this regard, the 1D spinor Fermi  gases with a short-range delta-function interaction and
spin-spin exchange interaction are  particularly interesting due to the existence of various BCS-like pairing phases
of quantum liquids associated with the BA \cite{Lee:2009,Essler:2009,Shlyapnikov:2011,Kuhn:2012a,Kuhn:2012b}.
Large spin fermionic systems also exhibit many new phases  of matter which do not appear  in the usual spin-1/2 systems.
In particular, spin-3/2 systems possessing  a generic $SO(5)$ symmetry have attracted much
theoretical interest \cite{Wu:2003,Wu:2005,Wu:2006,Lecheminant:2005}.
In these systems, s-wave scattering acquires interaction  in total spin singlet and quintet channels.
The two interacting channels present a hidden $SO(5)$ symmetry without fine tuning.
The spin-3/2 systems exhibit a quintet pairing phase with total spin-2 and quartetting
order as a four-fermion counterpart of the Cooper pairing.

The Hamiltonian describing these $SO(5)$-invariant  systems reads \cite{Wu:2003,Wu:2005,Wu:2006}
\begin{eqnarray}
 H=\int d^d \left\{ \sum_{\alpha=\pm \frac{3}{2},\pm\frac{1}{2}}\psi^{\dagger} _{\alpha }({\bf r})\left(-\frac{\hbar^2}{2m} \Delta^2 -\mu \right) \psi_{\alpha}({\bf r})\right. \nonumber\\
+\left. g_0P^{\dagger}_{0,0}({\bf r})P_{0,0}({\bf r})+g_2\sum_{\ell=\pm 2,\pm1,0}P^{\dagger}_{2,\ell}({\bf r})P_{2,\ell}({\bf r})\right\}\label{Ham-3-2}
\end{eqnarray}
where $d$ is the dimensionality and $\mu$ is the chemical potential.
The operators $P^{\dagger}_{0,0}$ and  $P^{\dagger}_{2,m}$ denote the spin singlet and quintet pairing operators given
by $P^{\dagger}_{F,m}({\bf r})=\sum_{\alpha,\beta}\langle \frac{3}{2}\frac{3}{2};F,m|\frac{3}{2}\frac{3}{2};\alpha \beta\rangle \psi_{\alpha}^{\dagger}({\bf r}) \psi_{\beta }^{\dagger}({\bf r})$
with $F=1,2$ and $m=-F,-F+1,\ldots, F$.
Using the Dirac matrices, the $SO(5)$ algebra has been constructed explicitly \cite{Wu:2003}.
The spin singlet pair and quintet pair have been constructed in terms of the $SO(5)$ scalar and vector operators.
The $SO(5)$ generators commute with the Hamiltonian (\ref{Ham-3-2}).  The models restore $SU(4)$ symmetry when $g_0=g_2$.
The Fermi liquid theory of $SO(5)$ symmetry  describes different competing orders in the spin-3/2 continuum and lattice models.
The lattice version of the Hamiltonian (\ref{Ham-3-2}) can be viewed as the one-band generalized Hubbard model exhibiting $SO(5)$
symmetry at arbitrary filling and $SO(7)$ symmetry at half-filling \cite{Wu:2003,Wu:2005}.

\begin{figure}[t]
{{\includegraphics [width=1.0\linewidth]{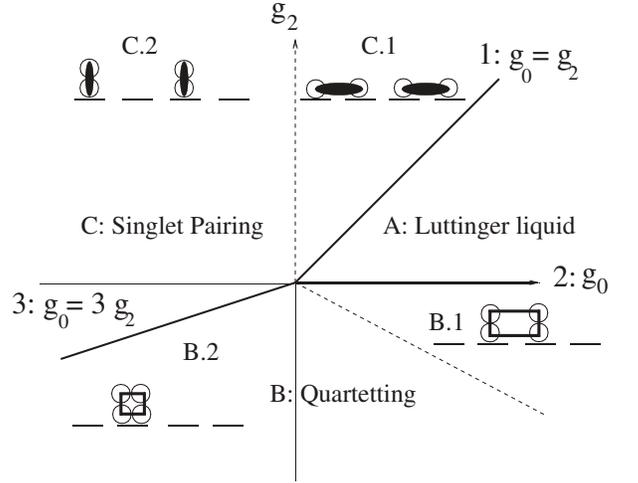}}}
\caption{Phase diagram of the 1D  spin-3/2 Fermi gas with $SO(5)$ symmetry in terms of the singlet and quintet interaction channel
parameters $g_0$ and $g_2$. Various competing orders of singlet and CDW  pairing,  as well as  QROS and CDW
of quartets are formed. From Wu (2005). }
\label{fig:Wu-phase}
\end{figure}

The $SO(5)$ invariance can apply equally well in the one-dimensional continuum model and the lattice model \cite{Wu:2005,Wu:2006,Lecheminant:2005}.
Writing the left and right moving currents  in terms of the $SO(5)$ scalar, vector and tensor currents \cite{Wu:2005}, the  low energy physics can be
described by an effective Hamiltonian density, which can be treated by bosonization.
The result indicates  two spin gap phases and various order parameters, see Fig.~\ref{fig:Wu-phase}.
The TLL phase exists in the repulsive region $g_0\ge g_2\ge 0$ where $K_c<1$.
The quartetting phase $(B)$  has two  orders -- the quasi-long range ordered superfluidity (QROS) $(B.1)$
and charge density wave (CDW)  of quartets $(B.2)$.
The  pairing phase $(C)$ separates into a spin singlet pairing  phase $(C.2)$ and dimenerization of spin Peierls order $(C.1)$.
 Here the competition between the quartetting and pairing phases is characterized by the Ising duality.
 In this scenario, the low energy physics of 1D arbitrary half-spin fermions of ultracold atoms has been studied by bosonization
  \cite{Lecheminant:2005,Szirmai:2011,Nonne:2010}.
  See also a recent study on competing orders in 1D half-filed fermionic ultracold atoms in an optical lattice \cite{Nonne:2011}.
  Two different superfluid orders, a confined BCS pairing phase and a confined molecular superfluid, were found for $F=N-1/2$ fermionic ultracold atoms.

\subsubsection{Integrable  spin-3/2 fermions with $SO(5)$ symmetry}

 Controzzi and Tsvelik (2006) write that ``{although  high symmetries do not occur frequently in nature,
 they deserve attention since every new symmetry brings with itself a possibility of new physics}"
 indicates a perspective of large spin fermionic atoms.
 Fortunately the 1D realization of spin-3/2 fermions with $SO(5)$ symmetry  is
 exactly solved under  a suitable condition \cite{Controzzi:2006,Jiang:2009}.
In particular, Jiang {\em et al.} (2009)  found that  the Hamiltonian (\ref{Ham-3-2})  reduces to an integrable  many-body Hamiltonian
\begin{equation}
{\cal H}=-\sum_{j=1}^N\frac{\partial ^2}{\partial x_j^2}+\sum_{\ell <j}^N\left(c_0+c_2{\bf S}_j {\bf S}_{\ell }\right) \delta(x_j-x_{\ell})\label{Ham-1D-SO5}
\end{equation}
on a line $c_0/c_2=-3/4$.
Here the coupling constants are given by $c_0=(g_0+2g_2)/3$ and $c_2=(g_2-g_0)/3$.
The integrable Hamiltonian (\ref{Ham-1D-SO5}) possesses $SO(5)$ symmetry.
But the individual spin components are no longer conserved.
However,  $I_1=N_{\frac{3}{2}}+N_{\frac{3}{2}}+N_{\frac{1}{2}}+N_{\frac{1}{2}}+N_{-\frac{3}{2}}$, $I_2=N_{\frac{3}{2}}-N_{-\frac{3}{2}}$ and $I_3=N_{\frac{1}{2}}-N_{-\frac{1}{2}}$
are three independent conserved quantities.
The integrability is guaranteed by the two-body scattering matrix
\begin{eqnarray}
S_{jl}=\frac{k_j-k_l-\mathrm{i}\frac{3c}{2}}{k_j-k_l+\mathrm{i}\frac{3c}{2}}P_{jl}^0+P_{jl}^1+\frac{k_j-k_l-\mathrm{i}\frac{c}{2}}{k_j-k_l+\mathrm{i}\frac{c}{2}}P_{jl}^2+P^{3}_{jl}\nonumber
\end{eqnarray}
which satisfies the YBE
\begin{eqnarray}
&&S_{12}(k_1-k_2)S_{13}(k_1-k_3)S_{23}(k_2-k_3)=\nonumber\\
&&S_{23}(k_2-k_3)S_{13}(k_1-k_3)S_{12}(k_2-k_3).
\end{eqnarray}
In the above equation, $P_{jl}^m$ is the projection operator onto the spin $m$ channel.

\begin{figure}[t]
{{\includegraphics [width=1.0\linewidth]{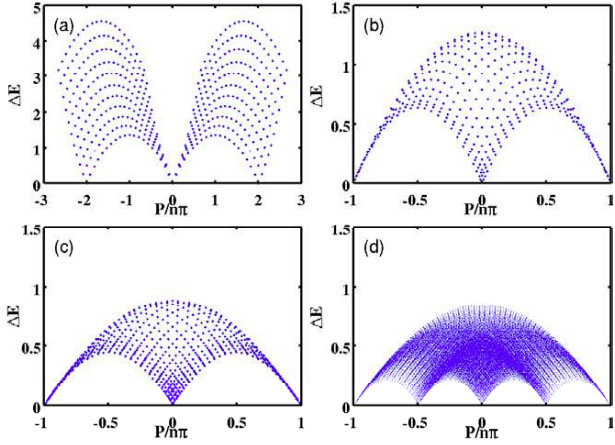}}}
\caption{Charge and spin excitations: (a) charge particle-hole excitations; (b) spin excitations of two real $\lambda$-holes
in the spin-$\lambda$  sector; (c) spin excitations of  two real $\mu$-holes   in the spin-$\mu$  sector and
(d) spin excitations of  two string-2 $\lambda$ holes  in the spin-$\lambda$  sector.  From  Jiang {\em et al.}  (2009). }
\label{fig:Jiang-excitation}
\end{figure}

The  solution has been derived in terms of the BA  \cite{Jiang:2009}.
In the repulsive regime,  the low energy physics can be described by the spin-charge separation theory.
The elementary spin excitations, including a spin-3/2 spinon, a neutral spinon  and a spin-1/2 spinon,
has been studied \cite{Jiang:2009}, see Fig.~\ref{fig:Jiang-excitation}.
Here we see that the charge excitations (a) indicate a particle-hole type.
The spin excitations (b) and (c) involve two real $\lambda$ and $\mu$ holes, respectively.
The spin excitations (d) give two string-2 hole excitations.
These spin  excitations are different from the case of $SU(4)$ symmetric fermions.
For an  attractive interaction, competing pairing orders lead to quantum phases of pairs and quartets.
However, the BA equations give very complicated root patterns.
The study of the attractive integrable $SO(5)$ symmetric model still remains an open problem.

Although integrable spin systems with higher symmetries have been extensively studied in the literature, exactly solved
models of fermionic ultracold atoms  with higher spin symmetry are still restricted to $SU(2s+1)$ \cite{Sutherland:1968}
and $Sp(2s+1)$  \cite{Jiang:2011}.
For spin-dependent interaction, the spin exchange interaction between  the $i$- and $j$-th particles  can be written
as a summation of spin projection operators $P^{m}_{ij}$ in the channels with even total spin $m=0,\,2,\ldots, \, 2s$.
The integrable $Sp(2s+1)$-invariant models of atomic fermions are artificially written as
\begin{eqnarray}
{\cal H} =-\sum_{i=1}^N\frac{\partial ^2}{\partial x_i^2}+\sum _{i\ne j}V_{ij}\delta (x_i-x_j),  \label{Ham-SP}
\end{eqnarray}
where the interaction potential   reads
\begin{eqnarray}
V_{ij}=(-1)^{2s+1}\left[ s+1/2-(-1)^{2s}\right]cP^0_{ij}+c\sum_{m=2,4,\ldots 2s}P^{m}_{ij}.\nonumber
\end{eqnarray}
 The BA equations for the model (\ref{Ham-SP}) have been derived  \cite{Jiang:2011}.
In addition, Melo and Martins (2007) derived the BA solution of a many-body problem of
interacting spin-$s$ particles with an arbitrary $U(1)$ factorized $S$-matrix.

The stability of spinor Fermi gases in tight waveguides was discussed in del Campo {\em et al.} (2007).

\subsection { Unified results for $SU(\kappa)$  Fermi gases}
\label{Fermi-N}

Fermionic alkaline-earth atoms provide unique  opportunities to study exotic many-body physics in the context of  higher spin symmetry,
e.g., $SU(\kappa)$ with $\kappa=2I+1$, where $I$ is the nuclear
spin \cite{Gorshkov:2010,Cazalilla:2009,Hermele:2009}.\footnote{In this section we use the symbol $\kappa$ to avoid any confusion with the number of particles $N$.}
De Salvo {\em et al.} (2010)  have created a degenerate gas  of ultracold fermionic  atoms $^{87}$Sr($F=I=9/2$) in an optical trap.
Taie  {\em et al.} (2010) have  reported  the realization of a degenerate Fermi mixture of two isotopes of ytterbium atoms
$^{171}$Yb (I=1/2) and $^{173}$Yb (I=5/2) with $SU(2)\times SU(6) $ symmetry.
More recently they \cite{Taie:2012} have successfully realized the $SU(6)$ symmetric Mott-insulator state with the atomic Fermi gas of  $^{173}$Yb in a 3D optical lattice.
It was found that loading fermions adiabatically into a higher symmetry Mott insulating state can achieve lower temperatures than the
$SU(2)$ symmetry state owing to differences in the entropy carried by isolated hyperfine spins.

These alkaline-earth atoms have a particular filled electron shell structure such that their nuclear  spins decouple from the electronic
angular momentum $J$ in these two states.
This decoupling implies that the nuclear spins give the hyperfine spins, i.e., $F=I$.
For these atomic systems, the $2I+1$ hyperfine levels are likely to display  $SU(2I+1)$ symmetry where the s-wave scattering lengths are independent of the nuclear spins.
Such fermionic systems with  enlarged symmetries are motivated to simulate quantum many-body phenomena  \cite{Gorshkov:2010,Hermele:2009,Xu:2010}
which may shed light on physics of strongly correlated transition-metal oxides, heavy-fermion materials and spin-liquids phases.
In contrast to the weaker quantum spin effect for a composite object with larger spin in solid state, large hyperfine spins can be essential to the
quantum magnetic states \cite{Wu:2003,Gorshkov:2010,Cazalilla:2009,Xu:2010,Wu:2010} because spin fluctuations are accommodated
in a large number  of hyperfine spin states.
Large-hyperfine fermions may also be used to study Cooper pairing phenomena within hyperfine spins \cite{Ho:1999} and quantum information \cite{Daley:2008,Gorshkov:2009}.

The Hamiltonian for the 1D $N$-body $\delta$-function interacting fermion problem \cite{Sutherland:1968} is again as defined in Eq.~(\ref{Ham}).
There are now  $\kappa$ possible hyperfine states $|1\rangle, |2\rangle, \ldots,
|\kappa\rangle$ that the fermions can occupy.   This system has
$SU(\kappa)$ spin symmetry and $U(1)$ charge symmetry.
 For an irreducible representation $[\kappa^{N_{\rm \kappa}},(\kappa-1)^{N_{\rm \kappa-1}},\ldots,2^{N_{\rm 2}},1^{N_{\rm 1}}]$,
 the Young diagram  has $\kappa$ columns with the quantum numbers  $N_i=N^{i}-N^{i+1}$.
Here $N^{\rm i}$ is the number of fermions at the $i$-th hyperfine level  such that $N^{\rm 1}\geq N^{\rm 2}\geq\ldots\geq
N^{\kappa}$.  The groundstate properties and thermodynamics of 1D $SU(\kappa)$-invariant Fermi gases  have  been studied
by means of the  BA solution, e.g., the three-component Fermi gas   \cite{Guan:2008a,Liu:2008a,He:2010},
$SU(4)$-invariant spin-3/2 fermions  \cite{Guan:2009,Schlottmann:2012a,Schlottmann:2012b,Schlottmann:2012c},
and $\kappa$-component fermions  \cite{Schlottmann:1993,Schlottmann:1994,Guan:2010,Lee:2011a,Yang:2011,Yin:2011a}.

\subsubsection{Groundstate energy}

For the groundstate of the $SU(\kappa)$-invariant Fermi gas,
the generalized Fredholm equations for $c>0$ are given by \cite{Sutherland:1968}
\begin{eqnarray}
r_{0}(k)&=&\beta_{0}+\displaystyle
\int_{-B_{1}}^{B_{1}}K_{1}(k-k') r_{1}(k')dk',\nonumber \\
r_{m}(k)&=&\displaystyle\sum_{\alpha =m-1}^{m+1} \int_{-B_{\alpha}}^{B_{\alpha }}
K_{1+\delta_{\alpha m}}(k-\lambda)r_{\alpha } (\lambda)d\lambda \label{FE-R}
\end{eqnarray}
where $1\leq m \leq \kappa-1$  and $\beta_{0}=1/(2\pi)$, $r_{0}(k)$ is the particle
quasimomentum distribution function, whereas $r_{m}(k)$ with $m\ge 1$
are the distribution functions for  the $\kappa-1$ spin  rapidities.
The groundstate energy $E$ per unit length is given  by $E=\int_{-B_0}^{B_0}k^2 r_{0}(k) d k$.
For the  balanced case, the integration boundaries $B_m$ with $m\ge 1$ are infinitely large.
Thus the distributions are given by \cite{Guan:2012b}
$r_{m}(\lambda )=\frac{1}{2\pi}\int_{-\infty}^{\infty}\tilde{r}_m(\omega) e^{-\mathrm{i} \omega \lambda }d\omega$,
where
\begin{equation}
\tilde{r}_m(\omega)=\frac{\tilde{r}_{0_{\rm in}} (\omega)\sinh\left[ \frac{1}{2}(\kappa -m)|\omega |c\right]}{\sinh\left[\frac{1}{2}\kappa |\omega |c \right]} \label{r-m-o}
\end{equation}
with $m=1,\ldots, \kappa-1$.

For strong repulsion,  i.e.,  $cL/N\gg 1$,  the groundstate energy of the balanced $\kappa$-component Fermi gas follows as \cite{Guan:2012b}
\begin{eqnarray}
 E \approx  \frac{n^3\pi^{2}}{3}\left[1-\frac{4 Z_{1}}{\gamma }+
\frac{12 Z_{1}^{2}}{\gamma^2 }- \frac{32 }{\gamma^3}\left(  Z_{1}^{3}-
\frac{Z_{3}\pi^{2}}{15}\right) \right] \label{E-R1}
\end{eqnarray}
with the constants  $Z_1=- \frac{1}{\kappa} \left[\psi(\frac{1}{\kappa})+C\right]$ and $Z_3= \kappa^{-3}\left[ \zeta(3,\frac{1}{\kappa})-\zeta(3)\right]$.
Here $\zeta(z,q)$ and $\zeta(z)$ are  the Riemann  zeta functions,   $\psi(p)$ denotes the Euler psi function and $C$ is the Euler constant.
For $\kappa \to \infty$,   it is seen that $ \lim_{\kappa\rightarrow \infty}Z_{1}= \lim_{\kappa\rightarrow \infty}Z_{3}=1$.
This  indicates an insight into the hyperfine spin effect -- the groundstate energy (\ref{E-R1})
reduces to the energy of the spinless Bose gas as $c\to \infty$.
This result was first noticed by Yang and You (2011) by means of  the Fredholm equations (\ref{FE-R}).
The suppression of the hyperfine spin effect  is further manifest in the groundstate energy per unit length
for the highly polarized  case, which reads
\begin{eqnarray}
E&=&\frac{ n^{3}\pi^{2}}{3}\left\{1
- \frac{8m_{1}}{c }+
\frac{48m_{1}^{2}}{c^{2}}-
\frac{256m_{1}^{3}}{c^{3}}\right.\nonumber\\
&& \left.+ \frac{32\pi^{2}m_{1}n^{2}}
{5c^{3}}\right\}+O(c^{-4}),
 \label{E-HP}
 \end{eqnarray}
 where $m_1=M_1/L$ with $M_1=\sum_{j=1}^{\kappa-1}N^{\rm j+1}\ll N$.
 This result shows that  spin variation does not play an essential  role in the groundstate due to the strong repulsion.
 However, for small polarization, i.e., a small external field lifting the $SU(\kappa)$ symmetry,
 the integral boundaries $B_m$ with $m\ge 1$ are very large.
 In this case, logarithmic singularities arise in the zero temperature susceptibility.
 This configuration is drastically changed as the interaction is decreased.

For the weak coupling limit,
the mean field  result  for the groundstate energy per unit length is
\begin{eqnarray}
E =\frac{1}{3}\sum_{i=1}^{\kappa}p_i^3\pi^2+2c \sum_{i=1}^{\kappa -1}\sum_{j=i+1}^{\kappa}p_ip_j+O(c^{2}).\label{E-R5r2}
\end{eqnarray}
Here $p_i=N^{\rm i}/N$ with $i=1,2,\ldots, \kappa-1$ denote the polarizations, and  $N^i$ is the number of fermions in the $i$th level.
The first part is the kinetic energy of the $\kappa$-component  fermions whereas the second parts is the interaction energy.
This result is valid for arbitrary spin imbalance in the weakly repulsive and attractive regime.
The higher order corrections have not been obtained.
 For the balanced case, i.e., $N_i=N/\kappa$,
the energy is  $E=\displaystyle \frac{\pi^{2}n^{3}}{3\kappa^{2}} +c(\kappa-1)n^2/\kappa +O(c^{2})$
which is  the same as for spinless bosons with a weak repulsion as $\kappa \to \infty$.
The groundstate properties for the limits  $c= 0^+$ and $c\to \infty$ have been discussed by Schlottmann (1997).

In the attractive regime,  the complex string solutions for $k_{j}$  form $m$-atom bound states  up to length
$2, \ldots, \kappa$ with the binding energy for a bound state
 $\varepsilon_{\rm b}^{(\ell)}= \ell (\ell^{2}-1)c^{2}/12$.
 Such bound states were studied by Gu and Yang (1989).
A bound state in quasimomentum space of length $m$ takes on the form
$k_{\alpha}^{m,j}=\lambda_{\alpha}^{(m-1)}+ \mathrm{i}(m+1-2j)|c'|+O(\exp(-\delta L))$,
where $j=1,\ldots,m$. The number of  bound
states with length $1\leq m \leq\kappa$  is  denoted by  $N_{m}$.
The real part is $\lambda_{\alpha}^{(m-1)}$.
The unpaired atoms have  real quasimomenta $k_i$.
Takahashi (1970b) derived the Fredholm equations for the attractive Fermi gas with an arbitrary number of components
\begin{eqnarray}
\rho_{m}(\lambda)&=&m\beta_{0}+
\sum_{r=1}^{m-1} \sum_{s=r}^{\kappa}
\int_{-Q_{s}}^{Q_{s}}K_{s+m-2r}(\lambda-\Lambda) \rho_{s}(\Lambda)
d\Lambda\nonumber\\
&&+\sum_{s=m+1}^{\kappa}
\int_{-Q_{s}}^{Q_{s}} K_{s-m}(\lambda-\Lambda)
\rho_{s}(\Lambda)d\Lambda,   \label{FE-A}
\end{eqnarray}
where  $\rho_{1}(k)$ is the density distribution function of single fermions,
$\rho_{m}(k)$ is  the density distribution function  for  the $m$-atom bound state with $1<  m \le  \kappa$.
The  total number of fermions is given by $N = \sum_{m=1}^{\kappa} mN_m$.
 The integration boundaries  $Q_m$,  characterizing the Fermi points in each Fermi sea, are determined by  $ n_m:= \frac{N_m}{L}=\int_{-Q_m}^{Q_m}\rho_m(k)dk$.
The groundstate  energy per unit length is given by
$E=\sum_{m=1}^{\kappa}\int_{-Q_m}^{Q_m}\left(mk^{2}-\frac{m(m^{2}-1)}{12}c^{2}\right)\rho_{m}(k)dk$.

For weak attraction $|c|L/N\ll 1$,  the two sets of the Fredholm equations (\ref{FE-R}) and (\ref{FE-A})
for the repulsive and attractive regimes preserve  the symmetry
\begin{eqnarray}
\rho_{m} &\rightarrow&  r_{\kappa-m},\qquad  Q_{m}\rightarrow B_{\kappa-m},\nonumber\\
\int_{-Q_{m}}^{Q_{m}} & \rightarrow  &\int_{-\infty}^{\infty}-\int_{-B_{\kappa-m}}^{B_{\kappa-m}},\qquad c \rightarrow -c .
\label{symmetry}
\end{eqnarray}
Indeed,  the groundstate energy of the $\kappa$-component Fermi gas with weak attraction  has the same closed form
as  (\ref {E-R5r2})  with the replacement $c\to -c$.
This means that the groundstate energy of the $\kappa$-component  gas with arbitrary polarization
continuously connects at $c=0$.

For strong attraction $|c|L/N\gg 1$, the bound states of different sizes form tightly bound molecules of different sizes.
 In this regime, all the Fermi momenta of the molecules are finite, i.e., $|c|\gg Q_m$ with $m=1,\ldots, \kappa$.
 Here $Q_1$ characterizes the Fermi momentum of the single spin-aligned  atoms.
 Therefore, in this regime  the strong coupling condition allows one to expand  the Fredholm equations (\ref{FE-A})
 in powers of $1/|c|$.
 The  groundstate energy per unit length  of the gas with arbitrary polarization  in the strong attractive regime is given by
$E=\sum_{m=1}^{\kappa}(E_m-n_m\varepsilon_{\rm b}^{(m)})$.
The energy $E_\ell$ of the cluster state of an $\ell$-atom  is  given by \cite{Guan:2012b}
\begin{eqnarray}
E_{m}&\approx &\frac{\pi^{2}N_{m}^{3}}{3m L^{3}}
\left\{1+ \frac{8}{mL|c|}F_{m} + \frac{48}{m^{2}L^{2}|c|^{2}}F_{m}^{2} \right.\label{E-A5} \\
&& \left.+ \frac{256}{m^{3}L^{3}|c|^{3}}F_{m}^{3} +\frac{16\pi^{2}}{m^{3}L^{3} |c|^{3}}\left[-G_{m}+\overline{G}_{m}/15\right]\right\}. \nonumber
\end{eqnarray}
The coefficients $F_m$, $G_m$ and $\overline{G}_{m}$ can be found in Guan {\em et al.} (2012).

This  closed form for the groundstate energy with arbitrary polarization is very accurate for a finitely  strong attraction (for $|c|> 5$).
This high precision groundstate energy has been given for the two-component attractive Fermi gas \cite{He:2009,Iida:2007,Zhou:2012}.
Up to order $1/\gamma^2$, the above groundstate energy of 1D
$\kappa$-component fermions with arbitrary population imbalance
 can be further simplified to  the general form  given in (\ref {E}).

From the result (\ref{E-A5})  one can obtain full phase diagrams and magnetism at zero temperature. In particular,
for  the three-component Fermi gases  \cite{Lee:2011a,Guan:2008a,Kuhn:2012c}, spin-3/2 Fermi gas with
 $SU(3)$ symmetry \cite{Guan:2009,Schlottmann:2012a}, spin-5/2, 7/2 and 9/2  attractive Fermi gases
 \cite{Schlottmann:2012b,Schlottmann:2012c}, and $SU(\kappa)$
 Fermi gases  \cite{Schlottmann:1993,Schlottmann:1994,Guan:2010,Lee:2011a,Yang:2011,Yin:2011a}.

The $SU(\kappa)$ symmetry requires each hyperfine spin state to be conserved.
Therefore, there are $\kappa$ chemical potentials associated with each spin state.
For convenience in analyzing the quantum phases of the multi-component attractive Fermi gas,
the Zeemann energy is chosen as $E_{z}/L = -\sum_{\ell=1}^{\kappa-1}n_{\ell}H_{\ell}$
where $H_\ell$ is an effective external field  for the cluster state of size $\ell$-atom.
Here $H_1$ denotes  the chemical potential for unpaired fermions.
The particle numbers are changed by $\mu N$, with $\mu$ the total chemical potential.
These effective fields result in unequally spacing Zeeman splitting via $\Delta_{i+1,i}=-H_{i-1}+2H_i-H_{i+1}$.
Here $H_{\kappa}=0$ because of the spin singlet state.
Defining the effective chemical potential of the $m$-atom bound state by $\mu_\ell=\mu+(H_\ell +\varepsilon_{b}^\ell )/\ell$,
the result  $ \mu_{\ell } = \frac{1}{\ell }\frac{\partial}{\partial n_{\ell}}\left(\frac{E}{L}+\sum_{\alpha  =1}^{\kappa}n_{\alpha }\varepsilon_{b}^{\alpha }\right)$ with $\ell =1,\ldots \kappa$ can be obtained from the groundstate energy.

\begin{figure}[t]
{{\includegraphics [width=1.0\linewidth]{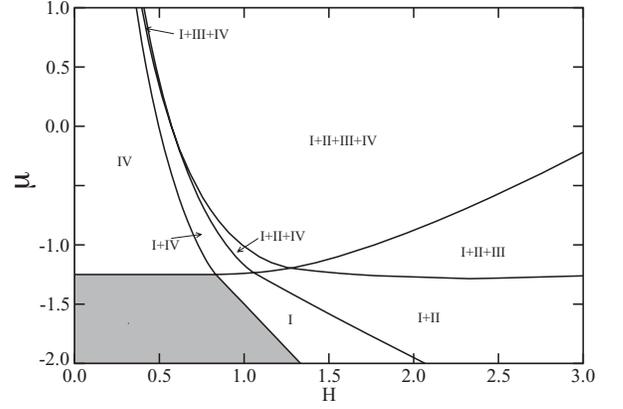}}}
\caption{Phase diagram of the 1D integrable   spin-3/2 Fermi gas  with an attractive interaction
and pure Zeeman splitting.  Quantum phases of single excess fermions (I), colour BCS pairs (II), trions (III) and quartets (IV)
are displayed   in the $\mu-H$ plane for $|c|=1$. The shaded area corresponds to the vacuum.
From Schlottmann and Zvyagin (2012a). }
\label{fig:Phase-4r}
\end{figure}

The energy-field transfer  relation between $H_{m}$ and the effective chemical
potentials $\mu_m$  \cite{Guan:2010}
\begin{equation}
H_{\ell }=\ell\, (\mu_{\ell}-\mu_{\kappa})+\frac{m\epsilon_{\kappa}}{\kappa}-\epsilon_{\ell}
\label{externalField}
\end{equation}
determines full phase diagrams of the system in terms of the effective fields $H_m$ and chemical potentials.
In the special case of pure Zeeman splitting
where $\Delta_{\ell+1,\ell}=\Delta$ for all $\ell $, the system has three
distinct magnetic phases for a strong attractive regime.
For weak coupling, even for pure Zeeman splitting, the phase diagram is very sophisticated,  see Fig.~\ref{fig:Phase-4r}.
Some subtle phase diagrams for the spin-3/2, 5/2, 7/2 and 9/2  attractive Fermi gases have been studied  \cite{Guan:2009,Schlottmann:2012b,Schlottmann:2012c}.

Furthermore, for the polarized phase, the cluster states  form multi-component TLLs in the low energy physics.
The low-lying excitations are described by the
linear dispersion relations $\omega^{r}(k)=v_r(k-k_F^r)$, where the velocities can be calculated by
\begin{equation}
 v_r=\sqrt{\frac{L}{mn_r}\frac{1}{r}\left[ \frac{\partial ^2E_r}{\partial ^2L} \right]} \label{volocity-v-r}
 \end{equation}
for a system featuring Galilean invariance. For a strong attractive interaction,
 the velocities for unpaired fermions and charge bound state of $r$-fermions
are given by
$v_r\approx \frac{\hbar \pi n_r}{mr}(1+\frac{2}{|c|}A_r+\frac{3}{c^2}A_r^2)$ with $r=1,\ldots, \kappa$.
The function  $A_{r}$  is given by (\ref{A3}). These dispersion relations  naturally lead  to the universal form for
finite-size corrections to the groundstate energy $E_0^{\infty}$  \cite{Guan:2010},
\begin{equation}
E(L,N)\approx LE_0^{\infty}  -\frac{\pi \hbar
C}{6L}\sum_{r=1}^{\kappa}v_r.\label{FSC}
\end{equation}
The central charge $C=1$ for $U(1)$ symmetry.   Here the universal
 finite-size corrections (\ref{FSC}) indicate  the TLL signature of the many-body physics.
 In this case the low energy excitations of the system are described by the CFT of the Gaussian model.
 In the next subsection, we shall discuss the thermodynamics of these cluster states.

\subsubsection{Universal thermodynamics of $SU(\kappa)$-invariant fermions}
\label{Fermi-N}

Schlottmann (1993) derived the TBA equations for $SU(\kappa)$ fermions with repulsive and attractive interaction.
Lee {\em et al.} (2011b) derived  a different  set of TBA equations
 which are more convenient for analysis of thermodynamics and phase transitions for the attractive  Fermi  gas.
To understand the thermodynamics of this model, it is
crucial to separate different physical regimes, i.e., 1) groundstate for $T\to 0$;   2) TLL phases, $T<| \mu-\mu_c|$ and  $T<|H-H_c|$;
3) quantum criticality, $T>| \mu-\mu_c|$ and  $T>|H-H_c|$;  4)  high temperatures, $T \gg c^2 $.
The TBA equations involve an infinite number of coupled nonlinear integral equations.
At zero temperature, the phase diagrams can be analytically or numerically obtained through the dressed energy equations
which can be derived from the TBA equations in the limit
$T\to 0$ \cite{Schlottmann:1993,Schlottmann:1994,Guan:2010,Lee:2011a,Yang:2011,Yin:2011a,Schlottmann:2012a,Schlottmann:2012b}.

In the strongly attractive regime, the effective ferromagnetic spin-spin coupling constants  are  given by
$J^{(r)}\approx \frac{2}{r|c|}p^{(r)}$ for $r=1,\,2,\ldots,\kappa-1$.
$p^{(r)}$ is the pressure for charge $r$-atom bound states.
In this sense, we may simply view the non-neutral charge bound state as a molecule with spin $s=\frac{\kappa+1}{2}-r$,
which could flip its spin to form the spin wave bound states (spin strings) due to thermal fluctuations.
However,   in the physically interesting regime where
$T\ll \epsilon_r$,  $T\ll \Delta_{i+1\,i}$ and $ \gamma\gg 1$
 the breaking of charge  bound states
and spin wave fluctuations  are strongly suppressed.
The spin string contributions to thermal fluctuations in this  regime can be
asymptotically calculated from the TBA i.e., $f_s^{(r)}\approx
Te^{-\frac{\Delta_{r+1\,r}}{T}}e^{-\frac{J^{(r)}}{T}}I_0(\frac{J^{(r)}}{T})$.
It is obvious that $f_s^{(r)}$ becomes exponentially small as $T \to 0$.
Thus each dressed energy can be written in a single particle form
$\epsilon^r(k)=\hbar^2rk^2/2m-\bar{\mu}^{(r)}+O(1/\gamma^3)$,
where the marginal scattering energies among composites and unpaired
fermions as well as  spin-wave thermal fluctuations are considered in the
chemical potentials $\bar{\mu}^{(r)}$.

With the help of  this simplification, the thermodynamics at finite temperatures  have been  given as \cite{Guan:2010}
\begin{eqnarray}
p^{(r)}&\approx
&-\sqrt{\frac{rm}{2\pi\hbar^2}} \, T^{\frac{3}{2}} \, {\rm Li}_{\frac{3}{2}}\left(-\mathrm{e}^{\bar{\mu}^{(r)}/T}\right),
 \label{TBA}\\
\bar{\mu}^{(r)}&\approx &r\mu^{(r)}-\sum^r_{j=1}\sum^{N}_{\scriptscriptstyle{\small
    \begin{array}{c}i=j\\i\ne
2j-r\end{array}}}\frac{4p^{(i)}}{i(i+r-2j)|c|}+f_s^{(r)},\nonumber
\end{eqnarray}
for $r=1,\ldots,N$. The total pressure of  the system is given by $p=\sum_{r=1}^{N}p^{(r)}$.
Here ${\rm Li}_{s}(x)$ is the standard polylog function.
Furthermore, the suppression of spin fluctuations leads to
a universality class of a multi-component TLL  in each gapless
phase, where the charge bound states of $r$-atoms form hard-core
composite particles, i.e.,
the leading low temperature corrections to the free energy reads
\begin{equation}
f \approx f_0-\frac{\pi
  T^2}{6\hbar}\sum_{r=1}^{N}\frac{1}{v_{r}}. \label{FreeE}
\end{equation}
The velocity $v_r$  of the $r$-atom bound state is given by (\ref{volocity-v-r}).
This result  is  consistent with the finite-size correction result  (\ref{FSC}).
In the above
equation $f_0= E_0^{\infty}-\sum_{r=1}^{N-1}n_rH_r$.
This result proves the existence of TLL phases in 1D gapped systems at
low temperatures.  The existence of the TLL leads to a crossover
from a relativistic dispersion to a nonrelativistic dispersion between
different regimes at low temperatures \cite{Maeda:2007}.
Linear Zeeman splitting may result in a two-component Luttinger
liquid in a large portion of Zeeman parameter space at low temperatures.

\section{Correlation functions}
\label{CFT-CF}

It is well established that the universality classes of critical behaviour of two-dimensional systems
are described by a rational conformal field theory (CFT), where the Hilbert space of states contains a
direct sum of irreducible representations of a Virasoro algebra.
Using the CFT one can calculate the critical exponents that characterize power law decay of correlation
functions at large distance \cite{Henkel:1999,Voit:1994,Essler:2005}.
In general, CFT predicts  that the two-point correlation function  for primary fields with conformal dimension $\Delta^{\pm}$  is of the form
\begin{equation}
\langle \phi (\tau,y)\phi(0,0)\rangle=\frac{\exp(2\mathrm{i}\Delta Dk_Fy)}{(v\tau +\mathrm{i} y)^{2\Delta^+}(v\tau -\mathrm{i} y)^{2\Delta^-}}.
\end{equation}
Here $\tau$  is the Euclidean time ($-\infty < \tau  <\infty,\,  -L \le  y \le 0$) and $v$  is the velocity of light.
The conformal dimensions  $\Delta ^{\pm}$ can be read off from the finite-size corrections of low-lying excitations of 1D systems  via
\begin{eqnarray}
E_Q-E_0&=&\frac{2\pi v}{L}\left(x+ N^+ +N^-\right),\nonumber\\
P_Q-P_0&=&\frac{2\pi }{L}\left(s+ N^+ -N^-\right)+2\Delta Dk_F,
\end{eqnarray}
where  $x=\Delta^++\Delta^-$ is  the conformal dimension and  $s=\Delta^+-\Delta^-$ is the conformal spin.
The non-negative integers $N^+$ and $N^-$ label the level of the descendant.
$\Delta D$  represents the number of particles backscattered.

Moreover,  it was shown by Affleck (1986)
that  conformal invariance gives a universal form for the finite temperature effects in the free energy
by replacing $1/L$ with $T$ in the conformal map $z=\exp(2\pi\omega/L)$.
Considering a conformal mapping of the complex plane without the origin (corresponding to $T = 0$)
onto a strip of width $1/T$  in the imaginary time direction, the two point correlation function at finite temperatures reads
\begin{equation}
\langle \phi (\tau,y)\phi(0,0)\rangle=e^{2\mathrm{i}\Delta Dk_Fy}f_T^+(y-\mathrm{i}v\tau )f_T^-(y+\mathrm{i}v\tau )
\end{equation}
with $f_T^{\pm} (x)=\left(\pi T/v\right)^{2\Delta^{\pm} }/\left[\sinh\frac{\pi T}{v}x\right]^{2\Delta^{\pm} }$.
For such critical phenomena, the critical Hamiltonian of the gapless sector exhibits not only
global scale invariance but also local  conformal invariance, i.e., a local version of the scale invariance.
 Thus the critical Hamiltonian can be approximately described  by the conformal Hamiltonian
 which is described by the generators  of the underlying Virasoro algebra  with central charge $C$.

The QISM provides a systematic  way to calculate the critical exponents for BA integrable models.
Bogoliubov {\em et al.} (1986) derived explicit critical exponents for  the $XXX$ and $XXZ$ chains.
In this approach, the dressed charge matrix  $Z_{\alpha\beta}$ formalism presents a unified framework to
calculate the conformal towers through  the finite-size corrections to the eigenspectrum of  multi-component BA systems \cite{Izergin:1989}.
The universality  class of critical exponents is uniquely determined by the symmetry of the models associated with the quantum $R$-matrix.
Consequently, the asymptotic behaviour of correlation functions for  integrable models, such as
impenetrable Bose gas \cite{Its:1990}, the supersymmetric $t-J$ model
\cite{Kawakami:1991} and the Hubbard model
\cite{Frahm:1990,Frahm:1991,Woynarovich:1987,Woynarovich:1989},  has
been studied  in the framework of the QISM.
In addition, exact results for the form factors have been derived for several cases \cite{Slavnov:1989,Slavnov:1990,Kojima:1997,Kitanine:1999}.
Further developments in the theoretical study of correlation functions have been
reported  \cite{Calabrese:2007,Caux:2006,Caux:2007,Gangardt:2003a,Gangardt:2003b,Kitanine:2009,Kormos:2009}.

\subsection{Correlation functions and the nature of FFLO pairing }

As remarked in Sec.~I,   for  a system with  partial polarization,  the Fermi energies of spin-up and spin-down
electrons are unlikely to match.
This leads to a non-standard form of pairing known as the FFLO state.
The power-law decay of the pair correlation $n^{\mathrm{pair}}\propto\cos(k_{\mathrm{FFLO}}|x|)/|x|^{\alpha}$ with the
spatial oscillations depending solely on  the mismatch of the Fermi surfaces $k_{\mathrm{FFLO}}=\pi(n_{\uparrow}-n_{\downarrow})$
has been identified numerically.
This spatial oscillation is a typical characteristic of the FFLO pairing.

The FFLO-like pair correlations and spin correlation for the attractive Hubbard model  were investigated   by several  groups via various methods,
such as DMRG  \cite{Feiguin:2007,Tezuka:2008,Rizzi:2008a,Luscher:2008,Tezuka:2010},
QMC \cite{Batrouni:2008,Baur:2010,Wolak:2010},
mean field theory and other methods \cite{Parish:2007,Liu:2007,Liu:2008b,Zhao:2008,Edge:2009,Datta:2009,Pei:2010,Edge:2010,Devreese:2011,Kajala:2011,Chen:2012}.
The FFLO signature is displayed by the on-site pair correlation function
$O_{\rm on-site}(z_i,z_j) :=  \langle \Psi_0 | \hat{c}^{\dagger }_{i,\downarrow}\hat{c}^{\dagger }_{i,\uparrow}\hat{c}^{ }_{j,\uparrow}\hat{c}^{ }_{j,\downarrow} |\rangle$
in the 1D attractive Hubbard model  \cite{Tezuka:2008} where $|\Psi_0 \rangle$ is the eigenstate for  the groundstate of the system.

\begin{figure}[h]
{{\includegraphics [width=1.0\linewidth]{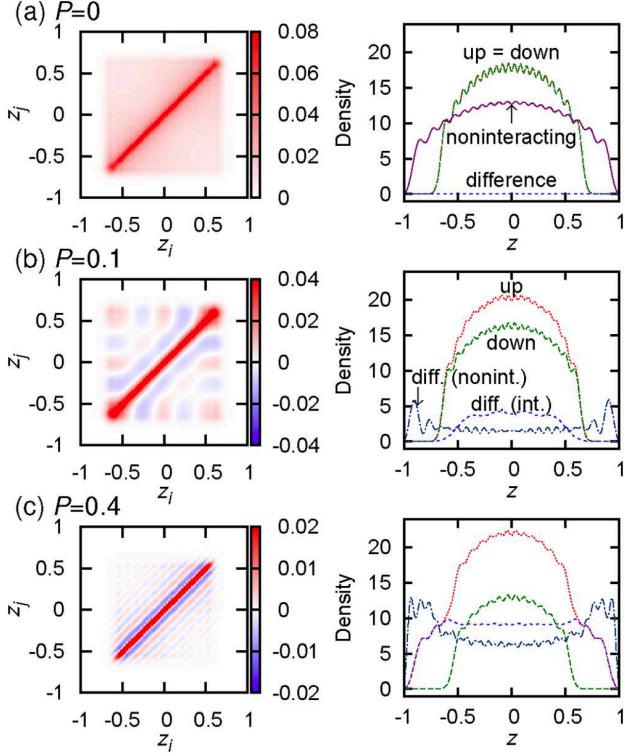}}}
\caption{Left panels:  the on-site pair correlation function $O_{\rm on-site}(z_i,z_j)$ in the $z_i-z_j$ plane for polarization
$(N_{\uparrow},N_{\downarrow})=(a) (20,20), (b) (22,18)$ and $(c) (28,12)$.
Right panels: the axial density distribution profiles in a harmonic trap.
Numerical setting with interacting strength $U/t=-4$.
From Tezuka and Ueda (2008). }
\label{fig:pair-correlation}
\end{figure}

\begin{figure}[h]
{{\includegraphics [width=1.0\linewidth]{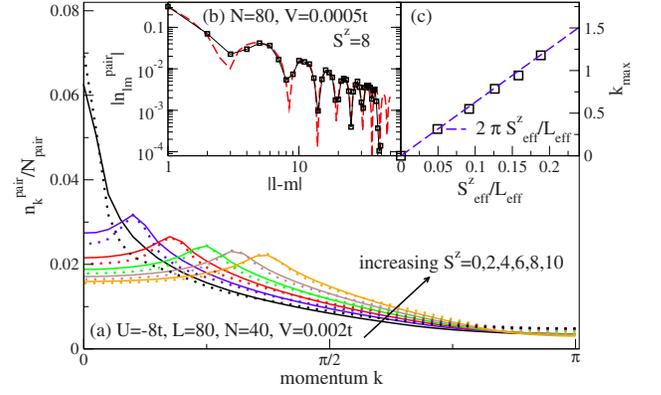}}}
\caption{(a) Momentum distribution function of pairs in the 1D attractive Hubbard model with $U/t=-8$.
 (b) Power-law decay behaviour in real space and comparison of the DMRG result (symbols)  to the bosonization (lines).
 (c) The positions of the peaks  in  the momentum distribution function $n_k^{\rm pair}$. From Feiguin and Heidrich-Meisner (2007). }
\label{fig:pair-correlation-p}
\end{figure}

In Fig.~\ref{fig:pair-correlation},  for polarization $P=0$, the on-site pair correlation indicates a power-law decay without changing the
sign (i.e., spatial oscillation).
For low polarization $P=0.1$, the pair correlation function still has a  power-law decay but periodically  changes its sign.
For $P=0.4$,  the oscillation frequency  of the pair correlation becomes faster due to a large mismatch of the two Fermi energies.
The wave vector $q$ of oscillations $q=\Delta k_{\mathrm FFLO}$.
Feiguin and Heidrich-Meisner (2007)  showed that the pair momentum distribution function has a peak at the mismatch of
the Fermi surfaces $k=\Delta k_{\mathrm FFLO}$,  i.e.,
\begin{eqnarray}
n_k^{\rm pair} =\frac{1}{L}\sum_{lm}\exp\left(\mathrm{i} k(l-m) \right)\rho_{lm}^{\rm pair}
\end{eqnarray}
where the pair correlation $\rho_{lm} ^{\rm pair} \propto |\cos (\Delta k_{\mathrm FFLO}|l-m|)/|l-m|^{\Delta(P)}$, see Fig.~\ref{fig:pair-correlation-p}.
The correlation exponent $\Delta(P)$  depends on  both the polarization $P$ and interaction strength.
Such a FFLO pairing wave number was also confirmed by the occurrence of a peak in the pair momentum distribution
corresponding to the difference between the Fermi momenta of
individual species \cite{Rizzi:2008a,Batrouni:2008,Heidrich-Meisner:2010b,Feiguin:2009}.

On the other hand, the critical behaviour of 1D many-body systems with
linear dispersion in the vicinities of their Fermi points can be
described by conformal field theory.
The critical behaviour of the Hubbard model with attractive interaction was
investigated by Bogoliubov and Korepin (1988,1989,1990,1992).
As demonstrated in previous Sections,  the low-energy physics of
the homogeneous  1D Fermi gases  with polarization  is described by the
TLLs of bound pairs  and excess unpaired fermions in the charge sector and ferromagnetic spin-spin
interactions in the spin sector \cite{Zhao:2009,Guan:2010}.
This paves a way to study asymptotic behaviour of  various correlation functions by using CFT.
The method used to study correlation functions of the spin-1/2 Fermi gas with attractive interaction follows
closely the method set out in the literature \cite{Woynarovich:1989,Kawakami:1991,Frahm:1990,Frahm:1991,Essler:2005}.
Consequently, the asymptotic correlation functions and FFLO signature of the 1D  attractive spin-1/2 Fermi gas have been
analytically studied  by the dressed charge formalism \cite{Lee:2011a}.
This study can be carried out naturally for 1D attractive  multi-component Fermi gases \cite{Schlottmann:2012b}.

In the gapless phase,  the bound
pairs and excess unpaired fermions form two Fermi seas which can be
described by a two-component TLL,
where the spin fluctuations are strongly suppressed at low temperatures.
The conformal dimensions of two-point correlation functions of the 1D attractive spin-1/2 Fermi gas can be
calculated from the elements of the dressed charge matrix $\mathbf{Z}$
describing the finite-size corrections for the low-lying excitations.
The long distance asymptotics of various correlation
functions are then examined through the dressed charge formalism at the $T=0$.
The finite-size corrections to the groundstate energy were computed from the BA equations, with result \cite{Lee:2011a}
\begin{equation}
E_{0}\approx \varepsilon_{0}^{\infty}-\frac{\pi}{6L^{2}}\sum_{\alpha=u,b}v_{\alpha}
\end{equation}
where $v_{u}$ and $v_{b}$ are the velocities of unpaired fermions
and bound pairs, respectively.

Three types of low-lying excitations are considered in
the calculations of finite-size corrections: a Type 1 excitation is characterized by moving a particle close to the
right or left Fermi points outside the Fermi sea;
a Type 2 excitation arises from the change in total number of unpaired
fermions or bound pairs;  a Type 3 excitation is caused by moving a particle
from the left Fermi point to the right Fermi point and vice versa.
Such kinds of excitation are also known as backscattering.
All three types of excitations can be unified in the following form
of the finite-size corrections for the energy and total momentum of
the system \cite{Lee:2011a}
\begin{eqnarray}
\Delta E &=& \frac{2\pi}{L}\left(\frac{1}{4}\phantom{.}^{t}(\Delta
N)^{t}(\mathbf{Z}^{-1})\mathbf{VZ}^{-1}\Delta N\right. \nonumber\\
&&\left. + \phantom{.}^{t}(\Delta D)\mathbf{ZV}\mathbf{Z}^t\Delta
D+\sum_{\alpha=u,b}v_{\alpha}(N_{\alpha}^{+}+N_{\alpha}^{-})\right),
\nonumber \\
 \Delta P &=& \frac{2\pi}{L}\left(\phantom{.}^{t}\Delta N\Delta D+N_{u}\Delta
D_{u}+N_{b}\Delta
D_{b}\right. \nonumber\\
&& \left. +\sum_{\alpha=u,b}(N_{\alpha}^{+}-N_{\alpha}^{-})\right),\label{E-finite}
\end{eqnarray}
with the notation
\begin{eqnarray}
&& \nonumber \Delta N=\left(
           \begin{array}{c}
             \Delta N_{u} \\
             \Delta N_{b} \\
           \end{array}
         \right),\qquad \Delta D=\left(
                                   \begin{array}{c}
                                     \Delta D_{u} \\
                                     \Delta D_{b} \\
                                   \end{array}
                                 \right),\\ &&\mathbf{V}=\left(
                                                             \begin{array}{cc}
                                                               v_{u} & 0 \\
                                                               0 & v_{b} \\
                                                             \end{array}
                                                           \right),\qquad
                                                           \mathbf{Z}=\left(
                                                                        \begin{array}{cc}
                                                                          Z_{uu}(Q_{u}) & Z_{ub}(Q_{b}) \\
                                                                          Z_{bu}(Q_{u}) & Z_{bb}(Q_{b}) \\
                                                                        \end{array}
                                                                      \right).\nonumber
\end{eqnarray}
Here the quantum numbers $\Delta N_{u,b}$ are  characterized by the change in quantum numbers \cite{Lee:2011a}
\begin{equation}
\Delta D_{u}:=\frac{\Delta N_{u}+\Delta
N_{b}}{2}\,\, \mathrm{(mod\phantom{a}1)},\,\, \Delta
D_{b}:=\frac{\Delta N_{u}}{2}\,\,\mathrm{(mod\phantom{a}1)}.\label{QN-D}
\end{equation}
The dressed charges $Z_{uu}(Q_{u})$, $Z_{ub}(Q_{b})$, $Z_{bu}(Q_{u})$
and $Z_{bb}(Q_{b})$ satisfy  a set of dressed charge  equations \cite{Lee:2011a}.

The finite-size spectrum is  described by a critical theory based on the product of
two Virasoro algebras each of which has a central charge $C=1$.
The finite-size scaling form of the energy (\ref{E-finite})  determines  the critical
exponents of two-point correlation functions between primary fields
$\langle O^{\dagger}(x,t)O(x',t')\rangle$.
At zero temperature, the  two-point correlation functions  take the form
\begin{eqnarray}
&&\langle O(x,t)O(0,0)\rangle= \\
&&\frac{\exp(-2\mathrm{i}(N_{u}\Delta
D_{u}+N_{b}\Delta D_{b})\frac{\pi}{L}x)}
{(x-\mathrm{i}v_{u}t)^{2\Delta^{+}_{u}}(x+\mathrm{i}v_{u}t)^{2\Delta^{-}_{u}}
(x-\mathrm{i}v_{b}t)^{2\Delta^{+}_{b}}(x+\mathrm{i}v_{b}t)^{2\Delta^{-}_{b}}}.\nonumber
\label{eq:corrfunc_gen}
\end{eqnarray}
The conformal dimensions are given by
\begin{eqnarray}
2\Delta_{u}^{\pm} &=& \left(Z_{uu}\Delta D_{u}+Z_{bu}\Delta D_{b}\pm
\frac{Z_{bb}\Delta
N_{u}-Z_{ub}\Delta N_{b}}{2\det Z}\right)^{2}\nonumber\\
&&+2N_{u}^{\pm},\nonumber  \\
2\Delta_{b}^{\pm} &=& \left(Z_{ub}\Delta D_{u}+Z_{bb}\Delta D_{b}\pm
\frac{Z_{uu}\Delta N_{b}-Z_{bu}\Delta N_{u}}{2\det
Z}\right)^{2}\nonumber\\
&&+2N_{b}^{\pm}.\nonumber
\end{eqnarray}
Here $N_{\alpha}^{\pm}$ ($\alpha=u,b$) characterize the descendent
fields from the primary fields.

In this way the quantum numbers for the low-lying excitations completely
determine the nature of the asymptotic behaviour of these correlations.
The exponential oscillating term in the asymptotic behaviour comes
from  the backscattering process.
Various  correlation functions, e.g.,  the single particle Green's function
$G_{\uparrow}(x,t)$, charge density correlation function
$G_{nn}(x,t)$, spin correlation function $G^{z}(x,t)$ and pair
correlation function $G_{p}(x,t)$,  can be derived based on the choice of
$(\Delta N_{u},\Delta N_{b})$ which define the quantum numbers (\ref{QN-D}) \cite{Lee:2011a}.
We now discuss these correlation functions in detail.

The asymptotic form of the single particle Green's function $G_{\uparrow}(x,t) =
\langle\psi_{\uparrow}^{\dagger}(x,t)\psi_{\uparrow}(0,0)\rangle$  is  given by
\begin{eqnarray}
\nonumber G_{\uparrow}(x,t) & \approx & \frac{A_{\uparrow,1}\cos\left(\pi(n_{\uparrow}-2n_{\downarrow})x\right)}
{|x+\mathrm{i}v_{u}t|^{\theta_{1}}|x+\mathrm{i}v_{b}t|^{\theta_{2}}}\nonumber\\
&& +\frac{A_{\uparrow,2}\cos\left(\pi
n_{\uparrow}x\right)}{|x+\mathrm{i}v_{u}t|^{\theta_{3}}|x+\mathrm{i}v_{b}t|^{\theta_{4}}},
\end{eqnarray}
where the critical exponents for the strong coupling regime are given in terms of the polarization  by
\begin{eqnarray}
\begin{array}{ll}
\theta_{1} \approx 1+\frac{(1-P)}{|\gamma|}, &  \theta_{2} \approx
\frac{1}{2}-\frac{(1-P)}{2|\gamma|}+\frac{4P}{|\gamma|}, \\
\theta_{3} \approx 1-\frac{(1-P)}{|\gamma|}, &  \theta_{4} \approx
\frac{1}{2}-\frac{(1-P)}{2|\gamma|}-\frac{4P}{|\gamma|}.\end{array} \nonumber
\end{eqnarray}
The constants
$A_{\uparrow,1}$ and $A_{\uparrow,2}$ cannot be derived from the finite-size corrections for low-lying excitations.

The  asymptotic form of the  pair correlation function $G_{p}(x,t) = \langle\psi_{\uparrow}^{\dagger}(x,t)\psi_{\downarrow}^{\dagger}(x,t)\psi_{\uparrow}(0,0)\psi_{\downarrow}(0,0)\rangle $ is given by
\begin{eqnarray}
G_{p}(x,t)  &\approx& \frac{A_{p,1}\cos\left(\pi(n_{\uparrow}-n_{\downarrow})x\right)}{|x+\mathrm{i}v_{u}t|^{\theta_{1}}|x+\mathrm{i}v_{b}t|^{\theta_{2}}}\nonumber\\
&&  +\frac{A_{p,2}\cos\left(\pi(n_{\uparrow}-3n_{\downarrow})x\right)}{|x+\mathrm{i}v_{u}t|^{\theta_{3}}|x+\mathrm{i}v_{b}t|^{\theta_{4}}},
\end{eqnarray}
where the critical exponents in the strong coupling regime are
\begin{eqnarray}
\begin{array}{ll}
\theta_{1} \approx \frac{1}{2}, & \theta_{2} \approx
\frac{1}{2}+\frac{(1-P)}{2|\gamma|},  \\
 \theta_{3} \approx
\frac{1}{2}+\frac{2(1-P)}{|\gamma|},  &\theta_{4} \approx
\frac{5}{2}-\frac{(19P-3)}{2|\gamma|}.
\end{array}\nonumber
\end{eqnarray}
It is clearly apparent that the leading order for the long distance asymptotics of the pair
correlation function $G_{p}(x,t)$ oscillates with wave number
$\Delta k_F$, where $\Delta k_F =\pi(n_{\uparrow}-n_{\downarrow})$.

The Cooper pair correlation has been further discussed with a connection to CFT \cite{Schlottmann:2012b}.
Moreover, the leading order for the charge density correlation function
$G_{nn}(x,t)= \langle n(x,t)n(0,0)\rangle$ and the spin correlation function
$G^{z}(x,t)$  oscillates twice as fast with wave number $2\Delta k_F$, namely
\begin{eqnarray}
&& G_{nn}(x,t) \approx
n^{2}+\frac{A_{nn,1}\cos\left(2\pi(n_{\uparrow}-n_{\downarrow})x\right)}{|x+\mathrm{i}v_{u}t|^{\theta_{1}}} \nonumber \\
&&
+\frac{A_{nn,2}\cos\left(2\pi N_{\downarrow}x\right)}{|x+\mathrm{i}v_{b}t|^{\theta_{2}}} +\frac{A_{nn,3}\cos\left(2\pi(n_{\uparrow}-2n_{\downarrow})x\right)}{|x+\mathrm{i}v_{u}t|^{\theta_{3}}|x+\mathrm{i}v_{b}t|^{\theta_{4}}}
 \nonumber\\
  && G^{z}(x,t) \approx  (m^{z})^{2}
+\frac{A_{z,1}\cos\left(2\pi(n_{\uparrow}-n_{\downarrow})x\right)}{|x+\mathrm{i}v_{u}t|^{\theta_{1}}}
\nonumber \\
&&
+\frac{A_{z,2}\cos\left(2\pi
n_{\downarrow}x\right)}{|x+\mathrm{i}v_{b}t|^{\theta_{2}}} +\frac{A_{z,3}\cos\left(2\pi(n_{\uparrow}-2n_{\downarrow})x\right)}{|x+\mathrm{i}v_{u}t|^{\theta_{3}}|x+\mathrm{i}v_{b}t|^{\theta_{4}}}.
\end{eqnarray}
Here $A_{z,i}$ and $A_{nn}$ are constants and the correlation exponents are given by
\begin{eqnarray}
\begin{array}{ll}
\theta_{1} \approx 2, & \theta_{2} \approx 2-\frac{2(1-P)}{|\gamma|}, \\
\theta_{3} \approx 2+\frac{4(1-P)}{|\gamma|}, & \theta_{4} \approx
2-\frac{2(1-P)}{|\gamma|}+\frac{16P}{|\gamma|}.\end{array} \nonumber
\end{eqnarray}

The oscillations in $G_{p}(x,t)$, $G_{nn}(x,t)$ and $G^{z}(x,t)$ are caused by  the mismatch in
Fermi surfaces between both species of fermions. These spatial
oscillations give a novel signature of the Larkin-Ovchinikov pairing phase \cite{Larkin:1965}.
This finding is consistent with the numerical results from
DMRG  \cite{Feiguin:2007,Tezuka:2008,Rizzi:2008a,Luscher:2008,Tezuka:2010} and QMC \cite{Batrouni:2008,Baur:2010,Wolak:2010}.
It is remarkable to see that the asymptotics of the spatial oscillation terms in the
pair and spin correlations are a consequence of Type 3 excitations,
i.e., backscattering for bound pairs and unpaired fermions.
The asymptotic behaviour of the Fermi field, Cooper pair and charge density wave
correlation functions for the 1D attractive Hubbard model
were computed for the magnetic field $H\to H_{c1}+0^+$ at the half-filled band \cite{Bogoliubov:1990}.

Furthermore, the correlation functions in momentum space can be given by taking
the Fourier transform of their counterparts in position space, i.e., the Fourier transform of
equal-time correlation functions reads
\begin{equation}
g(x,t=0^{+}) =
\frac{\exp(ik_{0}x)}{(x-\mathrm{i}0)^{2\Delta^{+}}(x+\mathrm{i}0)^{2\Delta^{-}}},
\end{equation}
where $\Delta^{\pm}=\Delta_{u}^{\pm}+\Delta_{b}^{\pm}$ is given by
\begin{equation}
\widetilde{g}(k\approx
k_{0})\sim[\mathrm{sign}(k-k_{0})]^{2s}|k-k_{0}|^{\nu}.
\end{equation}
Here the conformal spin of the operator is $s=\Delta^{+}-\Delta^{-}$ and
the exponent $\nu$ is expressed in terms of the conformal dimensions
by $\nu=2(\Delta^{+}+\Delta^{-})-1$.
However, experimental observation of this momentum distribution remains very difficult because
the transverse expansion dominates the expansion along the axial direction \cite{Bolech:2012}.
In this regard, it is practicable to consider the long-time behaviour of the distribution during the
sudden expansion of spin-imbalanced ultracold lattice fermions with an attraction after turning
off the longitudinal confining potential.
However,  the existence of the FFLO signature in the expanding  gas is still in question \cite{Bolech:2012,Lu:2012}.

For fields above the lower critical field $H_{c1}$, these correlators decrease as power-laws.
Thus  the pairs lose their dominance, i.e.,  three types of ordering coexist -- superconductivity,
charge density waves and spin density waves.
Indeed the FFLO correlation is more robust in the  1D  homogenous attractive Fermi gas.
In the experimental setting discussed in Sec.~VII, the quasi-1D system of ultracold fermionic
atoms is formed by loading into a bunch of quasi-1D tubes created by two counter-propagating laser beams.
By tuning the intensity of the lasers one could adiabatically tune the tunnelling between two neighbour tubes.
Such Josephson-like  tunnelling is likely to  lead to a 3D long-range order of the superconducting phase.
A quasi-1D model consisting of attractive Hubbard chains with interchain tunnelling $t_{\perp}$ \cite{Bogoliubov:1989},
where $t_{\perp}$ is much less than the energy gap $\Delta$.
For a singlet pair state, the charge density wave is suppressed, thus the system shows an anomalous average value, i.e.,
$\langle \Psi_{i\uparrow}\Psi_{1\downarrow}\rangle \ne 0$.
This means that a superconductive current exists.
Through analysis of the instability of the normal state (refer to the TLL phase)
to the superconductor transition, the critical temperature is estimated to be $T_c \sim \Delta [q(t_{\perp}/\Delta )^2]^{1/(2-\gamma') } $,
where the $\gamma'$ is the critical pair correlation exponent of the 1D homogeneous  attractive Hubbard chain \cite{Bogoliubov:1989}.
For  $T\ge T_c$, the gapless excitation spectrum of the Cooper pair leads to power-law behaviour of pair  correlation function.

\subsection{1D two-component repulsive fermions }

In Sec.~\ref{S-C-Separation} we saw that for 1D repulsive spin-1/2 fermions, the low energy physics of the model can be
reformulated as  two massless bosonic theories for the charge and spin degrees of freedom with dispersions $\omega(q)=v_{c,s}|q|$.
Based on this spin-charge separation, the bosonization approach  has been used to  compute correlation functions of the
1D  Hubbard  model \cite{Tsvelik:1995,Schulz:1990,Schulz:1991,Ren:1993,Giamarchi:2004}.
Due to the fact  that the spin and charge excitations are independent of each other,   the description of critical phenomena
in the repulsive Hubbard model  has been made by the conformal field theory approach \cite{Woynarovich:1989,Kawakami:1991,Essler:2005}.
Using the critical theory based on the product of two Virasoro algebras with central charge $C=1$,
one can systematically compute asymptotics of correlation functions from the finite-size spectrum of low-lying excitations
in terms of the BA solution \cite{Frahm:1990,Frahm:1991}.
In the same fashion, the long distance asymptotics of various correlation functions of  the spin-1/2 Fermi gas
have been investigated in the strong repulsive regime \cite{Lee:2012a}. The model is gapless and thus critical at
zero temperature. At $T=0$  the  correlation functions decay as some power of distance governed by the critical exponents.
For $T>0$ the decay is exponential.

In the context of spin-charge separation, the general two-point correlation function for primary fields
$\varphi$ with conformal dimensions $\Delta^{\pm}_{c,s}$ at $T=0$
and $T>0$ are given by
\begin{equation}
\langle\varphi(x,t)\varphi(0,0)\rangle=\frac{e^{-2iD_{c}(k_{F\uparrow}+k_{F\downarrow})x}e^{-2iD_{s}k_{F\downarrow}x}}
{\prod_{a=c,s}(x-iv_{a}t)^{2\Delta^{+}_{a}}(x+iv_{a}t)^{2\Delta^{-}_{a}}}\label{correlation-0}
\end{equation}
and
\begin{eqnarray}
&&\langle\varphi(x,t)\varphi(0,0)\rangle_{T} =\label{correlation-T} \\
&&
\frac{ \left(\pi T/v_{a} \right)^{2(\Delta_{a}^{+}+\Delta_{a}^{-})} e^{-2iD_{c}(k_{F\uparrow}+k_{F\downarrow})x}e^{-2iD_{s}k_{F\downarrow}x} }
{\prod_{a=c,s} ^{} \sinh^{2\Delta_{a}^{+}} \left[\frac{\pi
T}{v_a}(x-\mathrm{i} v_{a}t)\right]
 \sinh^{2\Delta_{a}^{-}}\left[\frac{\pi T}{v_a}(x+\mathrm{i} v_{a}t)\right] }\nonumber
\end{eqnarray}
where $k_{F\downarrow,\uparrow}$ are the Fermi momenta, $0<x\leq L$
and $-\infty<t<\infty$ is Euclidean time. The conformal dimensions
of the fields can be written in terms of the elements of the dressed
charge matrix as
\begin{eqnarray}
2\Delta_{c}^{\pm}&=&\left(Z_{cc}D_{c}+Z_{sc}D_{s}\pm
\frac{Z_{ss}\Delta N_{c}-Z_{cs}\Delta N_{s}}{2\det
Z}\right)^{2}\nonumber\\
&& +2N_{c}^{\pm}\nonumber \\
2\Delta_{s}^{\pm}&=&\left(Z_{cs}D_{c}+Z_{ss}D_{s}\pm
\frac{Z_{cc}\Delta N_{s}-Z_{sc}\Delta N_{c}}{2\det
Z}\right)^{2}\nonumber \\
&& +2N_{s}^{\pm}.\nonumber
\end{eqnarray}
Here the dressed charge matrix is denoted by
\begin{eqnarray}
   \mathbf{Z}=\left(
                                                                        \begin{array}{cc}
                                                                          Z_{cc}(Q_{c}) & Z_{cs}(Q_{s}) \\
                                                                          Z_{sc}(Q_{c}) & Z_{ss}(Q_{c}) \\
                                                                        \end{array}
                                                                      \right),\nonumber
\end{eqnarray}
which can be obtained from the dressed charge equations, see \cite{Lee:2012a}.
$Q_{c,s}$ are the Fermi boundaries for charge and spin degrees of freedom.
The non-negative integers $\Delta N_{\alpha}$, $N_{\alpha}^{\pm}$
and the parameter $D_{\alpha}$ where $\alpha=c,s$ characterize  the
three types of low-lying excitations. $\Delta N_{\alpha}$
denotes the change in the number of down-spin fermions.
$N_{\alpha}^{\pm}$ characterizes particle-hole excitations where
$N_{\alpha}^{+}$ ($N_{\alpha}^{-}$) is the number of  particles at the right (left) Fermi level jumps to.
$D_{\alpha}$ represents
fermions that are backscattered from one Fermi point to the other, i.e.,
\begin{equation}
D_{c}\equiv\frac{\Delta N_{s}+\Delta N_{s}}{2}\quad (\mathrm{mod}1),
\,\,  D_{s}\equiv\frac{\Delta N_{c}}{2}\quad (\mathrm{mod}1).
\nonumber
\end{equation}
With the expressions (\ref{correlation-0}) and (\ref{correlation-T})
 the general two-point
correlation functions for the operator $O(x,t)$, namely $\langle
O(x,t)O^{\dagger}(0,0)\rangle$,  can be written as a
linear combination of primary fields with conformal dimensions
$\Delta^{\pm}_{c,s}$ and their descendent fields from the finite-size spectra of the model.

Various  correlation functions of operators can be
written in terms of the field operators $\psi_{\sigma}(x,t)$ and  $\psi_{\sigma}(x,t)$ where
$\sigma=\uparrow,\downarrow$.  E.g.,

(i) One particle Green's function:
\begin{equation}
G_{\sigma}(x,t)=\langle\psi_{\sigma}(x,t)\psi_{\sigma}^{\dagger}(0,0)\rangle.\nonumber
\end{equation}

(ii) Charge density correlation function:
\begin{equation}
G_{nn}(x,t)=\langle n(x,t)n(0,0)\rangle \nonumber
\end{equation}
where $
n(x,t)=n_{\uparrow}(x,t)+n_{\downarrow}(x,t)$ and
$n_{\sigma}(x,t)=\psi_{\sigma}^{\dagger}(x,t)\psi_{\sigma}(x,t)$.

(iii) Longitudinal spin-spin correlation function:
\begin{equation}
G^{z}(x,t)=\langle S^{z}(x,t)S^{z}(0,0)\rangle\nonumber
\end{equation}
where $S^{z}(x,t)=\frac{1}{2}(n_{\uparrow}(x,t)-n_{\downarrow}(x,t))$.

(iv) Transverse spin-spin correlation function:
\begin{equation}
G^{\perp}(x,t)=\langle S^{+}(x,t)S^{-}(0,0)\rangle
\end{equation}
where $ S^{+}(x,t)=\psi_{\uparrow}^{\dagger}(x,t)\psi_{\downarrow}(x,t)$, and
$S^{-}(x,t)=\psi_{\downarrow}^{\dagger}(x,t)\psi_{\uparrow}(x,t)$.

(v) Pair correlation function:
\begin{equation}
G_{p}(x,t)=\langle\psi_{\downarrow}(x,t)\psi_{\uparrow}(x,t)
\psi_{\uparrow}^{\dagger}(0,0)\psi_{\downarrow}^{\dagger}(0,0)\rangle.
\end{equation}

The critical exponents for each of the above correlation functions are determined by the values of
the quantum state with
\begin{eqnarray}
\begin{array}{ll}
 G_{\uparrow}(x,t) :& (\Delta N_{c}=1,\Delta
N_{s}=0,D_{c}\in\mathbb{Z}+\frac{1}{2},D_{s}\in\mathbb{Z}+\frac{1}{2}) \\
 G_{\downarrow}(x,t) : &(\Delta N_{c}=1,\Delta
N_{s}=1,D_{c}\in\mathbb{Z},D_{s}\in\mathbb{Z}+\frac{1}{2}) \\
G_{nn}(x,t) :& (\Delta N_{c}=0,\Delta
N_{s}=0,D_{c}\in\mathbb{Z},D_{s}\in\mathbb{Z}) \\
G^{z}(x,t) : &(\Delta N_{c}=0,\Delta
N_{s}=0,D_{c}\in\mathbb{Z},D_{s}\in\mathbb{Z}) \\
G^{\perp}(x,t) :& (\Delta N_{c}=0,\Delta
N_{s}=1,D_{c}\in\mathbb{Z}+\frac{1}{2},D_{s}\in\mathbb{Z}) \\
G_{p}(x,t) :& (\Delta N_{c}=2,\Delta
N_{s}=1,D_{c}\in\mathbb{Z}+\frac{1}{2},D_{s}\in\mathbb{Z}).
\end{array}\nonumber
\end{eqnarray}
with $N^{\pm}_{c,s}\in\mathbb{Z}_{\geq 0}$ for every case.
The explicit results for  these correlation functions for $H\ll 1$ and $H\to H_c$ are given by solving the dressed charge equations \cite{Lee:2012a}.
In the zero field limit $H\ll 1$ the dressed charge equations were solved by  the Wiener-Hopf  method, explicitly,
\begin{eqnarray}
\begin{array}{lll}
Z_{ss}(Q_s) &\approx&  \frac{1}{\sqrt{2}}\left(1+\frac{4n_{\downarrow }H}{cH_c}+\frac{1}{4\ln (H_0/H)} \right),\\
Z_{sc}(Q_c)&\approx& \frac{1}{2}+\frac{2n_{\downarrow }\ln 2}{c}-\frac{2H}{\pi^2H_c}, \\
Z_{cs}(Q_s)&\approx & \frac{2\sqrt{2}H}{\gamma H_c},\\
Z_{cc}(Q_c) &\approx&  1+\frac{2\ln2 }{\gamma}-\frac{4H^2}{\pi^2 \gamma H_c^2}.\end{array} \nonumber
\end{eqnarray}
Here $H_0=\sqrt{\pi^3/(2e)}H_c$ with the critical field value $H_c \approx 8n^3\pi^2/(3c)$ for strong repulsion.

For the approach to the critical field $h\to H_c$ the dressed charge equations can be solved by
asymptotic expansion \cite{Lee:2012a,Essler:2005}, with result
\begin{eqnarray}
\begin{array}{lll}
Z_{ss}(Q_s) &\approx & 1-\frac{1}{\pi}\sqrt{1-\frac{H}{H_c}}+\frac{8}{\pi \gamma }\sqrt{1-\frac{H}{H_c}},\\
Z_{sc}(Q_c)&\approx & \frac{2}{\pi}\sqrt{1-\frac{H}{H_c}},\\
Z_{cs}(Q_s)&\approx & \frac{4}{\gamma}\left(1-\frac{1}{\pi}\sqrt{1-\frac{H}{H_{c}}}\right),\\
Z_{cc}(Q_c) &\approx  & 1+\frac{8}{\pi\gamma}\sqrt{1-\frac{H}{H_{c}}}.\end{array} \nonumber
\end{eqnarray}
A few terms of the asymptotic expansion for the dressed charges together with the above values for the
low-lying excitations determine the asymptotics of the correlation functions.
As an illustration of this approach,  in the zero field limit, the asymptotics of the density-density correlation function are given by
\begin{eqnarray}
&& G_{nn}(x,t) \approx n^{2}
+\frac{A_{1}\cos(2k_{F\downarrow}x)}{|x+iv_{c}t|^{\theta_{c1}}|x+iv_{s}t|^{\theta_{s1}}} \label{G-nn}\\
&& +\frac{A_{2}\cos(2k_{F\uparrow}x)}{|x+iv_{c}t|^{\theta_{c2}}|x+iv_{s}t|^{\theta_{s2}}}+\frac{A_{3}\cos(2(k_{F\downarrow}+k_{F\uparrow})x)}{|x+iv_{c}t|^{\theta_{c3}}},\nonumber
\end{eqnarray}
where the critical exponents are
\begin{eqnarray}
\begin{array}{lll}
\theta_{c1} &= &\frac{1}{2}+\frac{2\ln
2}{\gamma}-\frac{4}{\pi^{2}}\left(\frac{H}{H_{c}}\right)-\frac{8\ln
2}{\pi^{2}\gamma}\left(\frac{H}{H_{c}}\right)  \\
\nonumber \theta_{c2} &= & \frac{1}{2}+\frac{2\ln
2}{\gamma}+\frac{4}{\pi^{2}}\left(\frac{H}{H_{c}}\right)+\frac{8\ln
2}{\pi^{2}\gamma}\left(\frac{H}{H_{c}}\right)  \\
\theta_{c3} &= &2+\frac{8\ln 2}{\gamma}  \\
\theta_{s1} &=&
1+\frac{1}{2\ln(H_{0}/H)}+\frac{4}{\gamma}\left(\frac{H}{H_{c}}\right)\nonumber
\\
\theta_{s2} &= & 1+\frac{1}{2\ln(H_{0}/H)}-\frac{4}{\gamma}\left(\frac{H}{H_{c}}\right).\end{array}\nonumber
\end{eqnarray}

The presence of the external field does not change the form of the correlator  (\ref{G-nn}).
In the large field limit, i.e., $H\to H_c$, the exponents are given by \cite{Lee:2012a}
\begin{eqnarray}
\begin{array}{lll}
 \theta_{c1} &=&
2-\frac{8}{\pi}\sqrt{1-\frac{H}{H_{c}}}+\frac{32}{\pi\gamma}\sqrt{1-\frac{H}{H_{c}}}
\\
\theta_{c2} &= & 2+\frac{32}{\pi\gamma}\sqrt{1-\frac{H}{H_{c}}} \\
\theta_{s1} &=&
2-\frac{4}{\pi}\sqrt{1-\frac{H}{H_{c}}}+\frac{32}{\pi\gamma}\sqrt{1-\frac{H}{H_{c}}} \\
\theta_{s2} &=&
2-\frac{16}{\gamma}-\frac{4}{\pi}\sqrt{1-\frac{H}{H_{c}}}+\frac{64}{\pi\gamma}\sqrt{1-\frac{H}{H_{c}}}.\end{array}
\end{eqnarray}
The constants $A_i$ with $i=1,2,3$ depend on the interaction.
For $k_{F,\uparrow}=k_{F,\downarrow}=\pi n/2 \equiv k_F$, we see that the density-density correlation contains $2k_F$ and $4k_F$-oscillations.
However, in the strong coupling limit, the system behaves like non-interacting spinless free fermions, thus the $2k_F$ terms should vanish,
i.e., $A_1=A_2 \sim 0$.
The leading  contributions  to  the asymptotics of the density-density correlation function are from the $4k_F$-oscillation term
because this oscillation is a consequence of interactions  \cite{Frahm:1990,Essler:2005}.

Furthermore, the large distance behaviour of the two-point correlation function determines the singularities of spectral functions near $\omega \approx \pm v_{c,s} (k-k_F)$.
The correlation functions in momentum space can be determined by Fourier transforming the asymptotics of the above  correlators.
 Of particular interest are the dynamical response functions  such as the spectral function $A(\omega,k)$ which can be  obtained from the
 imaginary part of the retarded single particle Green's function.
 The interacting spectral function often has a non-zero width.
 There are singularities in the spectral function $A(\omega,k_F+q)$  for $\omega \to v_{c,s} q$ \cite{Essler:2010,Essler:2005}, with
\begin{eqnarray}
A(\omega,k_F+q) \sim \left\{\begin{array}{ll}(\omega-v_cq)^{\frac{\alpha_1-1}{2}} &{\rm for } \omega \to v_cq,\\
(\omega+v_cq)^{\frac{\alpha_1}{2}} &{\rm for } \omega \to -v_cq,\\
(\omega-v_sq)^{\alpha_1-\frac{1}{2}} &{\rm for } \omega \to v_sq,\\
(\omega+v_sq)^{\alpha_1} &{\rm for } \omega \to- v_sq.\end{array} \right.
\end{eqnarray}
Here the exponent $ \alpha_1$,  which can be obtained from Fourier transformation,  is always greater than zero.
Using TLL theory, the Fourier transforms of the zero-temperature single particle Green's function have been computed \cite{Voit:1993,Meden:1992}.
The Fourier transform of the $2k_F$ Luttinger liquid density correlation function was calculated recently in \cite{Iucci:2007}.
In general,  the explicit calculation of the spectral function is much more involved \cite{Frahm:1990,Frahm:2005,Essler:2005}.

\subsection{1D multi-component fermions}

In general, $SU(\kappa)$ Wess-Zumino-Witten (WZW) theory can be used to capture the low energy behaviour
of a family of critical quantum spin chains \cite{Affleck:1987}.
The $SU(\kappa)$ WZW models of level $\ell$ describe integrable higher spin models with critical points
characterised by the central charge $C=\ell (\kappa^2-1)/(\ell +\kappa)$ and the scaling dimensions of the primary fields \cite{Affleck:1987}.
The $U(1)\otimes SU(\kappa)$ symmetry interacting fermions in 1D have $\kappa$  branches of states
characterised by one  charge degree of freedom and $\kappa-1$ spin rapidities,  see Eq.~(\ref{FE-R}).
The low-lying excitations are described by the linear dispersion	relations	$\omega_r(k)=v_r(k-k^r_F)$
with the charge velocity $v_c$ and spin velocities $v_r$ for $r=1,\ldots, \kappa-1$, respectively.
The BA result \cite{Guan:2010}  predicts that the energy has a universal finite-size scaling in the
low energy excitation spectrum, i.e., $E_{0,L}-E_{0,\infty} =-\frac{\pi C }{6L} \sum_{r} v_r +O(1/L^2)$.
The low energy spectrum can be interpreted in terms of a  product of $\kappa$ independent Virasoro
algebras with the same central charge $C=1$.
For vanishing magnetic field, the spin velocities are equal.
Thus we have a spin-charge separation of the $C=1$ Gaussian field theory (in the charge sector  with $U(1)$ symmetry)
and $C=\kappa-1$ WZW theory (in  the spin sector with $SU(\kappa)$ symmetry).

Using the BA solution, Frahm and Schadschneider (1993) and Kawakami (1993)
have calculated finite-size corrections and the spectrum of the low-lying excitations in the
1D multi-component  degenerate Hubbard model.
The asymptotic behaviour of various correlation functions has been determined from these low-lying spectrum.
In the same fashion, the asymptotics of correlation functions for the Bose-Fermi mixtures have been studied \cite{Frahm:2005}.
The general structure of the  critical exponents for the multiple nested BA solvable models  has been well understood, see a review by
 Schlottmann (1997).

The low-lying excitations above the groundstate energy of the critical Fermi gases with $SU(\kappa)$ symmetry
have been obtained \cite{Frahm:1993,Schlottmann:1997,Schlottmann:2012a,Schlottmann:2012b,Kawakami:1993}
\begin{eqnarray}
\Delta E &=& \frac{2\pi}{L}\left(\frac{1}{4}\phantom{.}^{t}(\Delta
{\bf N})^{t}(\mathbf{Z}^{-1})\mathbf{VZ}^{-1}\Delta N\right. \nonumber\\
&&\left. +
\phantom{.}^{t}(\Delta {\bf D})\mathbf{ZV}\mathbf{Z}^t\Delta
{\bf D}+\sum_{r=0}^{\kappa-1}v_{r}(N_{r}^{+}+N_{r}^{-})\right),\label{E-N-finite}\\
 \Delta P &=& \frac{2\pi}{L}\left(\phantom{.}^{t}\Delta {\bf N}\Delta {\bf D}+ \sum_{r=0}^{\kappa-1} N_{r}\Delta
{\bf D}_{r} +\sum_{r=0}^{\kappa-1 } (N_{r}^{+}-N_{r}^{-})\right),\nonumber
\end{eqnarray}
where ${\bf V}={\rm diag}(v_c,v_1,\ldots, v_{\kappa-1})$ and  $N_{r}^{\pm}$  are positive integers characterizing the excited states
at  the branches $r=0,1,\ldots, \kappa-1$.
Here $r=0$ stands for the charge degree of freedom.
$\Delta {\bf N} $  and $\Delta {\bf D} $ are  vectors,  where $\Delta N_r $ in vector $\Delta {\bf N}$
denotes the changes of the total numbers in each branch.
The values of $\Delta D_r $ in vector $\Delta {\bf D}$ are integer or half-odd integer depending on $\Delta N_r$, i.e.,
\begin{eqnarray}
\Delta D_{c}& := &\frac{\Delta N_{c}+\Delta
N_{1}}{2}\,\, \mathrm{(mod\phantom{a}1)},\nonumber \\
 \Delta D_{r}& := & \frac{\Delta N_{r-1}+\Delta N_{r+1}}{2}\,\,\mathrm{(mod\phantom{a}1)}\label{QN-D-N}
\end{eqnarray}
with  $r=1,\ldots,\kappa-1$ and $\Delta N_0=\Delta N_c, \, \Delta N_{\kappa}=0$.

The conformal dimensions $\Delta _{r}^{\pm}$ of the primary field are given in terms of the dressed charge matrix ${\bf D}$ \cite{Frahm:1993}
\begin{eqnarray}
2\Delta_r^{\pm}=\left(({\bf Z}^t \Delta {\bf D})_r\pm \frac{1}{2}({\bf Z}^{-1} \Delta {\bf N})_r \right)^2+2{ N}_r^{\pm}.
\end{eqnarray}
The dressed charges $Z_{ij}\equiv Z_{ij}(Q_j)$ can be obtained by solving the dressed charge equations obtained from the
dressed energy equations by definition \cite{Bogoliubov:1990,Frahm:1993,Izergin:1989}.
Consequently, different dressed charge equations are accordingly derived for the multi-component Fermi gases  in the repulsive and attractive regimes.
The dressed charges for the degenerate $SU(\kappa)$ Hubbard model  was calculated explicitly in the absence of magnetic field \cite{Frahm:1993}.
The single particle Green's function and charge density-density correlation were thus studied.
In general, using the CFT result,  the asymptotics  of correlators for primary fields  at $T=0$ are given by
\begin{equation}
\langle\varphi(x,t)\varphi(0,0)\rangle=\frac{\exp{\left[-2\mathrm{i}\left(\sum_{r=0}^{\kappa-1}\Delta D_rk_F^r\right)x\right]}}
{\prod_{r=0}^{\kappa-1} (x-iv_{r}t)^{2\Delta^{+}_{r}}(x+iv_{r}t)^{2\Delta^{-}_{r}}}.\label{correlation-N}
\end{equation}
Here $k_F^r=\pi n_r $ is the Fermi momentum of each branch.
The critical exponents of the correlation functions of 1D multi-component Fermi gases at zero temperature in external magnetic fields
can be calculated in a straightforward way.
Some response functions in multi-component Luttinger liquids were computed analytically \cite{Orignac:2012}.
The quantum numbers of low-lying excitations, such as $\Delta N_r, \Delta D_r$ and $N_r^{\pm}$, completely determine the
nature of the asymptotic behaviour of the correlation functions.
There are many superfluid orders in multi-component attractive Fermi gases, for which
the operators for the superfluidity  $\varphi(x)$ are written in terms of fermion anniliation operators of length-$r$.
The instability of superfluidity and normal phase can be determined by the correlation function (\ref{correlation-N}).
In the context of ultracold gases with higher spin symmetries, Schlottmann and Zvyagin (2012a,2012b)
have  studied various superfluidity ordering in spin-3/2 and spin-5/2 attractive Fermi gases,
in which  the relevant superfluidity  operators of few-body bound states were labeled.
However, the study of colour superfluidity, the FFLO-like modulated ordering in 1D interacting fermions
with large spins still remains preliminary.
It remains to further investigate magnetism, superfluidity, charge and spin density  waves by using CFT and the BA  solutions .

\subsection{Universal contact in 1D}

The universal nature and phenomena of interacting fermions, such as Landau's Fermi liquid theory, Luttinger liquid theory, and quantum criticality,
have always attracted great attention from theory and experiment.
Tan (2008a, 2008b,2008c) has shown that a few universal relations for two-component interacting fermions involve an extensive quantity
called the universal contact ${\cal C}$.
The first Tan relation is for the tails of  the momentum distribution which exhibits a universal $n_{\sigma }(k) \sim {\cal C}/k^4$
decay as the momentum tends to infinity.
Here we denote $ \sigma =\uparrow,\downarrow$.
The constant ${\cal C}$  measures the probability of two fermions with opposite spin at the same position.
Secondly, this contact also reflects the rate of change of the energy or free energy  due to a small change in the inverse scattering
length $1/a$ for fixed entropy $s$  or  temperature $T$
\begin{equation}
\left(\frac{d E}{da^{-1}} \right)_s=-\frac{\hbar^2 }{4\pi m} {\cal C},\,\, \left(\frac{d F}{da^{-1}} \right)_T=-\frac{\hbar^2 }{4\pi m} {\cal C}.
\end{equation}
Tan's  additional relation states that the pressure and the energy density are related by $P=\frac{2}{3}{\cal E}+\frac{\hbar^2}{12\pi m a }{\cal C}$.

The Tan relations  were found as a consequence of operator identities following from a Wilson operator product expansion of the one-particle density matrix \cite{Braaten:2008}.
These relations hold more generally as long as the interaction range $r_0$ is much smaller than any
other characteristic length scale like the average interparticle distance and the thermal wavelength \cite{Zhang:2009}.
Prior to Tan's results,  the derivative of the energy $E$ with respect to the inverse scattering length $a$ has been
experimentally examined from measurements of the photoassociation rate of a trapped gas of ${}^6$Li atoms \cite{Partridge:2005}.
The static and dynamic structure factor for the trapped gas of  ${}^6$Li atoms  has been studied the  by using Bragg spectroscopy \cite{Hu:2010}.
In particular, Tan's universal contact and thermodynamic relations have been confirmed for a trapped gas  of ${}^{40}$K atoms \cite{Stewart:2010}.
Recent developments  on the Tan relations for fermions with large scattering length have been reviewed by Braaten (2012).

The generalization of Tan relations to 1D interacting fermions has been studied by Barth and Zwerger (2011)
using the Operator Product Expansion method developed by Kadanoff (1969) and Wilson (1969).
The Tan adiabatic theorem has been generalized to the 1D Gaudin-Yang model for which  the 1D analog of the Tan
adiabatic theorem  is given in the form $\frac{d}{d a_{\rm 1D}}E =\frac{\cal C}{4m}$.
Tan's universal contact also exists in the asymptotic behaviour of the tail momentum distribution of the 1D Fermi gas.
In terms of the fields $\psi^{\dagger} _{\sigma} (R)$ and $\psi_{\sigma} (R)$ the momentum distribution given by
$\tilde{n}_{\sigma}(k) = \int dR \int dx \,\, \mathrm{e}^{-\mathrm{i} kx}  \langle \psi^{\dagger} _{\sigma }(R)\psi_{\sigma }(R+x) \rangle$.
Using the Operator Product Expansion method, Barth and Zwerger (2011) found that the momentum distribution
of the 1D two-component Fermi gas behaves as $\tilde{n}_{\sigma}\to {\cal C}/k^4$ as $k\to \infty$.
Furthermore, using  scaling analysis, they also derived the pressure relation which connects pressure and energy density ${\cal E}$
via the universal contact $p=2 {\cal E}+a_{1D} \frac{{\cal C}(a_1)}{4m}$ for the 1D Fermi gas.

Tan's universal contact also arises in the additional relation for the pair distribution function and the related static structure factor.
From the short distance expansion  of the two particle density matrix Barth and Zwerger (2011) found the total pair distribution function at short distance
\begin{equation}
g^{(2)}(x) =\frac{2n_{\uparrow}n_{\downarrow}}{n^2}g_{\uparrow \downarrow}^{(2)}(0)\left(1-\frac{2|x|}{a_{\rm 1D}}+O(x^2) \right).
\end{equation}
This short distance singularity gives rise to a $1/q^2$ power law tail  in the associated static structure factor $S(q)=1+n\int dx e^{-\mathrm{i} qx}(g^{(2)} (x) -1)$, i.e.,
\begin{equation}
S(q\to \infty)=1-4n_{\uparrow}n_{\downarrow}\gamma g_{\uparrow\downarrow}^{(2)}(0)/q^2
\end{equation}
with the dimensionless interaction  constant $\gamma =-2/(na_{\rm 1D})$. In fact,
for the  1D two-component Fermi  gas, the universal contact is obtained by calculating the change of the interaction energy
with respect to interaction strength by the Hellman-Feynman theorem, i.e., ${\cal C}=\frac{4}{a_{1D}^2}n_{\uparrow}n_{\downarrow}g^{(2)}_{\uparrow,\downarrow}(0)$.
The local pair correlation $g^{(2)}_{\uparrow,\downarrow}$ is accessible via the exact BA solution through the relation
$g^{(2)}_{\uparrow,\downarrow}(0) =\frac{1}{2n_{\uparrow}n_{\downarrow}}\partial E/\partial c$.
Here $c=-2/a_{\rm 1D}$ and $E$ is the groundstate energy per unit length.

In a similar way, for the 1D $\kappa$-component Fermi gas, there exists a 1D analog of the Tan adiabatic theorem where the universal contact
is given by the local pair correlations for two fermions with different spin states.
The two-body local pair correlation function is given by the expectation value of the four-operator  term in the second quantized Hamiltonian,
explicitly,
 $g^{(2)}_{\sigma,\sigma'}(0)=\frac{\kappa}{(\kappa-1)n^2}\frac{\partial E}{\partial c}$ with $\kappa>1$ \cite{Guan:2012b}.
 $E$ is again the groundstate energy per unit length.
From the asymptotic expansion result for the groundstate energy of the balanced Fermi gas obtained in Section \ref{Fermi-N}, we easily find
$g^{(2)}_{\sigma,\sigma'}(0)\to 1$ as $|c|\to 0$.
The local pair correlation can be obtained from the groundstate energy  (\ref{E-R1}) of  the balanced gas with strong repulsion,
\begin{eqnarray}
 g^{(2)}_{\sigma,\sigma'}(0)=\frac{4\kappa\pi }{3(\kappa-1)\gamma^2}\left[ Z_1-\frac{6Z_1^2}{\gamma }+\frac{24}{\gamma^2}\left( Z_1^3-\frac{Z_3\pi^2}{15}\right)\right]. \label{g2-rs}
\end{eqnarray}

\begin{figure}[t]
\includegraphics[width=1.050\linewidth]{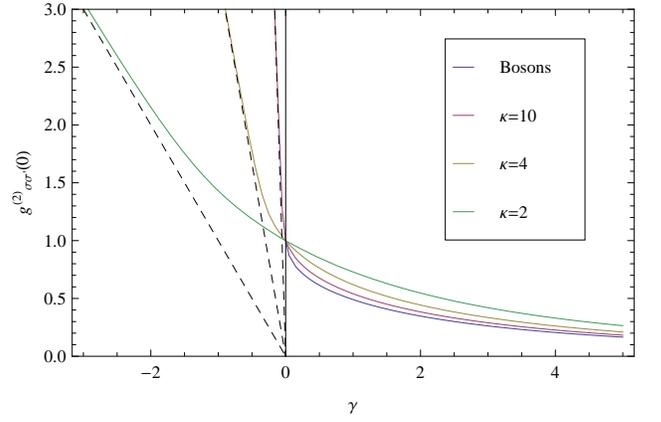}
\caption{The local pair correlation $g^{(2)}_{\sigma,\sigma'}(0)$  vs  $\gamma =-2/(na_{\rm 1D})$  for the balanced  multi-component
Fermi gas for $\kappa =2, 4, 10$ and for the spinless Bose gas.  The solid lines are the numerical solutions  obtained from the
two sets of Fredholm equations. The dashed lines are  the asymptotic limits obtained from the first term in (\ref{g2as}).
From  Guan {\em et al.} (2012). }
\label{fig:g2-N}
\end{figure}

This local pair correlation reduces to that for the spinless Bose gas \cite{Guan:2011b} as $\kappa \to \infty$.
For strong attractive interaction,  the local pair correlation for two fermions with different spin states  is given by \cite{Guan:2012b}
\begin{eqnarray}
g^{(2)}(0)&=&\frac{\kappa(\kappa+1)|\gamma|}{6}+\frac{4\pi^2}{3\kappa^5(\kappa-1)\gamma^2}\left[ A_\kappa+\frac{6A_{\kappa}^2}{\kappa^2|\gamma|}\right.\nonumber\\
&&\left. +\frac{24}{\kappa^4 \gamma^2}\left(A_\kappa^3-\frac{B_\kappa}{15}\right)\right]+O(1/\gamma^4).\label{g2as}
\end{eqnarray}
Here $A_\kappa=\sum_{r=1}^{\kappa-1}1/r$ and $B_\kappa=\sum_{r=1}^{\kappa-1}1/r^3$.
It is noted that  for  the whole interaction regime, the local  pair correlations for  the balanced $\kappa$-component Fermi gas tend
 to the limiting value for the spinless bosons as $\kappa \to \infty$.
 Fig.~\ref{fig:g2-N} shows  the local pair correlation for two fermions with different spin states in the multi-component Fermi gas.

\section{Experimental progress}
\label{chap:Experiments}

In order to realize 1D systems in experiments, one has to apply strong confinement in two transverse directions and allow free motion along the longitudinal direction.
1D systems have been simulated in various settings from solid electronic materials to ultracold atomic gases.
Although some 1D features have been observed in solid electronic materials, it is still challenging to control material parameters and separate imperfect effects caused by defects and Coulomb interactions etc.
In contrast, ultracold atomic gases trapped in external potentials have high controllability and clean environment.
In this section, we briefly review the experimental developments in confining quantum systems of ultracold atoms in 1D.

\subsection{Realization of 1D quantum atomic gases}

Neutral atoms can be trapped by coupling their permanent or induced dipole moments to electromagnetic field gradients.
By using a laser field (or an inhomogenous magnetic field), it is possible to trap neutral atoms via the interaction between the induced electric dipole moment and the laser electric field (or via the interaction between the permanent magnetic dipole moment and the magnetic field).
Correspondingly, there are two typical techniques for realizing 1D quantum atomic gases: optical lattices and atom chips.

\begin{figure}[htb]
\includegraphics[width=0.9\columnwidth]{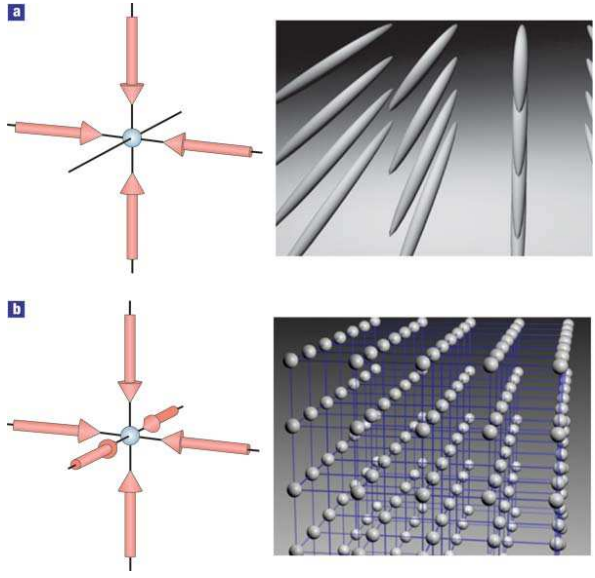}
\caption{Optical lattices of different spatial configurations. (a) For a 2D optical lattice formed by superimposing two orthogonal standing waves, the atoms are confined to an array of 1D potential tubes. (b) For a 3D optical lattice formed by superimposing three orthogonal standing waves, the atoms are approximately confined by a 3D simple cubic array of harmonic oscillator potentials at each lattice site. From Bloch (2005).}
\label{fig:OL}
\end{figure}

\subsubsection{Optical lattices}

Optical lattices are created by superimposing one or more pairs of counter-propagating laser beams \cite{Bloch:2005}, see Fig.~\ref{fig:OL}.
The laser-atom interaction results in an ac-stark shift dependent on the laser intensity and the detuning from resonance $\delta=\omega_{L}-\omega_{a}$, in which $\omega_{L}$ is the laser frequency and $\omega_{a}$ is the atomic transition frequency.
Therefore, due to the reason that optical lattices have periodic laser intensities, their ac-stark shifts provide periodic potentials for confining atoms.
For a negative detuning (red detuning), atoms are attracted to the region of large laser intensities. On the contrary, for a positive detuning (blue detuning), atoms are attracted to the region of small laser intensity.

Similar to the electrons in a conductor, even if the lattice well depth exceeds the kinetic energy of the atoms, the atoms confined in optical lattices can move among neighbouring lattice sites via quantum tunnelling.
In the case of Bose atoms, the quantum phase transition between superfluid and Mott insulator phases has been experimentally observed~\cite{Greiner:2002}.
The groundstate appears as a superfluid when the system is dominated by the quantum tunnelling among neighbouring sites.
Whereas the groundstate appears as a Mott insulator if the system is dominated by the on-site interaction between atoms.
For fermionic atoms, the metal-insulator transition and Neel anti-ferromagnetic states etc have been demonstrated in recent experiments~\cite{Jordens:2008,Schneider:2008}.

Ultracold atomic systems confined in optical lattices have highly controllable parameters such as the inter-site hopping strength, on-site interaction strength and dimensionality.
The inter-site hopping strength can be adjusted by tuning the laser intensity of the optical lattices.
The on-site interaction strength can be tuned from positive infinity to negative infinity  by using Feshbach resonances.
In particular, the dimensionality can be tuned from 3D to 1D by controlling the spatial configuration of the optical lattices, see Fig.~\ref{fig:OL}.

The 1D lattice can be created by a pair of counter-propagating laser beams to form of a standing-wave,
\begin{equation}
V_{1DL} (x)=V_0 \sin^2 \left(k_L x\right).
\end{equation}
The 2D square lattice can be formed by two orthogonal standing waves,
\begin{equation}
V_{2DL}(x,y)=V_0 \left[ \sin^2 \left(k_L x\right)+\sin^2 \left(k_L y\right)\right].
\end{equation}
The 3D simple cubic lattice can be produced by three orthogonal standing waves,
\begin{equation}
V_{3DL}(x,y,z)=V_0 \left[ \sin^2 \left(k_L x\right)+\sin^2 \left(k_L y\right)+ \sin^2 \left(k_L z\right)\right].
\end{equation}
By imposing laser beams under different spatial configurations, it is also possible to make more complex lattices such as the triangle, honeycomb and kagome lattices.

It has been  demonstrated that an array of 1D systems can be created by superposing a harmonic potential onto a 2D optical lattice.
The potential for such a system is of the form
\begin{equation}
V(x,y,z)=V_{2DL}(x,y)+V_{har}(x,y,z).
\end{equation}
Under the tight-binding condition, each lattice well can be regarded as an independently harmonic potential.
That is, the potential $V_{ij}(x,y,z)$ around the lattice site $(x_i, y_j)$ can be expressed approximately as
\begin{equation}
V_{ij}(x,y,z)=\frac{1}{2}m \omega_{xy}^{2} \left[ \left(x-x_i\right)^2 +\left(y-y_j\right)^2 \right] + \frac{1}{2}m \omega_{z}^{2}z^2.
\end{equation}
If the lattice depth is sufficiently large, we have $\omega_{xy} \gg \omega_z$ and so all atoms will always stay in the lowest transverse vibrational state along the $x$ and $y$ directions.
The absence of transverse excitations means that each lattice tube along the $z$-axis can be viewed  as a quasi-1D system.
Experimentally, the longitudinal frequency $\omega_z$ is around $2\pi \times 10 \thicksim 200$~Hz and the transverse frequency $\omega_{xy}$ is around $2\pi \times 10 \thicksim 40$~kHz \cite{Bloch:2005}.

\subsubsection{Atom chips}

\begin{figure}[h]
\includegraphics[width=0.9\columnwidth]{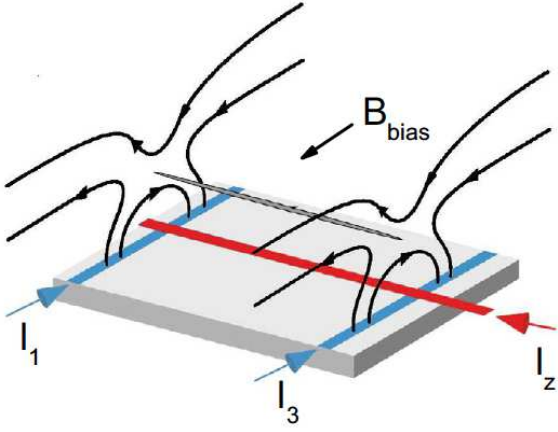}
\caption{1D magnetic potential generated by an atom chip. The transverse confinements are formed by the bias magnetic field $B_{bias}$ and the chip wire $I_z$. The longitudinal confinement is generated by the other two chip wires $I_1$ and $I_3$. From the PhD thesis of A. van Amerongen (University of Amsterdam), which can be downloaded from http://staff.science.uva.nl/~walraven/walraven/Theses.htm.}
\label{fig:AtomChip}
\end{figure}

Atom chips are a kind of nanofabricated atom-optical circuit used to trap and manipulate neutral atoms.
The basic idea of atom chips is to build atomic traps through a bias magnetic field and a system of electric wires fabricated on chip surfaces.
Thus the magnetic potential of a straight wire has a hole along the wire.
To close the magnetic potential with endcaps, there are two simple schemes: one scheme is achieved by combining a straight wire with an inhomogeneous bias field, the other scheme is achieved by combining a bent wire with a homogeneous bias field, see Fig.~\ref{fig:AtomChip}.
The basic property, practical design and experimental procedure of atom chips have been reviewed in the literature~\cite{Reichel:2002,Folman:2002,Fortagh:2007}.
Here, we briefly review how to use atom chips to realize 1D traps with neutral atoms.

One can obtain a highly elongated trap if the longitudinal confinement is very weak compared to the transverse confinements.
As illustrated in Fig.~\ref{fig:AtomChip}, the wire with current $I_{z}$ and the homogenous bias field $B_{\textrm{bias}}$ form a waveguide along the longitudinal direction.
This waveguide is closed by two perpendicular wires with currents $I_{1}$ and $I_{3}$.
Near the trap center, the magnetic field is approximated by
\begin{equation}
B =B_{0}+\frac{1}{2} \beta \left(z-z_0\right)^2 + \frac{1}{2}\left(\frac{\alpha^2}{B_0}-\frac{\beta}{2}\right) \rho^2
\end{equation}
with $B_0$ denoting the value of the magnetic field in the trap center.
The corresponding magnetic potential is given as
\begin{equation}
V(\rho,z)=\mu_{m}B_{0}+\frac{1}{2}m\omega_{z}^2 \left(z-z_0\right)^2+\frac{1}{2}m\omega_{\bot}^2 \rho^2
\end{equation}
with the trapping frequencies
\begin{equation}
\omega_{z} = \sqrt{\frac{\mu_{m}}{m} \beta},
\end{equation}
\begin{equation}
\omega_{\bot} = \sqrt{\frac{\mu_{m}}{m} \left(\frac{\alpha^2}{B_0}-\frac{\beta}{2}\right)},
\end{equation}
where $\mu_{m}$ is the atomic magnetic moment and $m$ is the single atomic mass.
At finite temperature, a real system enters its 1D regime when both the residual thermal energy (of order $k_{B}T$)
and the chemical potential $\mu$ are far less than the transverse excitation energy $\hbar\omega_{\bot}$, i.e., $\{k_{B}T, \mu \} \ll \hbar\omega_{\bot}$.
Experimentally, the atom chips with a Z-shaped wire can successfully generate quasi-1D potential traps of high frequency ratio
$\omega_{\bot}/\omega_{\parallel}$ up to a few hundred \cite{Trebbia:2006,Jo:2007a,Hofferberth:2007,van Amerongen:2008,Bouchoule:2011, Reichel:2004}.
Typically, the transverse frequency $\omega_{\bot}/(2\pi)$ is in an order of kHz and the longitudinal frequency $\omega_{\parallel}/(2\pi)$ is about 10 Hz.
It has also been demonstrated that a 1D-box trap, which has nearly constant potential at the trap minimum combined with tight harmonic confinement
in the transverse directions, can be formed by positioning two wiggles in a long straight wire~\cite{van Es:2010}.

\subsection{Tuning interaction via Feshbach resonance}

The 1D quantum gases of $\delta$-function contact interaction $U_{1D}(z)=g_{1D} \delta(z)$ are characterized by the dimensionless parameter $\gamma$~\cite{Olshanii:1998,Petrov:2000,Dunjko:2001,Bergeman:2003},
which is defined as the ratio between the interaction energy $\epsilon_{int}$ and the kinetic energy $\epsilon_{kin}$, $\gamma =\frac{\epsilon_{int}}{\epsilon_{kin}}=\frac{mg_{1D}}{\hbar^2 n_{1D}}$ .
Here, $g_{1D}$ denotes the coupling constant  and $n_{1D}$ is the 1D number density.
The coupling constant $g_{1D}$ is determined by the 3D scattering length $a$ and the transverse width of the wave function~\cite{Olshanii:1998,Bergeman:2003},
 \begin{equation}
g_{1D}=-\frac{2\hbar^2}{ma_{1D}}=\frac{2 \hbar^2 a}{m l_{\bot}^2}\frac{1}{1-A a/l_{\bot}},
\end{equation}
where the constant $A=1.0326$ and the transverse width $l_{\bot} = \sqrt{\hbar/\left(m \omega_{\bot}\right)}$.
If $\left| a\right| \ll l_{\bot}$, the atom-atom scattering acquires a 3D character, the interaction strength $g_{1D}$ is given as $g_{1D}=\frac{2\hbar^2 a}{m l_{\bot}^2}$.
Theoretically speaking, by varying the interaction strength $g_{1D}$, the gradual transitions from quasi BEC and the TG gas appear in 1D Bose systems,
the BCS-BEC-like crossover takes place in 1D two-component Fermi systems.
Experimentally, to explore different regimes of 1D quantum atomic gases, one may tune the 3D scattering length $a$ via the
well-developed techniques of Feshbach resonance.

Feshbach resonance takes place when the energy of a bound state (a state belonging to a closed channel)
for the inter-particle potential is close to the kinetic energy of the two colliding  particles.
The scattering length  of the two colliding  ultracold atoms undergo a Feshbach resonance if the energy of a
molecular state is close to the kinetic energy of the the colliding pair of atoms.
The interaction strength between a pair of ultracold atoms is proportional to the scattering length.
A  positive (negative)  scattering length  gives repulsive (attractive) interaction.
The scattering length $a$ near a magnetic Feshbach resonance point $B_0$ is given by
\begin{equation}
a(B) = a_{bg} \left(1- \frac{\Delta}{B-B_0}\right)
\end{equation}
with the background scattering length $a_{bg}$, the magnetic field $B$ and the resonance width $\Delta$.
Obviously, the sign and strength of the scattering length can be tuned by adjusting the magnetic field $B$.
Feshbach resonances provide an excellent tool for controlling the ultracold interaction between atoms and
have been widely used for exploring different regimes of quantum atomic gases.
The theory of Feshbach resonances in ultracold atomic gases has been introduced in~\cite{Duine:2004}.
The experimental developments in this field can be found in a recent review paper~\cite{Chin:2010}.

\subsection{Data extraction}

There are two usual probes for exploring the physics of quantum atomic gases.
One probe is the correlation function, which measures correlations between the atomic fields at different positions and times.
The correlation function has been extensively used for exploring many-body coherence and distinguishing order and disorder.
The other probe is the dynamical structure factor, which describes the total probability to populate any excited state by transferring
both momentum and energy from an initially equilibrium state.
The dynamical structure factor is a powerful quantity for characterizing the low-energy excitations.

\subsubsection{Detecting correlation functions via optical imaging}

The first-order correlation function (also called the single-particle correlation function) is
\begin{equation}
G_{i,j}^{(1)} (\mathbf{r}_{1},\mathbf{r}_{2}) = \langle \Psi^{+}_{i}(\mathbf{r}_{1})\Psi_{j}(\mathbf{r}_{2})\rangle.
\end{equation}
Here $\Psi_{i} (\mathbf{r})$ and $\Psi_{j} (\mathbf{r})$ denote the field operators for atoms in states $\left|i\right\rangle$ and $\left|j\right\rangle$.
Obviously, the first-order correlation function becomes the density
$n_i(\mathbf{r})=\langle \rho_i(\mathbf{r}) \rangle =\langle \Psi^{+}_{i}(\mathbf{r})\Psi_{i}(\mathbf{r})\rangle$
if $i=j$ and $\mathbf{r}=\mathbf{r}_{1}=\mathbf{r}_{2}$.
The density profile can be mapped out by the well-developed techniques of optical imaging,
in which information about an atomic gas is encoded onto the absorption, phase shift or polarization
information of the probe laser.
One popular optical imaging technique is absorption imaging, which is implemented by comparing imagines
taken with and without the atomic gas within the field of view and recording the fractional absorption of the probe laser.
However, absorption imaging shows large uncertainty in the measured column density in probing high density atomic gases.
Therefore, absorption imaging is usually taken after the atomic gas is released from the trap and expanded for a certain time.
This imaging method is known as time-of-flight imaging~\cite{Ketterle:1999,Altman:2004}.

In time-of-flight imaging, the fractional absorption is proportional to the expectation value of the atomic density operator
$n_i(\mathbf{r},t)=\langle \rho_i(\mathbf{r}, t) \rangle =\langle \Psi^{+}_{i}(\mathbf{r},t)\Psi_{i}(\mathbf{r},t)\rangle$.
By assuming negligible atomic collision effects in the ballistic expansion during the time of flight,
the density $n_i(\mathbf{r},t)$ is proportional to the initial momentum distribution $n_i(\mathbf{k})$
with the corresponding wave vector $\mathbf{k}=m\mathbf{r}/t$, i.e.,  the time-of-flight imaging gives the initial momentum distribution
\begin{equation}
n_i(\mathbf{k})=\langle \Psi^{+}_{i}(\mathbf{k})\Psi_{i}(\mathbf{k})\rangle \propto n_i(\mathbf{r},t).
\end{equation}
However, if the time-of-flight is not long enough and the atomic collision effects cannot be ignored during the time of flight,
the measured density distribution $n_i(\mathbf{r},t)$ will not be proportional to the initial momentum distribution
$n_i(\mathbf{k})$ with $\mathbf{k}=m\mathbf{r}/t$~\cite{Gerbier:2008,Pedri:2001}.
Recently, different from free expansion via time-of-flight, a controlled expansion via  shortcuts to adiabaticity
has been proposed \cite{Campo:2011, Campo:2012}.

By analysing the correlation of different time-of-flight images, one may reconstruct the density-density correlation function
$\langle n_{i}(\mathbf{r}_{1})n_{j}(\mathbf{r}_{2})\rangle= \langle \Psi^{+}_{i}(\mathbf{r}_{1})\Psi^{+}_{j}(\mathbf{r}_{2}) \Psi_{j}(\mathbf{r}_{2})\Psi_{i}(\mathbf{r}_{1})\rangle$.
This technique is known as noise spectroscopy~\cite{Altman:2004,Folling:2005,Greiner:2005a,Chuu:2005}.
In different runs of experiments, the time-of-flight images have technical noises and quantum fluctuations at the same time.
Each time-of-flight imaging is a measurement of the density operator.
To explore the quantum correlation of the density operator, all technical noises should be reduced below the quantum fluctuations.
Therefore, the measurement times $M$ of imaging must be sufficiently large so that the statistical error $1/\sqrt{M}$ is small enough.
The density-density correlation function is a special case of the second-order correlation function
\begin{equation}
G_{i,j}^{(2)} (\mathbf{r}_{1},\mathbf{r}_{2},\mathbf{r}_{3},\mathbf{r}_{4}) =
\langle \Psi^{+}_{i}(\mathbf{r}_{1})\Psi^{+}_{j}(\mathbf{r}_{2}) \Psi_{j}(\mathbf{r}_{3})\Psi_{i}(\mathbf{r}_{4})\rangle,
\end{equation}
which is also called  the two-particle correlation function.

To reconstruct the full correlation functions, in addition to the measurement of density (diagonal correlations),
one has to measure the non-diagonal correlations.
It has been suggested that the non-diagonal correlations $\langle \Psi^{+}_{i}(\mathbf{k^{\prime}})\Psi_{i}(\mathbf{k})\rangle$
can be measured by atomic interferometry~\cite{Torii:2000, Stenger:1999,Polkovnikov:2006}.
The Fourier sampling of time-of-flight imagines may also be used to reconstruct the full one-particle and two-particle correlation functions~\cite{Duan:2006}.
In this detection scheme, two consecutive Raman pulses at the beginning of the free expansion are used to induce a tunable momentum difference
to the correlation terms so that both diagonal and nondiagonal correlations in the momentum space can be detected.
The first Raman operation is implemented by two travelling-wave beams of different wave vectors $\mathbf{k_1}$ and $\mathbf{k_2}$.
The corresponding effective Raman Rabi frequency has a spatial dependent phase with
$\Omega(\mathbf{r})=\Omega_0 \mathrm{e}^{\mathrm{i}(\delta \mathbf{k} \cdot \mathbf{r} +\varphi_1)}$,
where $\delta \mathbf{k} = \mathbf{k}_2 -\mathbf{k}_1$ and $\varphi_1$ is a constant phase.
The second Raman operation is implemented by two counter-propagating laser beams of the resonant frequency
for the transition between the two involved hyperfine levels.

In contrast to the first Raman operation, the effective Raman Rabi frequency for the second Raman operation
is a spatial independent constant $\Omega_0 e^{i\varphi_2}$.
Therefore, through time-of-flight imaging, the real and imaginary parts of the nondiagonal correlation function
$\left\langle \Psi^{+}_{\alpha} (\mathbf{k}) \Psi_{\alpha} (\mathbf{k} -\delta \mathbf{k}) \right\rangle$
can be measured by choosing the relative phase $\delta \varphi = \varphi_2 -\varphi_1 =0$ and $\pi/2$, respectively.
Combined with the techniques of noise spectroscopy, through analyzing the correlations of imagines corresponding to
$\left\langle \Psi^{+}_{\alpha} (\mathbf{k}) \Psi_{\alpha} (\mathbf{k} -\delta \mathbf{k}) \right\rangle$ and
$\left\langle \Psi^{+}_{\alpha} (\mathbf{k}^{\prime}) \Psi_{\alpha} (\mathbf{k}^{\prime} -\delta \mathbf{k}^{\prime}) \right\rangle$,
respectively, the two-particle correlation function
$\left\langle \Psi^{+}_{\alpha} (\mathbf{k}) \Psi_{\alpha} (\mathbf{k} -\delta \mathbf{k})
\Psi^{+}_{\alpha} (\mathbf{k}^{\prime}) \Psi_{\alpha} (\mathbf{k}^{\prime} -\delta \mathbf{k}^{\prime}) \right\rangle$ can be reconstructed.
If the atomic gases have two spin components $\alpha_{1}$ and $\alpha_{2}$, the spin-spatial correlations
$\left\langle \Psi^{+}_{\alpha_{1}} (\mathbf{k}_1) \Psi_{\alpha_2} (\mathbf{k}_2) \right\rangle$
can be reconstructed by a combination of Fourier sampling with a pair of Raman pulses which mixes the two spin components.
Therefore, in addition to the single-spin density-density correlation functions
$\left\langle \Psi^{+}_{\alpha} (\mathbf{k}) \Psi_{\alpha} (\mathbf{k} -\delta \mathbf{k})
\Psi^{+}_{\alpha} (\mathbf{k}^{\prime}) \Psi_{\alpha} (\mathbf{k}^{\prime} -\delta \mathbf{k}^{\prime}) \right\rangle$,
the opposite-spin density-density correlation functions
$\left\langle \Psi^{+}_{\alpha_1} (\mathbf{k}_1) \Psi_{\alpha_1} (\mathbf{k}_1 -\delta \mathbf{k}_1)
\Psi^{+}_{\alpha_2} (\mathbf{k}_2) \Psi_{\alpha_2} (\mathbf{k}_2 -\delta \mathbf{k}_2) \right\rangle$
can be reconstructed.

\subsubsection{Detecting the dynamical structure factor via Bragg scattering}

Based upon the backward-scattering of incident particles (such as photons, electrons and atoms) from a target sample,
Bragg scattering is a powerful tool for determining both the momentum and the energy absorbed by the target sample.
In experiments with ultracold atomic gases, unlike  photons are diffracted by an atom grating, atoms are diffracted on a
grating of coherent photons (lasers).
Bragg scattering provides an effective tool of spectroscopy, called Bragg spectroscopy,
which can be used to detect the dynamical structure factor and explore the intrinsic mechanism of the low-energy excitations.

The dynamical structure factor $S(\mathbf{q},\omega)$ describes the total probability to populate an excited state
by transferring a momentum $\hbar \mathbf{q}$ and energy $\hbar \omega$.
At nonzero temperatures, the dynamical structure factor is~\cite{Pitaevskii:2003}
\begin{eqnarray}
S(\mathbf{q},\omega)=&\frac{1}{\mathbb{Z}}&\sum_{i,f} \mathrm{e}^{-\beta E_{i}}
\left|\langle \phi_{f}|\psi^{+}(\mathbf{q}+\mathbf{k})\psi(\mathbf{k}) |\phi_{i}\rangle\right|^2\nonumber\\
&&\times \delta\left(E_f(\mathbf{q}+\mathbf{k}) -E_i(\mathbf{k})-\hbar \omega \right),
\end{eqnarray}
where $\left|\phi_{i}\right\rangle$ and $\left|\phi_{f}\right\rangle$ are initial and final states of the many-body system,
with energies $E_i$ and $E_f$.
The operator $\psi^{+} (\mathbf{k})$ creates a particle of momentum $\mathbf{k}$.
The function $\mathbb{Z}=\sum_{i} \mathrm{e}^{-\beta E_{i}}$ with $\beta=1/(k_B T)$ is the partition function.
At zero temperature, the equilibrium system can only occupy its groundstate and the dynamical structure factor is expressed as
\begin{eqnarray}
S(\mathbf{q},\omega)=&\frac{1}{\mathbb{Z}}&\sum_{f} \left|\langle \phi_{f}|\psi^{+}(\mathbf{q}+\mathbf{k})\psi(\mathbf{k}) |\textrm{GS}\rangle\right|^2\nonumber\\
&&\times \delta\left(E_f(\mathbf{q}+\mathbf{k}) -E_{\textrm{GS}}(\mathbf{k})-\hbar\omega\right).
\end{eqnarray}
In a two-photon Bragg transition, the momentum shift $\mathbf{q}=\mathbf{k}_2-\mathbf{k}_1$ is controlled by wave vectors of the
two Bragg lasers and the frequency $\omega=\omega_2-\omega_1$ is determined by the frequency difference.

The techniques of Bragg spectroscopy have been successfully applied to measure the dynamical structure factor
of Bose condensed atoms~\cite{Stenger:1999,Ozeri:2005} and strongly interacting 3D Bose~\cite{Papp:2008} and Fermi~\cite{Veeravalli:2008}
atomic gases near Feshbach resonance.
The techniques of Bragg spectroscopy have been employed to detect the elementary excitations in an array of 1D Bose gases~\cite{Clement:2009}.
By varying the detuning of the Bragg lasers, it is possible to probe both the spin and density dynamical structure factors for multi-component Fermi systems~\cite{Hoinka:2012}.

\subsection{Experiments with 1D quantum atomic gases}

Over the last decade, there have been many breakthrough experiments with 1D quantum atomic gases.
These experiments are implemented by using an ensemble of ultracold Bose or Fermi alkaline atoms occupying one or multiple hyperfine levels.
The key experiments in 1D quantum atomic gases are listed in Table~\ref{tab:exp_groups}.
\begin{table}[htb]
\caption{\label{tab:exp_groups}Key experiments in 1D quantum atomic gases.}
\begin{ruledtabular}
\begin{tabular}{lllll}

Group (Leader) & Research topics\\
\hline

Amsterdam & Yang-Yang thermodynamics (2008)\\
(van Druten) &
non-equilibrium spin dynamics (2010)\\
\hline

Cambridge (K\"{o}hl) & quantum transport (2009)\\
\hline

CNRS & density fluctuations (2006, 2011)\\
(Bouchoule) & phonon fluctuations (2012)\\
& 1D-3D crossover (2011)\\
& three-body correlations (2010)\\
& mean-field breakdown (2006)\\
\hline

ENS (Salomon) & matter-wave solitons (2002)\\
\hline

ETH & confinement induced molecules (2005)\\
(Esslinger, K\"{o}hl) & p-wave Feshbach resonance (2005)\\
& 1D-3D crossover (2004)\\
& Bragg spectroscopy (2004)\\
& collective oscillations (2003)\\
\hline

Hamburg (Sengstock) & matter-wave solitons (2008)\\
\hline

Innsbruck (N\"{a}gerl) & super-Tonks-Girardeau gases (2009)\\
& sine-Gordon phase transition (2010)\\
& confinement induced resonance (2010)\\
& three-body correlations (2011)\\
\hline

Kaiserslautern (Ott) & spatiotemporal fermionization (2012)\\
\hline

LENS (Inguscio) & Bragg spectroscopy (2009, 2011, 2012)\\
& low-energy excitations (2009)\\
& impurity dynamics (2012)\\
\hline

Mainz/MPQ (Bloch) & impurity dynamics (2013)\\
& relaxation dynamics (2012, 2013)\\
& squeezed Luttinger liquids (2008)\\
& Tonks-Girardeau gases (2004)\\
\hline

MIT (Ketterle) & atomic interferometry (2007)\\
& fluctuations and squeezing (2007)\\
\hline

NIST & dipole oscillations (2005)\\
(Phillips, Porto) & three-body recombination (2004)\\
\hline

Pennsylvania (Weiss) & Tonks-Girardeau gases (2004)\\
& local pair correlations (2005)\\
& quantum Newton's cradle (2006)\\
\hline

Rice Uni. (Hulet) & spin-imbalanced Fermi gases (2010)\\
& matter-wave solitons (2002)\\
\hline

Vienna & quantum correlations (2011, 2012)\\
(Schmiedmayer) & twin-atom beams (2011)\\
& atomic interferometry (2005, 2010)\\
& quantum and thermal noises (2008)\\
& non-equilibrium dynamics (2007)\\

\end{tabular}
\end{ruledtabular}
\end{table}

\subsubsection{Bose gases}

The 1D Lieb-Liniger Bose gas is a prototypical many-body system featuring rich many-body physics.
This gas features three different regimes in quasi-1D traps: a true condensate regime, a quasi-condensate regime and the Tonks-Girardeau regime~\cite{Petrov:2000}.
In the limit of weak interaction, $\gamma=mg_{1D}/(\hbar^2 n_{1D}) \ll 1$, i.e., in the quasi-condensate regime, 
where Bose-Einstein condensation may take place and mean-field theory well describes the  low-energy physics.
In this regime, due to the intrinsic nonlinearity from s-wave scattering between atoms, matter-wave solitons have been observed 
in several experiments~\cite{Khaykovich:2002,Strecker:2002,Stellmer:2008, Becker:2008}.
In the limit of strongly repulsive interaction, $\gamma=mg_{1D}/(\hbar^2 n_{1D}) \gg 1$, i.e., in the Tonks-Girardeau regime, fermionization of Bose atoms occurs.
By loading ultracold Bose atoms into a 2D optical lattices, the effective mass of quasi-particles can be increased by applying an additional periodic potential along the longitudinal direction~\cite{Paredes:2004} and therefore an array of Tonks-Girardeau gases of quasi-particles can be prepared by increasing the effective mass.
The dimensionless parameter $\gamma$ can also be changed without modulating the longitudinal trapping enabling the realization of a set of parallel
1D Bose atomic gases in the Tonks-Girardeau regime~\cite{Kinoshita:2004}.
In the quasi-condensate regime of intermediate values of $\gamma$, the experimental measurements suggest that both mean-field theory and fermionization fail to describe this crossover~\cite{Trebbia:2006}.
The {\em in situ} measurements of the  linear density of a nearly 1D trapped Bose gas on an atom chip indicates a good agreement with theoretical prediction from  the Yang-Yang thermodynamics equations~\cite{van Amerongen:2008}.

Breathing modes and dipole modes are two usual collective excitations in trapped quantum atomic gases.
The breathing modes can be excited by time-periodically modulating the longitudinal harmonic potential.
The dipole modes can be excited by suddenly displacing the longitudinal harmonic potential.
The ratio of the frequencies of the lowest breathing modes and the dipole modes $\omega_B/\omega_D$ has been measured in an array of 1D gases in 2D optical lattices~\cite{Moritz:2003}.
The experimental data show $\omega_B/\omega_D \simeq 3.1$ for a Lieb-Liniger gas and $\omega_B/\omega_D \simeq 4$ for a thermal gas, which are consistent with the theoretical results~\cite{Pedri:2003}.
The experimental observation of strongly damped dipole oscillations due to quantum fluctuations~\cite{Polkovnikov:2004,Gea-Banacloche:2006} has been reported~\cite{Fertig:2005}.

The 1D-3D crossover and quantum phase transitions from a 1D superfluid to a Mott insulator have been observed in quasi-1D systems of an additional lattice potential
along the longitudinal direction~\cite{Stoferle:2004,Haller:2010,Fabbri:2012}.
For weak interaction, the system is still a superfluid at finite lattice depth and the transition to the Mott insulator is induced by increasing the lattice depth.
For strong interaction, an arbitrary perturbation by a lattice potential may induce a sine-Gordon quantum phase transition from a superfluid Luttinger liquid to a Mott insulator.
These quantum phases can be well distinguished by detecting their low-energy excitations via the technique of Bragg spectroscopy~\cite{Haller:2010}.

The techniques of atomic interferometry have been widely used to explore the phase coherence and fluctuations between different 1D gases~\cite{Jo:2007a,Jo:2007b,Jo:2007c,Kruger:2010,Schumm:2005,Hofferberth:2008}.
The second-order correlation functions, which relate to the density fluctuations and spatial correlations, have been probed by {\em in situ}
measurements of density fluctuations~\cite{Esteve:2006,Jacqmin:2011,Perrin:2012} and {\em ex situ} measurements of photo-association rates~\cite{Kinoshita:2005}.
These experiments confirm that the local second-order correlation functions are about $2$, $1$ and $0$ for an ideal 1D Bose gas, a quasi-condensed 1D Bose gas and a Tonks-Girardeau gas, respectively.
In addition to the density correlations, phase correlations have been probed by matter wave interferometry~\cite{Betz:2011}.
Moreover, the three-body correlation functions of 1D Bose gases have been obtained by measuring the three-body recombination rate~\cite{Haller:2011,Laburthe Tolra:2004}
and the {\em in situ} measurements of third-order number fluctuations~\cite{Armijo:2010}.
In addition to the measurements of local correlation functions~\cite{Kinoshita:2005}, the temporal two-body correlation function has been
measured by the techniques of scanning electron microscopy~\cite{Guarrera:2012}.

Beyond investigating equilibrium behaviour, there have been  several experimental studies on non-equilibrium behaviour in 1D Bose gases.
The coherence dynamics in both isolated and coupled 1D Bose gases have been explored by using an atom chip~\cite{Hofferberth:2007}.
The absence of thermalization in a 1D Bose gas has been confirmed by the time evolution from an out-of-equilibrium state~\cite{Kinoshita:2006}.
The non-equilibrium dynamics of an impurity in 1D Bose gases, such as large density fluctuations, multiple scattering events and interaction dependent
quadruple oscillations, have been studied  in some recent experiments~\cite{Palzer:2009,Fabbri:2012}.
Most recently,  the experimental observations of quantum dynamics of interacting bosons  \cite{Ronzheimer:2013,Kuhnert:2013} and  fast relaxation towards equilibrium of quasi-local densities, currents and coherence in an isolated
strongly correlated 1D Bose gas~\cite{Trotzky:2012} give a precise understanding of quantum dynamics and correlations,
including  observation of  the spin dynamics in 1D two-component Bose gases~\cite{Widera:2008}.

As remarked, quasi-1D trapped systems are created by tight transverse confinement.
A confinement-induced-resonance (CIR) takes place when the incident channel of two incoming atoms and a transversally
excited molecular bound state become degenerate.
It has been demonstrated that a CIR takes place if the 3D scattering length approaches the length scale of the transverse trap.
The CIR has been used to drive a crossover from the Tonks-Girardeau gas with a strongly repulsive interaction to a super
Tonks-Girardeau gas with strongly  attractive interaction~\cite{Haller:2009,Haller:2010b}.
In particular, the stable highly excited gas-like phase known as the super Tonks-Girardeau  gas has been realized  in the strongly attractive regime of bosonic Cesium atoms.

\subsubsection{Fermi gases}

\begin{figure}[t]
\includegraphics[width=0.9\columnwidth]{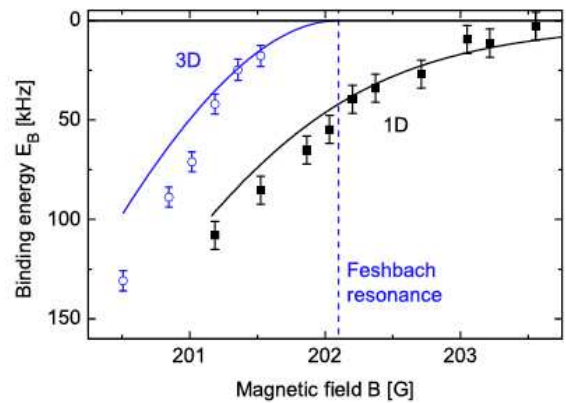}
\caption{Two-body bound states in 1D and 3D. In the 1D case, confinement induced molecules exist for arbitrary sign
of the scattering length. Whereas in the 3D case, there are no bound states at magnetic fields above the Feshbach
resonance (vertical dashed line). From Moritz {\em et al.}  (2005).}
\label{fig:binding-energy}
\end{figure}

In a 3D  trapped Fermi gas,  the  bound diatomic molecules  exist only when the scattering length between the atoms
is  positive under s-wave interaction, i.e., when $1/(k_Fa_s)\gg 0$~\cite{Regal:2003}.
In the limit $1/(k_Fa_s)\ll 0$, the system is a weakly attractive Fermi gas.
Thus the groundstate is a BCS pair state  for  a negative scattering length~\cite{Chin:2004,Greiner:2005b}.
However, in a 1D two-component trapped Fermi gas, the scattering properties of two colliding atoms are  altered by the tight transverse confinement.
The existence of a bound state does not rely on the sign of the scattering length.
Using ratio-frequency spectroscopy in an array of 1D Fermi atomic gases trapped within a 2D optical lattice,
Moritz and coworkers~\cite{Moritz:2005} reported this that
the two-body bound state exists irrespective of the sign of scattering length, see Fig.~\ref{fig:binding-energy}.
In this experiment,  they  further demonstrated that the bound states for negative scattering length can only be stabilized by the tight transverse confinement~\cite{Moritz:2005}.
Very recently, the particle and hole dynamics of fermionic atoms in amplitude-modulated 1D lattices was reported \cite{Heinze:2013}.

\begin{figure}[h]
\includegraphics[width=0.9\columnwidth]{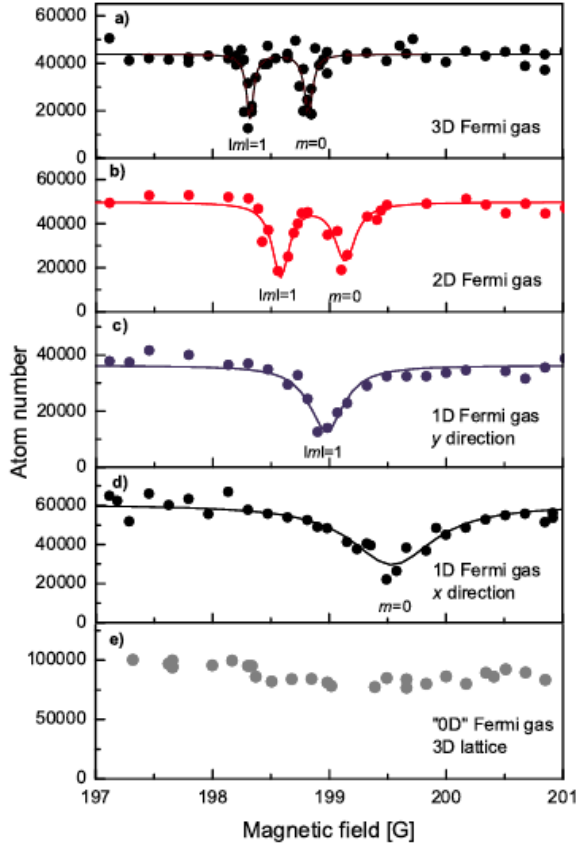}
\caption{Dimension dependence of the p-wave Feshbach resonance in spin-polarised Fermi atomic gases.
In both (a) 3D and (b) 2D Fermi gases, the coexistence of two collisional channels ($m=0$ and $|m|=1$)
give rise to a doublet feature. (c) In a 1D Fermi gas with the spin aligned orthogonal to the atomic motion,
the existence of only the collisional channel of $|m|=1$ gives rise to a single peak feature. (d) In a 1D Fermi gas
with the spin aligned orthogonal to the atomic motion, the existence of only the collisional channel of $m=0$
gives rise to another single peak feature. (e) In a 0D Fermi gas within a deep 3D optical lattice, there is no
resonant peak due to the absence of all collisional channels. From G\"unter {\em et al.} (2005).}
\label{fig:resonance-peak}
\end{figure}

The strong transverse confinement of a waveguide not only gives rise to an effective 1D s-wave interaction,
but also alters the p-wave interaction in spin-polarized fermions.
Due to the absence of s-wave interaction in the spin-polarized Fermi gas,
the p-wave interaction becomes dominant under resonant scattering conditions~\cite{Granger:2004,Imambekov:2010}.
In a spin-polarized Fermi gas, the angular part of asymptotic collision wave functions is either the spherical harmonic $Y_{l=0, m=0}$
if the scattering state is parallel to the quantization axis or $Y_{l=0, m=\pm 1}$ if the scattering state is perpendicular to the quantization axis.
Therefore, in both the 2D and 3D cases, due to the coexistence of two collision channels and the breakdown of degeneracy
between two collision channels~\cite{Ticknor:2004}, doublet structures of p-wave Feshbach resonance have been observed~\cite{Gunter:2005}.
However, for the 1D spin-polarized system, only one of two collision channels is involved.
Thus there is a single peak structure of p-wave Feshbach resonance in 1D spin-polarized  fermions.
The experimental observation has  confirmed such a particular signature~\cite{Gunter:2005}, see Fig.~\ref{fig:resonance-peak}.
By loading the spin-polarized Fermi atoms into a deep 3D optical lattice, one can prepare a band insulator of localized atoms in potential wells,
which is viewed as a zero-dimensional (0D) system.
There is no resonance feature in such a 0D system because the  p-wave scattering is completely inhibited.
When the geometry of the gas is tuned from 3D to 2D, the shift of the p-wave Feshbach resonance depends on the depth of the optical lattice.
In contrast, 1D Fermi gases show a further shift of resonance and give rise to  a broadening of the loss feature.

\begin{figure}[t]
\includegraphics[width=1.0\columnwidth]{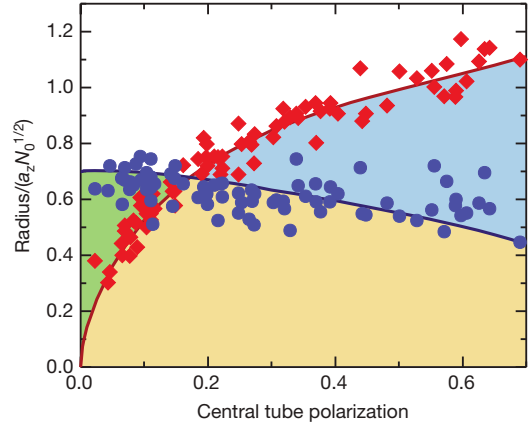}
\caption{Experimental phase diagram as a function of the central tube polarization. The red diamonds and blue circles denote
the scaled radii of the axial density difference and the minority state axial density, respectively. The solid lines are given by the
TBA equation (\ref{TBA-F}) at temperature $T=175\pm50$nK.  Experimental observation  is in reasonable
agreement with the theoretical prediction~\cite{Orso:2007,Hu:2007} for the zero temperature phase diagram of the trapped gas. 
This phase diagram experimentally verifies the coexistence of pairing and polarization at quantum criticality 
\cite{Guan:2007a,Feiguin:2007,Parish:2007,Zhao:2009,Yin:2011b}.
From Liao {\em et al.} (2010).}
\label{fig:phase-mu-h-E}
\end{figure}

The 1D two-component Fermi gas defined by the Gaudin-Yang model is an ideal system for exploring novel pairing mechanisms.
In particular, the formation of a FFLO-like state with a nonzero centre-of-mass momentum gives rise to a precise understanding
of the coexistence of BCS pairs and polarizations.
In 3D, the FFLO state occupies a tiny portion of the phase diagram and it is very difficult to observe in experiments.
In contrast, in the 1D Gaudin-Yang model~\cite{Guan:2007a,Hu:2007,Orso:2007} the FFLO state becomes much more robust
due to the band fillings of two Fermi seas related with one another, and it occupies major parts of the phase diagram,
as demonstrated in Fig.~\ref{fig:phase-s} and Fig.~\ref{fig:phase-attraction}.
See also the experimentally measured phase diagram Fig.~\ref{fig:phase-mu-h-E}.
The system has spin population imbalance caused by a difference in the number of spin-up and spin-down atoms.
The key features of these $T=0$ phase diagrams have been experimentally confirmed using finite temperature density profiles of
trapped fermionic ${}^6$Li atoms~\cite{Liao:2010}.
Experimental observation reveals  that the system has a partially polarized core surrounding by either fully paired or fully polarized
wings at low temperatures  that  is  in agreement with theoretical prediction of the zero temperature phase diagram within the 
LDA \cite{Hu:2007,Orso:2007}.  
The quantum phases of pairs, excess fermions as well as the mixture of pairs and excess fermions in the  phase diagram 
Fig.~\ref{fig:phase-mu-h-E} are also consistent with the analysis of several other groups 
\cite{Guan:2007a,Feiguin:2007,Parish:2007,Kakashvili:2009}. 
The finite temperature phase boundaries do not indicate a solid phase transition, detailed analysis can be seen in Yin {\em et al.} (2011b). 
In a 3D trapped Fermi gas, the fully paired core is surrounded by a shell of excess fermions~\cite{Partridge:2006,Zwierlein:2006}.

\begin{figure}[t]
\includegraphics[width=1.0\columnwidth]{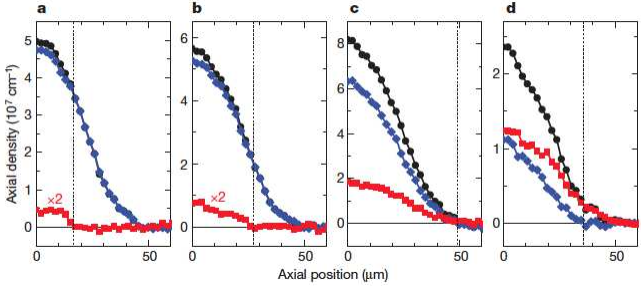}
\caption{Integrated axial density profiles of an array of 1D spin-imbalanced two-component Fermi gases for different central polarization $P$.
The black circles represent the majority, the blue diamonds represent the minority, and the red squares show the difference.
(a) At low $P (=0.015)$, the partially polarized core is enclosed by the fully paired edge. (b) For increasing $P (=0.055)$,
the partially polarized core grows and the fully paired edge shrinks. (c) Near $P_c (P=0.10)$, where almost the entire cloud
is partially polarized. (d) For the polarization exceeding the critical value $P_c (P=0.33)$,
the edge of the cloud becomes fully polarized. From Liao {\em et al.} (2010).}
\label{fig:1DFermiDensityProfile}
\end{figure}

The key method to map out the phase diagram  in Fig.~\ref{fig:phase-mu-h-E} in the experiment~\cite{Liao:2010} is to measure
the {\em in situ} densities of the two spin species.
The 1D spatial density profiles $n_{1,2}(z)$ can be expressed in terms of chemical potential $\mu=\mu_0-V(z)$ and
effective external field $h=h_0$.
Here $\mu_0$ is the chemical potential at the trapping centre and $h_0$ is the effective magnetic field of the homogeneous system.
They can be obtained from the relations (\ref{mu-h}) or the equation of state (\ref{EOS}).
Within the LDA (\ref{N_P}), theoretical density profiles can be used to fit the experimental data obtained by inverse Abel transformation
of the radial profiles, as per the Method section in Liao {\em et al.} (2010).
By changing the polarization, the threshold values of different phases in the density profiles can be read off the phase boundaries of the
phase diagram of Fig.~\ref{fig:phase-mu-h-E}.
From the density profiles given in Fig.~\ref{fig:1DFermiDensityProfile}, we see that at low polarization below the critical value,
the system has a partially polarized core surrounded by fully paired edges.
At the critical polarization, almost the whole region is partially polarized.
At high polarization above the critical value, the system has a partially polarized core surrounded by fully polarized edges.

\begin{figure}[t]
\includegraphics[width=1.0\columnwidth]{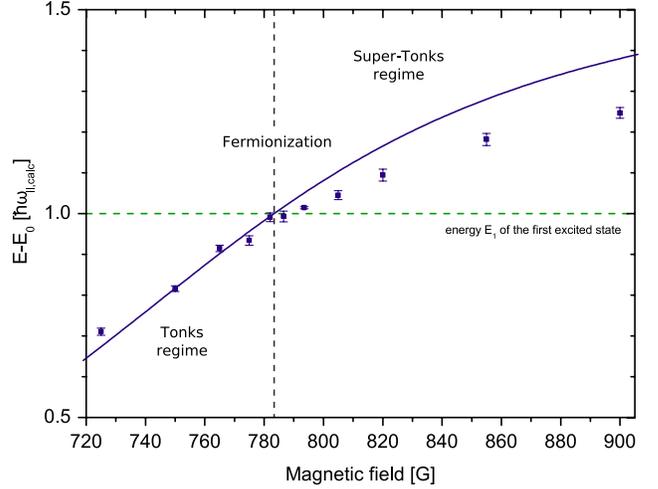}
\caption{Interaction energy of two distinguishable fermions trapped in 1D across a confined induced resonance.
The blue line is the theoretical result of the energy shift. The blue points show experimental data for the interacting energy
of the two distinguishable fermions. The black vertical line is the value of the magnetic field at the  confined induced resonance.
From Zurn {\em et al.} (2012).}
\label{fig:1DFermionization}
\end{figure}

Systems with a small number of trapped atoms are also experimentally feasible \cite{Serwane:2011}.
Most recently, the 1D fermionization of two distinguishable fermions has been experimentally studied by using two fermionic ${}^6$Li atoms~\cite{Zurn:2012}.
The energy of the two-particle system in the state $|\! \uparrow\downarrow \rangle$ is determined by tuning  the trapping potential barrier
through which the particles can  tunnel out.
The fermionization of two distinguishable fermions was identified by measuring the tunnelling time constants  for different values of the 1D interacting strength.
It is particularly interesting that for a magnetic field below the confined induced resonance two interacting fermions form a Tonks-Girardeau state
whereas a super Tonks-Girardeau gas is created when the magnetic field is above the resonance value, see Fig.~\ref{fig:1DFermionization}.
In this case, the two-particle super Tonks-Girardeau state is stable against three-body collisional losses since there is no third particle present.

\section{Conclusion and outlook}

In previous sections we have seen how results for exactly solved models of 1D Fermi gases provide
valuable insights into a wide range of many-body phenomena including Fermi polarons,
Fulde-Ferrel-Larkin-Ovchinnikov-like pairing, the few-body physics of trions, Tomonaga-Luttinger liquids,
spin-charge separation, universal contact,  quantum criticality and universal scaling.
This included discussion of the exotic many-body physics of 1D Fermi-Bose mixtures and
1D multi-component interacting fermions, in particular with $SU(3)$, $SO(5)$ and $SU(N)$ symmetries.
We reviewed experimental progress on the realisation of 1D quantum atomic gases and the experimental
confirmation of the phase diagram of the Gaudin-Yang model in a 1D  harmonic trap.
In the present section, we discuss some of the promising developments and provide an outlook for future research.
The key points we have identified are as follows.

(a) {\em High spin symmetries.} Very recent experimental exploration of highly symmetric Mott insulators \cite{Taie:2012}
gives insight into the Pomeranchuk (1950) type of cooling due to the fact that the spin degrees in the Mott insulating state
can hold more entropy than the Fermi liquid does.
In this Mott insulating state, the entropy per site  increases as the spin degrees of freedom increase.
It in turn leads to a temperature reduction through transfer of entropy from particle motion to spin degrees of freedom.
This exotic nature is different from large spins in solids where the quantum fluctuations of large spins are weak.
Such higher symmetry opens up further study of magnetically ordered phases in ultracold fermionic atoms.
In particular, loading large spin ultracold fermionic atoms into a 1D quantum wire geometry provides exciting opportunities to
test multi-component TLL theory and  competing superfluid ordering in 1D  interacting fermions with higher symmetries.
E.g., $SU(4)$, $SO(5)$ and $SO(4)$ symmetries in spin-3/2 fermions.
It is natural to expect that 1D interacting fermions with large spin resulting from higher mathematical symmetries
will greatly expand our understanding of many-body physics.
However, comprehensive understanding of these models still poses theoretical and experimental challenges due
to the more complicated grand canonical ensembles involved.

(b) {\em p-wave BCS  pairing and synthetic gauge fields.}  The  traditional  s-wave  BCS  pairing models
have various applications to problems in condensed matter physics \cite{Dukelsky:2004,Links:2003,Ying:2008a,Ying:2008b},
nuclear physics \cite{Pan:2002} and ultrasmall metallic grains \cite{vonDelft:2001} {\em etc}.
More work along this line has been achieved in the study of a $p+\mathrm{i}p$ pairing model \cite{Ibanez:2009,Dunning:2010,Rombouts:2010}.
In contrast to the s-wave pairing model, the $p+\mathrm{i}p$ pairing model has an exotic zero temperature phase diagram
in terms of density  and  attractive coupling strength \cite{Rombouts:2010}.
It shows two superfluid phases -- confined strong-pairing and deconfined weak-pairing, separated by a third-order
confinement-deconfinement quantum phase transition.
Moreover, a type of electron pairing model with spin-orbit interactions shows that the  pairing order parameter  can always have
$p+\mathrm{i}p$-wave  symmetry regardless of the strength of pairing interactions \cite{Liu-Jia:2011}.
More work is required along this line.

In particular, recent  developments in the study of ultracold atoms have opened up new opportunities to simulate
synthetic external abelian and non-abelian gauge fields  coupled to neutral atoms by controlling
atom-light interactions \cite{Lin:2009} and the ultracold-atom analog of the mesoscopic conductor \cite{Brantut:2012}.
An important application of this synthetic gauge field  is the realisation of spin-orbit coupling in degenerate
Fermi gases \cite{Wang:2012,Cheuk:2012}.
The  Fano-Feshbach resonances can be used to modify the strength of atomic interactions without
changing the characteristic range of the potential \cite{Williams:2012}.
Raman laser beams coupled to Zeeman states of ultracold fermionic atoms can create synthetic magnetic,
electric fields and spin-orbit coupling, see a brief review \cite{Bloch:2012,Zhai:2012}.
These new developments inspire further study of spin-orbit coupling, p-wave and d-wave pairing by
exact solutions of new mathematical models, with clear applications in physics.

(c) {\em Universal Wilson ratio.}  The low-energy physics of interacting Fermi systems exhibits universal phenomena,
such as TLL in 1D and Fermi liquids in higher dimensions.
It is highly desirable to find an intrinsic connection between these  low-energy theories.
The Wilson ratio is the ratio of susceptibility to specific heat.
Despite these two physical quantities having different low temperature behaviour,
the Wilson ratio is a constant at the fixed point of interacting fermionic systems  in 3D \cite{Wilson:1975}.
For noninteracting electrons, the Wilson ratio is unity.
The value of the ratio  indicates  interaction effects  and  quantifies spin fluctuations.
In contrast to the phenomenological TLL parameters,  this ratio gives a measurable physical quantity
which manifests universal TLL physics -- the effective fixed point model of the TLL universality class.
Experimental measurement of the  Wilson ratio of the TLL in a spin-1/2 Heisenberg ladder has been
recently reported \cite{Ninios:2012}.
It is naturally to believe that this ratio can be used to quantify the magnetic phases (the FFLO-like phases)
in 1D attractive Fermi gases of ultracold atoms with different symmetries.
We remark that at the critical point it is a constant solely dependent  on the  TLL parameter and onset
charge velocities of different  states.
This ratio is  experimentally measurable with ultracold Fermi atomic gases trapped in 1D.

(d) {\em Diffraction vs non-diffraction.} Non-diffraction in the many-particle scattering process is a unique
feature of integrable systems \cite{McGuire:1964,Gu:1989,Sutherland:2004,Lamacraft:2012}.
In principle, diffractive and non-diffractive scattering can be  tuned  via controlling
inter- and intra-species scattering lengths.
Experimental exploration of a violation of this  characteristic  would yield valuable insight into
understanding  the non-diffractive form of the Bethe ansatz wave function, i.e.,
how weak violations of non-diffractive scattering still give rise to the Bethe ansatz wave function
in the asymptotic region.
1D many-body systems with a finite range potential between atoms are likely to be an ideal
simulator for identifying the consequences of diffractive {\em vs} non-diffractive scattering.
In this regard,  across a  narrow resonance, the 1D effective potential is  determined by
not only  the scattering length, but also the effective range which is  introduced by  a
strong energy-dependent  scattering amplitude of two colliding atoms \cite{Gurarie:2007,Cui:2012a,Qi:2012}.
Beyond the fermionization of two distinguishable fermions \cite{Zurn:2012}, it is an immediate goal to
experimentally probe three distinguishable fermions in a 1D harmonic trap.
A possible test of the fundamental nature of diffraction {\em vs} non-diffraction for systems with three particles
may ultimately lead to experimental tests for Yang-Baxter integrability in quantum many-body systems.

\section*{Acknowledgments}

We thank
S. Chen, A. del Campo, A. Foerster, F. Goehmann, T.-L. Ho, Z.-Q. Ma, C.N. Yang, Y.-P. Wang and S.-Z. Zhang
for valuable comments on the manuscript.
We also thank J.-S. Caux for sharing his English translation of Gaudin's book.
This work has been supported by  the National Basic Research Program of China under Grant No. 2012CB821300 and No. 2012CB922101
as well as the Australia Research Council through grants DP1093353, DP1096713 and DP130102839.
C.~L. has also been partially supported by  the National Natural Science Foundation of China under Grant No. 11075223 and  the Program for New Century
Excellent Talents in University of the Ministry of Education of China under Grant No. NCET-10-0850.
X.-W.~G acknowledges the Institute for Advanced Study, Tsinghua University for kind hospitality when part of this paper was prepared.

This paper is dedicated to the memory of Miki Wadati (1945-2011).

\bibliographystyle{apsrmp}


\end{document}